\documentclass[fleqn,12pt,twoside]{article}
\usepackage{espcrc1}
\usepackage{times}

\usepackage{graphicx}
\usepackage[figuresright]{rotating}

\newcommand{\AmS}{{\protect\the\textfont2
  A\kern-.1667em\lower.5ex\hbox{M}\kern-.125emS}}
\title{ Restoration of chiral and $U(1)_A$ symmetries in excited
hadrons.}
\author{{ L. Ya. Glozman}\vskip5mm
  Institute for 
Physics, Theoretical Physics Branch, University of Graz, Universit\"atsplatz 5, A-8010
Graz, Austria}
\begin{document}
\maketitle

\newcommand{\be}{\begin{equation}}
\newcommand{\bea}{\begin{eqnarray}}
\newcommand{\ee}{\end{equation}}
\newcommand{\eea}{\end{eqnarray}}
\newcommand{\ds}{\displaystyle}
\newcommand{\low}[1]{\raisebox{-1mm}{$#1$}}
\newcommand{\loww}[1]{\raisebox{-1.5mm}{$#1$}}
\newcommand{\lmn}{\mathop{\sim}\limits_{n\gg 1}}
\newcommand{\vpint}{\int\makebox[0mm][r]{\bf --\hspace*{0.13cm}}}
\newcommand{\too}{\mathop{\to}\limits_{N_C\to\infty}}
\newcommand{\vp}{\varphi}
\newcommand{\vx}{{\vec x}}
\newcommand{\vy}{{\vec y}}
\newcommand{\vz}{{\vec z}}
\newcommand{\vk}{{\vec k}}
\newcommand{\vq}{{\vec q}}
\newcommand{\vpp}{{\vec p}}
\newcommand{\vn}{{\vec n}}
\newcommand{\vg}{{\vec \gamma}}

\begin{abstract}
The effective restoration of $SU(2)_L \times SU(2)_R$ and $U(1)_A$
chiral symmetries of QCD  in excited hadrons is reviewed.
While the low-lying hadron spectrum is mostly shaped by the
spontaneous breaking of chiral symmetry, in the high-lying
hadrons the role of the quark condensate of the vacuum becomes
negligible and the chiral symmetry is effectively restored. This
implies that the mass generation mechanisms in the low- and
high-lying hadrons are essentially different. The fundamental origin
of this phenomenon is a suppression of  quark quantum loop effects
in  high-lying hadrons relative to the classical contributions that
preserve both chiral and  $U(1)_A$ symmetries. Microscopically
the chiral symmetry breaking is induced by the dynamical Lorentz-scalar
mass of quarks due to their coupling with the quark condensate
of the vacuum. This
mass is strongly momentum-dependent, however, and vanishes in the
high-lying hadrons where the typical momentum of valence quarks is large.
This physics is illustrated within the solvable chirally-symmetric and
confining model. Effective Lagrangians for the approximate
chiral multiplets at the hadron level
 are constructed which can be used as phenomenological effective field
 theories in the effective chiral restoration regime. Different
 ramifications and implications of the effective chiral restoration
 for the string description of  excited hadrons, the
 decoupling of  excited hadrons from the Goldstone bosons, the
 glueball - quark-antiquark
 mixing and the OZI rule violations are discussed.
\end{abstract}


\bigskip
\bigskip
\bigskip

\noindent
{\bf Contents}
\bigskip

\noindent
1. Introduction

\noindent
2. Chiral symmetry of QCD

\noindent
3. Effective chiral and $U(1)_A$ symmetry restoration in excited hadrons
by definition

\noindent
4. Why should one expect a restoration of chiral and $U(1)_A$
symmetries in highly excited 

hadrons

\noindent
5. Simple quantum-mechanical example of effective symmetry restoration

\noindent
6. Can the quark-hadron duality and the Operator Product Expansion
predict the rate of the 

 symmetry restoration?

\noindent
7. Chiral multiplets of excited mesons

7.1 Representations of the parity-chiral group

7.2 $J=0$ mesons

7.3 $J>0$ mesons

\noindent
8. Chiral multiplets of excited baryons

\noindent
9. Generalized linear sigma-model. Mesons

\noindent
10. Generalized linear sigma-model. Baryons

\noindent
11. Solvable confining and chirally symmetric models

11.1 't Hooft model

11.2 Generalized Nambu and Jona-Lasinio model

11.3 Chiral symmetry breaking

11.4 Bethe-Salpeter equation for mesons and chiral symmetry
properties of the spectrum

11.5 The quantum origin of chiral symmetry breaking and the
semiclassical origin of effective 

\hspace{0.6cm} chiral restoration

11.6 The Lorentz structure of the effective confining single-quark
potential

11.7 Goldstone boson decoupling from the high-lying states

\noindent
12. Chiral multiplets and the string

\noindent
13. Chiral symmetry selection rules for heavy quarkonium decay into
two mesons

\noindent
14. Suppression of the glueball - $\bar n n$ mixing

\noindent
15. Conclusions and outlook

\newpage
\section{Introduction}

In the world of  light quarks there are two crucially important properties
of QCD - chiral symmetry  and confinement. While these two are subjects
of study for 30-40 years, their inter-relations and
mechanisms are not yet completely understood.

Another conceptual and closely related issue in QCD is the generation of 
the hadron mass. Historically
two different  views have been developing, 
starting from the preQCD time (and partly due to the same person!). 
The first one considers the hadron mass as originating from the spontaneous
(dynamical) breaking of chiral symmetry. The insight is best obtained
from the Gell-Mann - Levy sigma-model \cite{GL} and Nambu - Jona-Lasinio
model \cite{NJL}. In both models  a fermion (nucleon) which is  subject to the
Dirac equation, is initially  massless (in the Wigner-Weyl mode).
When  chiral symmetry is spontaneously
(dynamically) broken in the vacuum there appear  Goldstone bosons, 
$\vec \pi$,
and the chiral order parameter, the vacuum expectation value of
the sigma field, $\langle \sigma \rangle$, or the fermion condensate, 
$\langle \bar \psi \psi \rangle$ 
(in modern language it is the quark condensate). The nucleon acquires 
its dynamical mass via its coupling with the chiral order parameter. 
The coupling of the Goldstone boson
 to the nucleon is entirely due to the spontaneous breaking of chiral
 symmetry.
One of 
the most impressive
implications of this view is the Goldberger-Treiman relation,

\begin{equation}
g_{\pi N N} = \frac{g_A M_N}{f_\pi},
\label{gt}
\end{equation}

\noindent 
which connects the Goldstone boson - nucleon coupling constant with
the axial charge of the nucleon and its mass.
One has $g_{\pi N N} \simeq 13.4$ and
 $\frac{g_A M_N}{f_\pi} \simeq 12.7$, i.e. it is perfectly
 experimentally satisfied. 
 A subsequent development within QCD
 has shown that indeed the nucleon mass is mostly
induced by the quark condensate of the vacuum \cite{ioffe}.
This kind of physics has been confirmed by numerous successes of 
current algebra, chiral perturbation theory and lattice QCD calculations.

The alternative view on the mass generation is that the energy is accumulated
in the string connecting color charges.  While not so successful
for the lowest-lying hadrons in the light quark sector, this direction has
suggested an impressive phenomenology of Regge trajectories. In  case
of heavy quarks the string should be reduced to the static linear
quark-antiquark potential which is indeed observed on the lattice. When
the string connects the light quarks, chiral symmetry should be relevant,
and the dominating view was that this string could be mimicked to some
extent by the Lorentz-scalar confinement which manifestly breaks chiral
symmetry. If confinement has intrinsically the Lorentz-scalar nature,
the view shared also by the bag  and constituent quark models, the effect 
of chiral symmetry
breaking should increase higher in the light hadron spectrum.

Recent phenomenological and theoretical developments suggest, however,
that both these views are correct only partly. While the chiral
symmetry breaking effect is indeed crucially important for the lowest
hadrons, in
the highly excited light hadrons the chiral symmetry is almost restored,
even though it is strongly broken in the vacuum. The high-lying hadrons
almost entirely decouple from the quark condensate of the vacuum. This
phenomenon is referred to as effective restoration of chiral symmetry. If
correct, it implies that the mass generation mechanism for the lowest and
the high-lying hadrons is essentially different.

\begin{figure}
\begin{center}
\includegraphics[height=8cm,angle=-90,clip=]{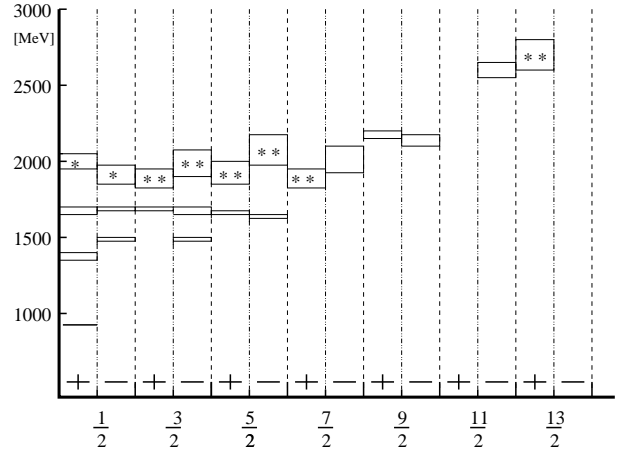}
\label{ns}
\caption{Low- and high-lying nucleons. Those states which are
not yet established are marked by ** or * according to the PDG classification.}
\label{ns}
\end{center}
\end{figure}

 When one looks carefully at the nucleon excitation spectrum, see Fig.
 \ref{ns}, one immediately notices regularities for the high-lying states
 beginning approximately from the 1.7 GeV region. The nucleon
 (and delta) high-lying states show obvious patterns of parity doubling:
 The states of the same spin but opposite parity are approximately
 degenerate. There are a few examples where such parity partners have 
 not yet been seen  experimentally. Such doublets are definitely absent
 in the low-lying spectrum. The high-lying hadron spectroscopy is
 a difficult experimental task and the high-lying spectra have never
 been systematically explored. However, it is conceptually important
 to answer the question whether the parity partners exist systematically 
 or not. If yes, and the existing data hint at it, then it would mean
 that some symmetry should be behind this parity doubling and this
 symmetry is not operative in the low-lying spectrum. What is this
 symmetry and why is it active only in the upper part of the spectrum?
It is amusing to note, that data which hint the onset of this symmetry regime
have existed for many years, but have never been in the focus.

This experimental fact has attracted the attention of Regge theorists
in the 60th. They have attempted to explain it as originating from the
so called McDowell symmetry \cite{collins} which requires that baryons
(but not mesons) lying on  linear Regge trajectories should be in 
parity doublets. On
the other hand, such parity doublets were clearly absent low in
the spectrum, a fact which could not been understood from the Regge
physics perspective. There have  been attempts to link the
parity doubling with dynamical \cite{iachello} and 
Poincar\'{e} \cite{marianna} symmetries.

Only quite recently it has been suggested that this parity doubling
in fact reflects restoration of the spontaneously broken chiral
symmetry of QCD \cite{G1}. The idea was that in the
high-lying baryons the typical momenta of valence quarks are larger
than the chiral symmetry breaking scale and hence these valence
quarks decouple from the quark condensate of the vacuum. If they decouple
from the quark condensate, then their dynamical (constituent) Lorentz-scalar
mass that breaks chiral symmetry should vanish and
the chiral symmetry should be approximately restored
in these baryons, even though it is strongly broken in the vacuum. Though
it was erroneously referred to as a "phase transition", it has been
emphasized that the present phenomenon should not be confused with
the chiral restoration phase transition at high temperature and /or
density. This mechanism of effective chiral
restoration is correct, as it can be seen from the present day perspective,
 though many important details were missing in
that paper.
This idea has obtained 
a further impetus in refs. \cite{CG1,CG2},
where a classification of the representations as well as possible
assignments of the high-lying baryons have been performed. Also an
attempt to relate effective chiral restoration in excited baryons
with the short-range part of the correlation functions has been done.
Still,
it was a puzzle why the effective chiral restoration was observed
in the high-lying baryons while it was not seen in the high-lying mesons.

That the low-lying hadrons could be arranged into linear chiral multiplets
which are  strongly split by the quark condensate of the vacuum,
has been discussed in the context of a certain class of effective
Lagrangians in refs. \cite{DETAR,TIT,JIDO,novak,bardeen}. However,
when the chiral symmetry breaking effect is very strong, as it is in the
low-lying hadrons, only the nonlinear realization of chiral symmetry
is relevant, which does not transform the hadrons of opposite parity
into each other and hence does  not require a one-to-one mapping
of hadrons of opposite parity \cite{jaffe,JPS}. Indeed, 
 such a mapping is missing in the
low-lying spectrum, which is well explored experimentally. This
implies that it is not possible to arrange the low-lying baryons into
systematic linear chiral representations.

Actually, it is understood both phenomenologically
and theoretically that the approximate symmetry of
the lowest baryons is $SU(6)_{FS} \times O(3)$. The contracted
$SU(6)_{FS}$ symmetry of the ground state baryons is an exact result
of QCD in the large $N_c$ limit \cite{DM}. From the microscopical
point of view the emergence of this approximate symmetry is related
to the dynamical chiral symmetry breaking. Indeed, at low momenta
the valence quarks are strongly coupled to the quark condensate
and consequently acquire a large Lorentz-scalar dynamical mass.
When the dynamical (constituent) mass of valence quarks is large they can
be considered as quasi-nonrelativistic quasi-particles. Then without
any residual interactions the symmetry is $SU(6)_{FS} \times O(3)$.
The axial vector current conservation requires that such
constituent quarks are strongly coupled to the Goldstone bosons.
 Then
one of the most important residual interactions between valence
quarks in the low-lying baryons is mediated by the flavor-spin dependent
Goldstone boson fields \cite{GR}. From the present perspective
it is important that the  $SU(6)_{FS} \times O(3)$ symmetry is not
compatible with the linear realization of chiral symmetry: There is no
one-to-one mapping of the positive and negative parity states within
$SU(6)_{FS} \times O(3)$. This implies that only the nonlinear
realization of the chiral symmetry can be considered in the low-lying
baryons. So the baryon spectrum represents a smooth
transition from the approximate
$SU(6)_{FS} \times O(3)$ symmetry in its lowest part, that is consistent
with the nonlinear realization of chiral symmetry,  to the approximate
linearly realized chiral symmetry in its upper part.

A dramatic development in the field happened when  results of
the partial wave analysis of the proton-antiproton annihilation into
mesons at LEAR in the energy range 1.9 - 2.4 GeV have been published
\cite{BUGG1,BUGG2}. This kinematic region had  been 
practically virgin
and a lot of new mesons have been discovered. It is not a surprise:
The formation experiment with the  partial wave analysis of decays of the
 intermediate mesons into many possible channels is the only {\it systematic}
 way to explore excited hadrons in  large kinematic intervals. 
 Historically exactly in the same way
 excited baryons have been obtained  in the 
 $\pi N$ formation experiments. These results have been analyzed in
 refs. \cite{G2,G3} and it turned out that the high-lying $\bar n n$
 mesons perfectly fit all possible linear chiral multiplets of both
 $SU(2)_L \times SU(2)_R$ and $U(1)_A$ groups with a few still
 missing states. Even though most of the states are quite reliable,
 because they are seen in a few different decay channels, an
 independent experiment must be performed to reconfirm these new
 states and to find still missing ones. Until then these new states
 cannot appear in the meson summary tables of PDG. One of the best
 possibilities  will be the low-energy antiproton ring at GSI with
 characteristics similar or better than LEAR.

\begin{figure}
\begin{center}
\includegraphics[height=8cm,,clip=]{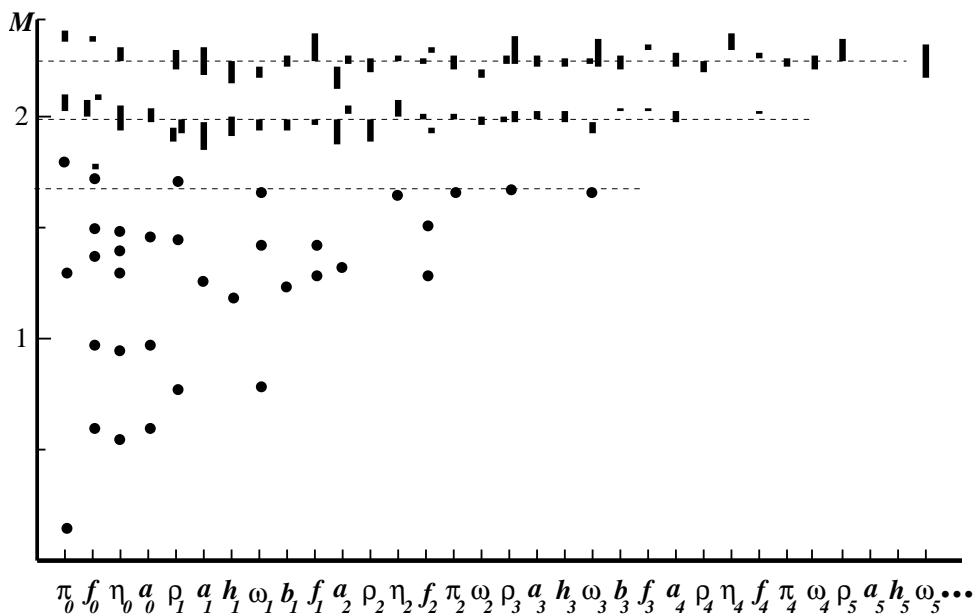}
\label{ms}
\caption{Masses (in GeV) of the well established  states from PDG 
(circles) and 
new $\bar n n$
states   from the proton-antiproton annihilation (strips). Note
that the well-established states include $f_0(1500), f_0(1710)$, which
are the glueball and $\bar s s$ states with some mixing and hence are
irrelevant from the chiral symmetry point of view. Similar, the
 $f_0(980), a_0(980)$ mesons most probably are not $\bar n n$ states and
 also should be excluded from the consideration. The same is true for
 $\eta(1475)$, which is the $\bar s s$ state and  $\eta(1405)$ with
 the unknown nature.}
\label{lear}
\end{center}
\end{figure}
 
These new results along with the well established states from the PDG
are shown in Fig. \ref{lear}. It is well seen that the states of the same spin
with different isospins and opposite parities are approximately
degenerate in the interval 1.7 - 2.4 GeV. Even more, the states
with different spins are also approximately degenerate. A similar feature
is also well seen in Fig. \ref{ns}.
This indicates
that a larger symmetry is observed that includes chiral 
$SU(2)_L \times SU(2)_R$ and $U(1)_A$ as subgroups.
Most probably this additional degeneracy reflects a dynamical
symmetry of the string. The latter conjecture is supported by the observation
that the $\bar n n$ meson spectrum can be separated into a few clusters and
the energy of each cluster is well described with a simple relation
$M^2(n) = an + b$ \cite{afonin}, that is typical for the open string. 
Such a spectrum neatly fits a view that the highly excited hadrons
can be considered as strings with quarks at the ends that have a definite 
chirality \cite{G4}.
While a tendency is well seen, there are a lot of missing states to be
discovered!

Actually it could had been expected that the physics of the low- and
highly-excited hadrons is very different, because it is quite
natural in complex many-body systems which hadrons are. Remember
that in Landau's fermi-liquid theory the quasiparticle degrees of
freedom are relevant only to the low-lying excitations while the
high-lying levels are excitations of bare particles.

It is important to precisely characterize what is implied under effective
restoration of chiral and $U(1)_A$ symmetry in excited hadrons. A 
mode of symmetry is defined only by the properties of the vacuum.
If a symmetry is spontaneously broken in the vacuum, then it is the
Nambu-Goldstone mode and the whole spectrum of excitations on the
top of the vacuum is in the Nambu-Goldstone mode. However, it may happen
that the role of the chiral symmetry breaking condensates becomes
progressively less important higher in the spectrum, because the
valence quarks decouple from the quark condensates. This means
that the chiral symmetry breaking effects become less and less
important in the highly excited states and asymptotically the
states approach the regime where their properties are determined
by the underlying unbroken chiral symmetry (i.e. by the symmetry
in the Wigner-Weyl mode). This effective restoration 
in excited hadrons should not
be confused with the chiral symmetry restoration in the vacuum at
high temperature/density. In the latter case the quark vacuum becomes
trivial and the system is in the Wigner-Weyl mode. In the former case
the symmetry is always broken in the vacuum, however this symmetry breaking
in the vacuum gets irrelevant in the highly excited states.

Turning to the mechanisms of effective restoration of chiral symmetry,
outline first the most fundamental one  \cite{G}. In the 
high-lying hadrons the semiclassical regime must be manifest. Semiclassically
the quantum loop effects are suppressed. Both chiral and $U(1)_A$
symmetry breakings are loop effects and hence must be suppressed in
the high-lying hadrons where classical contributions dominate. This
general claim has been illustrated \cite{GNR} within the manifestly chirally
symmetric and confining model, which is exactly solvable \cite{orsay}.
This model, which belongs  to the class of large $N_c$
models, can be used as a laboratory to get an insight. The
effective chiral restoration in the high-lying heavy-light mesons
with the quadratic confining potential as well as  a Lorentz 
structure of the effective confining
potential have been studied within this model in ref. \cite{KNR}. 
A complete spectrum of the light-light mesons
with the linear potential,
exhibiting the effective chiral restoration 
has been calculated in ref. \cite{WG1,WG2}. 
A form of the Regge
trajectories as well as  properties of the meson wave functions
have also been  studied. It is  possible to see directly a
mechanism of the effective chiral restoration. The chiral symmetry
breaking Lorentz-scalar dynamical mass of quarks arises via selfinteraction
loops  and vanishes at large momenta. When one increases excitation 
energy of a hadron, one also increases the typical momentum of
valence quarks. Consequently, the chiral symmetry violating dynamical
mass of quarks becomes small and chiral symmetry gets approximately
restored.

Restoration of the chiral symmetry in a hadron requires this hadron to
decouple from the Goldstone bosons, $g_{\pi N} \simeq 0$ 
\cite{G4,CG3,GN,jaffe,JPS}.
This in particular requires that the axial charge of the hadron
must also be very small,  $g_{A} \simeq 0$. This is a very interesting
prediction that could be tested experimentally. A hint for such a
decoupling is well seen experimentally: The coupling constant of the
excited hadron decay into the ground state hadron and the pion decreases
very fast higher in the spectrum (because the phase space factor
increases much faster than the width).

There are also other interesting implications of the chiral symmetry 
restoration. Among them is a suppression of the glueball - $\bar n n$ meson
mixing \cite{Gglue} and a strong violation of the OZI rule in $J/\Psi$
decays \cite{GOZI}.

The other direction is to formulate the problem on the lattice
\cite{DEGRAND,coh}. While this conceptually could be important, the
extraction of the high-lying hadrons on the lattice is an intrinsically
difficult problem. This is because the signals from  the excited states in
the Euclidean two-point correlation function are exponentially 
suppressed relative the low-lying states. While it is feasible,
though demanding, to
extract the  first excitations \cite{BURCH,LHC},
most likely the high-lying states in the light flavor
sectors could not be studied on the lattice. They would require
huge lattice volumes, because these hadrons are big, and unrealistically
 high statistics
in the vicinity of the chiral limit, because the signal from these states
is suppressed. This implies that the field requires other methods,
including analytical ones, and 
modeling.

This review is organized as follows. In Sec. 2, we give a short
overview of chiral symmetry in QCD. Sec. 3 contains a definition
of the effective chiral and $U(1)_A$ restorations in excited hadrons.
The most general origin of the effective
chiral and $U(1)_A$ restorations in excited hadrons is discussed
in Sec. 4. In Sec. 5,
we present a simple and pedagogical quantum-mechanical example
of effective symmetry restoration. In a very short Section 6, we
explain that an approximate knowledge of the two-point correlation
function from the operator product expansion at large space-like
momenta is not sufficient to obtain high-lying spectra. Chiral
classification and available experimental information
 of excited mesons and baryons are discussed in Sections 7 and 8,
 respectively. In sections 9 and 10, we construct simple effective
 Lagrangians at the hadron level that exhibit effective chiral
 restoration. These kinds of Lagrangians can be used as a basis for
 an effective field theory of approximate chiral multiplets.
We also discuss in these sections a fundamental symmetry reason
for Goldstone boson decoupling from the approximate chiral multiplets.
 Sec. 11 is devoted to solvable confining and chirally
 symmetric models that provide direct microscopical insight into the
 phenomenon of effective chiral restoration. In Sec. 12, we discuss
 possible implications of the string description of the high-lying
 hadrons. A violation of the OZI rule for heavy quarkonium decay
 and suppression of the glueball - usual meson mixings as originating 
 from the chiral restoration are discussed in Sections 13 and 14. Finally
 we present a general outlook in the concluding section.

\section{Chiral symmetry of QCD}

Consider the chiral limit where quarks are massless. It
is definitely justified for  $u$ and $d$ quarks since
their masses are quite small  compared to $\Lambda_{QCD}$ and
the typical hadronic scale of 1 GeV; in good approximation
they can be neglected. Define the right- and left-handed components
of quark fields

\begin{equation}
\psi_R = \frac{1}{2}\left( 1+\gamma_5 \right ) \psi,~~
\psi_L = \frac{1}{2}\left( 1-\gamma_5 \right ) \psi.
\label{RL}
\end{equation}

\begin{figure}
\begin{center}
\includegraphics*[width=5cm]{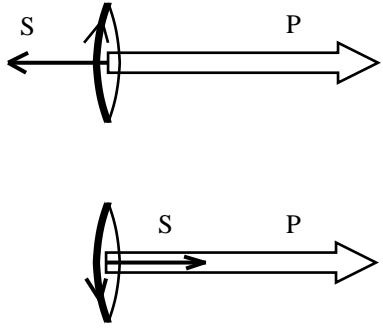}
\end{center}
\caption{Left-handed and right-handed massless fermions.}
\label{hel}
\end{figure}

\noindent
If there is no interaction, then the right- and left-handed
components of the quark field get decoupled, as it is well
seen from the kinetic energy term

\begin{equation}
{\cal L}_0 = i \bar \Psi \gamma_\mu \partial^\mu \Psi =
i \bar \Psi_L \gamma_\mu \partial^\mu \Psi_L +
i \bar \Psi_R \gamma_\mu \partial^\mu \Psi_R,
\label{L0}
\end{equation}

\noindent
see Fig. ~\ref{hel}. 

In QCD the quark-gluon interaction Lagrangian
is vectorial, 
$\bar \psi \gamma^\mu \psi A_\mu$, which does not mix the right- and
left-handed components of quark fields. Hence in the chiral limit
 the left- and right-handed components of
quarks are completely decoupled in the QCD Lagrangian. Then,
assuming only one flavor of quarks such a Lagrangian is invariant
under two independent global variations of phases of the
left-handed and right-handed quarks:

\begin{equation}
\psi_R \rightarrow 
\exp \left( \imath \theta_R \right)\psi_R; ~~
\psi_L \rightarrow 
\exp \left( \imath \theta_L\right)\psi_L.
\label{CH}
\end{equation}

\noindent
Such a transformation can be identically rewritten in terms
of the vectorial and axial transformations: 

\begin{equation}
\psi\rightarrow 
\exp \left( \imath \theta_V \right)\psi; ~~
\psi \rightarrow 
\exp \left( \imath \theta_A  \gamma_5 \right)\psi.
\label{VAS}
\end{equation}

\noindent
The symmetry group of these  phase transformations is

\begin{equation}
U(1)_L \times U(1)_R = U(1)_A \times U(1)_V.
\label{VASG}
\end{equation}

Consider now the chiral limit for two flavors, $u$ and $d$. 
The quark-gluon interaction Lagrangian is insensitive
to the specific flavor of quarks.  For example, one can substitute
 the $u$ and $d$ quarks by  properly normalized orthogonal 
linear combinations of $u$ and $d$
quarks ( i.e. one can perform a rotation in the isospin space)
and nothing will change. Since the left- and right-handed components
are completely decoupled, one can perform two independent isospin
rotations of the left- and right-handed components:

\begin{equation}
\psi_R \rightarrow 
\exp \left( \imath \frac{\theta^a_R \tau^a}{2}\right)\psi_R; ~~
\psi_L \rightarrow 
\exp \left( \imath \frac{\theta^a_L\tau^a}{2}\right)\psi_L,
\label{ROT}
\end{equation}

\noindent
where $\tau^a$ are the isospin Pauli matrices and the
angles $\theta^a_R$ and $\theta^a_L$ parameterize rotations
of the right- and left-handed components. These rotations leave
the QCD Lagrangian invariant. The symmetry group of these
transformations,

\begin{equation}
SU(2)_L \times SU(2)_R,
\label{chsymm}
\end{equation}

\noindent
is called chiral symmetry.

Actually in this case the Lagrangian is also invariant under
the variation of the common phase of the left-handed $u_L$ and $d_L$ quarks, 
which is the $U(1)_L$ symmetry and similarly - for the right-handed 
quarks. Hence the total chiral symmetry group of the QCD Lagrangian
is

\begin{equation}
U(2)_L \times U(2)_R 
= SU(2)_L \times SU(2)_R \times U(1)_V \times U(1)_A.
\label{tot}
\end{equation} 

\noindent

If one includes into this consideration the third flavor, the $s$ quark,
and neglects its mass, then one obtains the following chiral
symmetry group

\begin{equation}
U(3)_L \times U(3)_R 
= SU(3)_L \times SU(3)_R \times U(1)_V \times U(1)_A.
\label{totsu3}
\end{equation} 

\noindent
However,  the mass of the $s$ quark is of the order 100 MeV, i.e. it is
of the same order as $\Lambda_{QCD}$. Hence the $U(3)_L \times U(3)_R$
symmetry is strongly explicitly broken. Still this symmetry has 
important implications for the lowest pseudoscalar mesons. In this review
we limit ourselves to the two-flavor version of QCD, because the
$U(2)_L \times U(2)_R$ symmetry of the QCD Lagrangian is nearly perfect.
This is not the case if the $s$ quark is included, and a-priori it is
not clear whether one should regard this quark as light or "heavy" for our
present applications of symmetry in the high-lying hadrons. The second
reason is a practical one - there are good data on highly excited $u,d$
hadrons but such data are still missing for the strange excited hadrons.
Certainly it would be very interesting and important to extend the analysis
of the present report to the $U(3)_L \times U(3)_R$ case. We hope that
the present results will stimulate the experimental and theoretical work
in this direction.

The $U(2)_L \times U(2)_R$ is a symmetry of the QCD Lagrangian at the classical level,
i.e. before the second quantization of the theory.
The $U(1)_V$ symmetry is responsible for the baryon number conservation
and will not be discussed any longer.
The $U(1)_A$ symmetry of the classical Lagrangian is
explicitly broken  at the quantum level, 
i.e. once the second quantization of the theory is performed.
This is the famous 
axial anomaly, which appears due to quantum fluctuations 
of the quark field, specifically - due to the 
vacuum fermion loops \cite{ANOMALY}.

\begin{figure}
\begin{center}
\includegraphics*[width=3cm]{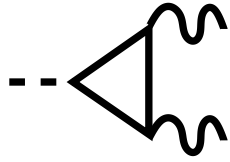}
\caption{Vacuum fermion loop contributing to the axial anomaly.}
\end{center}
\label{anfig}
\end{figure}

It is instructive to remind the reader of the essence of the axial
anomaly. For the free Dirac field the divergence of the flavor-singlet
axial current, $A_\mu(x) = \bar \psi (x) \gamma_\mu \gamma_5 \psi (x)$,
is given as

\begin{equation}
\partial^\mu A_\mu(x) = 2 i m \bar \psi (x) \gamma_5 \psi (x),
\label{pa}
\end{equation}

\noindent
where $m$ is the fermion mass. Hence the flavor-singlet axial
current for the free fermion fields is conserved in the chiral
limit.
However, once a coupling of the
fermion field and the gauge field is included, then the
vacuum triangle fermion loop of Fig. \ref{anfig} invalidates the equation
(\ref{pa}) and instead the divergence of the flavor-singlet
axial current becomes

\begin{equation}
\partial^\mu A_\mu(x) = 2 i m \bar \psi (x) \gamma_5 \psi (x)
+ \frac{\alpha_s}{(4\pi)} N_f  Tr F_{\lambda \nu}(x)
\tilde {F}_{\lambda \nu}(x),
\label{an}
\end{equation}

\noindent
where $F_{\lambda \nu}(x)$ is the gluon field tensor, $N_f$ is
the number of active flavors and $Tr$ is assumed in color space. 
The extra
term is produced by the renormalization. The anomaly 
is not affected by  higher-order
radiative corrections. The physical meaning of the anomaly is
that the flavor-singlet axial current is connected with the gauge
field and can mix with it.
Hence the $U(1)_A$ symmetry of the classical
Lagrangian is explicitly broken by the anomaly at the quantum
level and the flavor-singlet axial vector current
is not conserved in the chiral limit.
This explicit breaking of $U(1)_A$ is an effect of quantum
fluctuations of the fermion field, because the vacuum fermion loop
which produces the anomaly is a quantum fluctuation. At the formal
level this can be seen from the fact that the one-loop diagram
is proportional to $\hbar$. In the path-integral quantization formalism
the axial anomaly appears from the non-invariance of the path-integral
measure under the axial transformation \cite{FU}. The most visible effect of
the anomaly is that the lowest flavor-singlet pseudoscalar meson,
which in the 3-flavor case is the $\eta'$ meson, is not a (pseudo)Goldstone
boson, that would be required by the spontaneously broken
$U(3)_L \times U(3)_R$ symmetry \cite{WE1,H,WV}. Its mass is shifted
up due to the mixture of the flavor-singlet 
$\frac{1}{\sqrt 3}(\bar u u +\bar d d + \bar s s)$
valence quark pair with
the gluonic field via the vacuum fermionic loops.

Consider now the chiral group $SU(2)_L \times SU(2)_R$.
Generally if the Hamiltonian of a system is invariant under
some transformation group $G$, then one can expect
that one can find states which are simultaneously eigenstates
of the Hamiltonian and of the Casimir operators of the group, $C_i$.
If the ground
state of the theory, the vacuum, is invariant under the same
group,  i.e.  if for all $U \in G$
\begin{equation}
 U | 0 \rangle = | 0 \rangle ,
\label{vac}
\label{symvaccond}\end{equation}
then eigenstates of this Hamiltonian corresponding to excitations
above the vacuum can be grouped into degenerate multiplets corresponding
to the particular representations of $G$. 
 This mode of symmetry is
usually referred to as the Wigner-Weyl mode.  Conversely, if 
(\ref{symvaccond}) does not hold,  the excitations do not
generally form degenerate multiplets in this case.  This situation
is called spontaneous symmetry breaking.

If chiral $SU(2)_L \times SU(2)_R$
symmetry were realized in the Wigner-Weyl mode, then the excitations
would be grouped into representations of the chiral group. 
The representations of the chiral group are discussed in
detail in   the following sections. The important
feature is that the every representation except the trivial one 
 necessarily implies
parity doubling for hadrons with nonzero masses. 
In other words, for every baryon with
 given quantum numbers and parity, there must exist another
baryon with the same quantum numbers but opposite parity 
which must have the same mass.
In the case of mesons the chiral representations combine, e.g.
the pions with the $\bar n n= \frac{\bar u u + \bar d d}{\sqrt 2} $  
$f_0$ mesons, which should
be degenerate. This feature is definitely not observed
for the low-lying states in hadron spectra. This means that
Eq.~(\ref{symvaccond}) does not apply;
the continuous chiral symmetry of the QCD Lagrangian is spontaneously 
(dynamically)
broken in the vacuum, i.e.  it is hidden. Such a mode
of symmetry realization is referred to as the Nambu-Goldstone one.

The independent left and right rotations (\ref{ROT}) can be
represented equivalently  with independent isospin and axial 
rotations

\begin{equation}
\psi \rightarrow 
\exp \left( \imath \frac{\theta^a_V \tau^a}{2}\right)\psi; ~~
\psi \rightarrow 
\exp \left( \imath  \gamma_5 \frac{\theta^a_A\tau^a}{2}\right)\psi.
\label{VA}
\end{equation}

\noindent
The existence of 
 approximately degenerate isospin multiplets
in hadron spectra suggests that the vacuum is invariant under
the isospin transformation. Indeed, from the theoretical side the Vafa-Witten
theorem \cite{VW} guarantees that  in  local gauge theories
the vector part of the chiral symmetry cannot be spontaneously broken.
The axial  transformation mixes
states with opposite parity. The fact that the low-lying states
do not have degenerate chiral partners
implies that the vacuum is not invariant under the
axial  transformations. In other words the almost perfect
chiral symmetry of the QCD Lagrangian is dynamically broken
 by the vacuum down to the vectorial (isospin) subgroup

\begin{equation}
SU(2)_L \times SU(2)_R \rightarrow SU(2)_I.
\label{breaking}
\end{equation}

The non-invariance of the vacuum with respect to the three axial
 transformations requires existence of three massless
Goldstone bosons, which should be pseudoscalars and form an
isospin triplet. These are identified with the pions. The nonzero
mass of the pions is entirely due to the {\it explicit} chiral symmetry breaking
by the small masses of $u$ and $d$ quarks. These small masses can
be accounted for as a perturbation. As a result the squares of
the pion masses are proportional to the  $u$ and $d$ quark masses
\cite{GOR}

\begin{equation}
m_{\pi}^2 = -\frac{1}{f_{\pi}^2} \frac{m_u + m_d}{2} 
(\langle \bar u u \rangle + \langle \bar d d \rangle) + O (m_{u,d}^2).
\label{gor1}
\end{equation}

That the vacuum is not invariant under the axial transformation
is directly seen from the nonzero values of the quark condensates,
which are  order parameters for spontaneous chiral
symmetry breaking. These condensates are the vacuum expectation
values of the $\bar \psi \psi = \bar \psi_L \psi_R + \bar \psi_R \psi_L$
operator and at the renormalization scale of 1 GeV they approximately are

\begin{equation}
\langle \bar u u \rangle  \simeq \langle \bar d d \rangle  \simeq -
 (240 \pm 10 MeV)^3.
\label{con}
\end{equation}

\noindent
The values above are deduced from phenomenological considerations \cite{L}.
Lattice
gauge calculations also confirm the nonzero and rather large values
for quark condensates. However, the quark condensates above are not
the only order parameters for chiral symmetry breaking. There exist
chiral condensates of higher dimension (vacuum expectation values of
more complicated combinations of  $\bar \psi$  and $ \psi$ that are
not invariant under the axial transformations). Their numerical values
are difficult to extract from phenomenological data, however, and
they are still unknown.

In general
spontaneous breaking of chiral symmetry can  proceed
classically, like, e.g. in the linear $\sigma$-model, or at the
quantum level.
The spontaneous breaking of chiral symmetry in QCD is a pure
quantum effect based upon quantum fluctuations of the fermion
fields \cite{G}.  To see the latter we remind the reader that the chiral symmetry
breaking in QCD can be formulated via the Schwinger-Dyson (gap) equation.
It is not yet clear  at all which specific gluonic interactions
are the most important ones as a kernel of the Schwinger-Dyson
equation. It could be perturbative gluon exchanges \cite{A}, instantons
\cite{SD}, other topological configurations, a Lorentz-vector confining
potential \cite{orsay}, or a combination of different interactions.
But in any case it proceeds in QCD via the quantum fluctuations of the
fermion fields \cite{G}. Most generally it can be seen from
the definition of the quark condensate, which is a closed quark
loop. As any loop in field theory, it explicitly contains a factor
$\hbar$

\begin{equation}
\langle \bar \psi \psi \rangle = - Tr \lim_{x \rightarrow 0_+}
\langle 0 | T{\psi(0) \bar \psi (x)}| 0 \rangle ~ \sim \hbar.
\label{con}
\end{equation}

\noindent
The chiral symmetry breaking, which is necessarily a non-perturbative
effect, is actually a (nonlocal) coupling of a quark line with
the closed quark loop, which is a tadpole-like graph. 
Hence it
always contains an extra factor $\hbar$ as compared to the tree-level
quark line. A dependence of the solution of the Schwinger-Dyson
equation on  $\hbar$ is non-analytic, however \cite{GNR}. This will
be discussed in detail in one of the following sections.
The very fact that the spontaneous breaking of chiral symmetry in QCD
proceeds via the quantum fluctuations of the quark fields can be
seen microscopically in any known model for chiral symmetry breaking
in QCD. For example, the effective 't Hooft determinant interaction
\cite{H} is obtained upon integrating of the $U(1)_A$ anomaly,
which is manifestly an effect of quantum fluctuations of the quark field.
More explicitly,  chiral symmetry breaking by instantons proceeds
via annihilations by the instanton of the left-handed quarks and
creations instead of the right-handed antiquarks, and vice versa.

The quark condensate of the vacuum breaks not only the chiral
$SU(2)_L \times SU(2)_R$ symmetry, but also the $U(1)_A$ one.
Indeed,  the composite
$\bar \psi(x) \psi(x)$ field is not invariant
with respect to both
the
flavor-singlet axial rotation (\ref{VAS}) and the axial
rotation in the isospin  space (\ref{VA}) 

\begin{equation}
U(1)_A: ~~~~~~~~ \bar \psi(x) \psi(x) \rightarrow 
\cos 2\Theta_A \bar \psi(x) \psi(x) + i \sin 2\Theta_A
\bar \psi(x) \gamma_5 \psi(x),
\label{trax}
\end{equation}

\begin{equation}
``SU(2)_A": ~~~~~~~~ \bar \psi(x) \psi(x) \rightarrow 
\cos \Theta_A \bar \psi(x) \psi(x) + i \sin \Theta_A
\frac{\vec \Theta_A}{\Theta_A}  \cdot \bar \psi(x)\vec \tau \gamma_5 \psi(x),
\label{trax2}
\end{equation}

\noindent
where the inverted commas remind the reader that the axial
transformations in the isospin space do not form a group.
The vacuum expectation value of the pseudoscalar fields
$\bar \psi(x) \gamma_5 \psi(x)$ and $\bar \psi(x)\vec \tau \gamma_5 \psi(x)$
is necessarily zero because of the positive parity of the vacuum.
Then the transformation properties of the quark condensate
of the vacuum with respect to the singlet and isospin axial
rotations are

\begin{equation}
U(1)_A: ~~~~~~~~ \langle 0 | \bar \psi(x) \psi(x) | 0 \rangle
 \rightarrow 
\cos 2\Theta_A  \langle 0 | \bar \psi(x) \psi(x) | 0 \rangle,
\label{vtrax}
\end{equation}

\begin{equation}
``SU(2)_A": ~~~~~~~~ \langle 0 | \bar \psi(x) \psi(x) | 0 \rangle
 \rightarrow 
\cos \Theta_A  \langle 0 | \bar \psi(x) \psi(x) | 0 \rangle.
\label{vtrax2}
\end{equation}

\noindent
Obviously, these relations can be satisfied for an arbitrary
rotation angle only if the quark condensate of the vacuum
is identically zero. The nonzero quark condensate violates
{\it both} $``SU(2)_A"$ and $U(1)_A$ symmetries. This is important
to remember, because even if there were no axial anomaly,
or it were suppressed due to some reasons, like it is in the
large $N_c$ limit,
 then
there still would be no multiplets of the $U(1)_A$ group in the hadron
spectrum, because this symmetry is violated not only by the anomaly,
but also by the quark condensate of the vacuum.

\section{Effective chiral and $U(1)_A$ symmetry restoration in excited
hadrons by definition}

By definition effective symmetry restoration means the following. In QCD 
hadrons with  quantum numbers $\alpha$ are created
when one applies the  interpolating field (current) $J_\alpha(x)$
with such quantum numbers  on the vacuum 
$|0\rangle$. Then all 
hadrons that are created by the given interpolator
appear as intermediate states in the two-point correlator

\begin{equation}
\Pi_{J_\alpha}(q) = i \int d^4x~ e^{-iqx}
\langle 0 | T \left \{ J_\alpha(x) J_\alpha(0)^\dagger\right \} | 0 \rangle,
\label{corr}
\end{equation}

\noindent
where all possible Lorentz and Dirac indices ( specific for
a given interpolating field) have been omitted.
Consider two  interpolating fields  $J_1(x)$ and 
$J_2(x)$ which are connected by a chiral transformation\footnote{The
interpolators need not be necessarily the local operators. The same
definition can be trivially generalized to the nonlocal operators.}

\begin{equation}
J_1(x) = U J_2(x) U^\dagger,
\end{equation}

\noindent
where 

\begin{equation}
U \in SU(2)_L \times SU(2)_R
\end{equation}

\noindent
(or by the $U(1)_A$ transformation). Then if the theory were in 
the Wigner-Weyl mode, i.e. a vacuum were
invariant under the chiral group, 

\begin{equation}
U|0\rangle = |0\rangle,
\end{equation}

\noindent
it would follow from (\ref{corr}) that the spectra created by the
operators  $J_1(x)$ and  $J_2(x)$ would be identical. 
We know that in QCD in the Nambu-Goldstone mode  one finds

\begin{equation}
U|0\rangle \neq |0\rangle.
\end{equation}

\noindent
 As a consequence the spectra of the two operators
must be in general different 
 and we do not observe any chiral
or $U(1)_A$ multiplets in the low-lying hadron spectra.
 However,
it happens that the non-invariance of the vacuum becomes
unimportant (irrelevant) high in the spectrum. Then the masses
of the corresponding opposite parity hadrons, which 
are the members of the given parity-chiral multiplet,
become close  at large $s$ 
(and identical asymptotically high),

\begin{equation}
M_1 - M_2 \rightarrow 0.
\label{dm}
\end{equation}

We need to specify more precisely what it means that the masses
within the chiral multiplets become close. This is defined to occur
if (i) the states fall into approximate multiplets of 
$SU(2)_L \times SU(2)_R$ and $U(1)_A$ and the splittings between
the members of the multiplets vanish
at $J \rightarrow \infty$ and/or $n \rightarrow \infty$; (ii) the
splittings within the multiplets are much smaller than between 
two subsequent multiplets.

We stress that this effective 
 symmetry
restoration does not mean that the chiral symmetry breaking in
the vacuum disappears, but  that the role of the quark
condensates that break chiral symmetry in the vacuum becomes progressively
less important high in the spectrum. One could say that the valence
quarks in high-lying hadrons {\it decouple} from the quark condensate
of the QCD vacuum.

There are two complementary quantitative characteristics of the
effective symmetry restoration.
Define the {\it chiral asymmetry} as 

\begin{equation}
\chi = \frac{|M_1 - M_2|}{(M_1+M_2)},
\label{chir}
\end{equation}

\noindent
where $M_1$ and $M_2$ are masses of particles within the
same multiplet. This parameter  has the interpretation of the part
 of the hadron mass  due to the chiral
symmetry breaking. 
Then a "goodness" of the symmetry in the spectrum is specified by
the {\it spectral overlap} ($So$), which is defined as
the ratio of the splitting within the multiplet to the distance
between centers of gravity of two subsequent multiplets. Clearly,
the symmetry is "good" and can be easily recognized
 if this parameter is much smaller than 1.

As will be shown later on, for the high-lying hadrons the chiral
asymmetry is typically smaller than 0.02. At the same time
the  spectral overlap is typically below 0.2. These results
imply that practically the {\it whole} mass of these hadrons
is not due to the dynamical symmetry breaking in the
vacuum. The mass has a different origin.

This has to be contrasted with the low-lying hadrons,
where it is known that at least the largest part of the hadron mass 
is directly related
to the chiral symmetry breaking. 
When the chiral symmetry breaking effect is so strong 
it makes no sense to try to identify  chiral partners. The
whole notion of the chiral partner implies that the symmetry breaking
effect is weak and can be considered as a perturbation. Only in the
latter case  one can  expect a one-to-one correspondence in the
spectrum of the positive and negative parity states and identify them
as approximate linear chiral multiplets. In the low-lying part of
the hadron spectrum the chiral symmetry breaking effect is very strong,
consequently there is no the one-to-one mapping of states with opposite
parity and it is impossible to identify here systematic
linear chiral multiplets.
Only the nonlinear realization of chiral symmetry can be considered in 
this part of the spectrum which does not require the existence of the
linear chiral partners \cite{jaffe,JPS}.

\section{Why should one expect a restoration of chiral and $U(1)_A$
symmetries in highly excited hadrons}

Here we discuss the most general and fundamental reason for
chiral and $U(1)_A$ restorations in excited hadrons \cite{G}.

The whole content of a field theory and of  quantum mechanics (which is 
field theory in 0+1 dimensions) is determined by the path (functional)
integral. Most importantly the path integral prescribes the
quantum mechanical interference of all possible amplitudes (paths) that
could contribute into  the given quantity. The contribution of a given
path to the path integral is regulated by the classical action $S(\phi(x))$
along the path $\phi(x)$ through the weight factor 

\begin{equation}
e^{i \frac{S(\phi(x))}{\hbar}},
\label{weight}
\end{equation}

\noindent
where $\phi(x)$ denotes collectively all possible fields participating
in the given theory. Then the theory is completely specified by its
partition function

\begin{equation}
Z = \int D \phi e^{i \frac{S(\phi)}{\hbar}}.
\label{partition}
\end{equation}

One of the intrinsic advantages of the path integral formulation is
that it transparently exhibits the transition to the classical limit.
If the relevant typical action in the system, $S_0$, is much larger
than the Planck constant,

\begin{equation}
S_0 \gg \hbar, 
\label{sem}
\end{equation}

\noindent
then the path integral can be evaluated with the stationary phase
approximation (in Euclidean it is a steepest descent or saddle point
approximation). The essence of this approximation is that in the limiting
case $\hbar \rightarrow 0$ only the classical path, $\phi_{cl}(x)$, contributes, 
which provides an extremum of the action. This classical path is a
solution of the classical Euler-Lagrange equation

\begin{equation}
\partial_\mu \frac{\delta \cal L}{\delta(\partial_\mu \phi_{cl})} -
\frac{\delta \cal L}{\delta \phi_{cl}} = 0.
\label{LE}
\end{equation}

\noindent
This classical path is the only allowed trajectory in the classical
limit. All other paths, which are usually called "quantum fluctuations",
cancel each other exactly.

If the ratio $S_0/\hbar$ is large but finite, then the semiclassical
expansion can be performed. Typically it is valid in excited quantum
systems with  large angular momentum or radial quantum number.
In this case a crucial contribution
is provided by the classical path and all those paths that are 
infinitesimally close to the classical path. Contributions of
the quantum fluctuations are suppressed and can be taken into account
as power $\hbar/S_0$ corrections. Then the generating
functional can be expanded as

\begin{equation}
W(J) = W_0(J) + \hbar W_1(J) + ...,
\label{f}
\end{equation}

\noindent
where $W_0(J) = S(\phi_{cl}) + J\phi_{cl}$, $J$ is a source,
 and $W_1(J)$
represents contributions of the lowest order quantum fluctuations around the
classical solution (determinant of the classical
solution).

All this is a subject of  modern texts on quantum field theory
and quantum mechanics.
Now comes the key point. As we have emphasized in  Sec. 2,
the dynamical chiral symmetry breaking in QCD as well as the axial anomaly are
effects of the quantum fluctuations of the fermion fields. They necessarily 
vanish at the classical
level. Both the dynamical chiral symmetry breaking and the axial anomaly are
quark loop effects and hence their contribution can appear only at
the loop level. The loop contributions in the semiclassical expansion
start from the $\hbar W_1(J)$ order and hence are suppressed
relative to the classical term $W_0(J)$ by the factor $\hbar/S_0$.
The classical path $\phi_{cl}$ is a tree-level contribution and keeps
the tree-level classical symmetries, in particular the chiral and $U(1)_A$
symmetries of the classical Lagrangian.  Hence in a hadron with large
intrinsic action $S_0$ one should expect that the chiral and $U(1)_A$
symmetry breaking effects are suppressed relative to the classical
contributions that preserve both symmetries. In other words, these
symmetries should be effectively  restored in the highly
excited hadrons, i.e. in the hadrons with large $J$ or radial
quantum number $n$, where the action is large.

\begin{figure}
\begin{center}
\includegraphics[height=5cm,clip=]{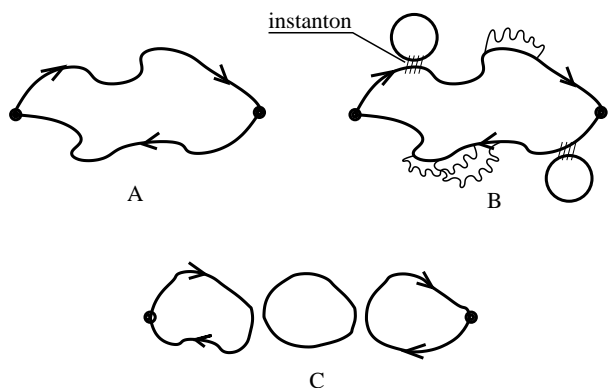}
\caption{The classical (A) and typical loop contributions (B and C)
into the two-point function.}
\label{fluk}
\end{center}
\end{figure}

For clarity, let us consider specifically mesons. They appear as
intermediate states in the two-point function

\begin{equation}
\Pi(x,y) = \frac{1}{Z}\int DA_\mu D \Psi D \bar \Psi
e^{\frac{i S(\bar \Psi,\Psi, A)}{\hbar}} J(x) J(y)^\dagger,
\label{mtp}
\end{equation}

\noindent
where $J(x)$  is a source with the required quantum numbers. There
are different kinds of contributions to the two-point function,
schematically depicted in Fig. \ref{fluk}.

Diagram A is the classical contribution. There are no quantum
fluctuations of the valence quark lines and the only possible gluonic 
contribution is a smooth string-like field connecting valence quarks
or other possible classical gluonic fields. Most importantly, there
are no creations or annihilations of quarks anywhere except at the
points $x,y$ where the source is applied. Diagram B represents
typical mechanisms of dynamical chiral symmetry breaking. In this
case there are quantum fluctuations (i.e. creations and
annihilations) of the valence quarks due to their quantum interactions
with the gluonic fields. Due to these self-energy interactions the
valence quarks acquire a dynamical Lorentz-scalar mass and the chiral
symmetry gets broken. Graph C shows the vacuum loop contributions
which are crucial, e.g. for the anomaly and  provide the $\eta - \eta \prime$
splitting, etc. There are also other types of  vacuum
fermion loop graphs which represent, e.g. the "two-meson" component of the
meson wave function, etc. 

In the low lying mesons, where the action is
small, $S_0 \sim \hbar$, according to the quantum-mechanical interference
paradigm, {\it all} these amplitudes are equally or almost equally
important and contribute. Upon increasing the action in the meson,
the relative importance of these diagrams changes and eventually
in the highly-excited hadron
the semiclassical expansion of the path integral requires
the contribution A to strongly dominate over all other contributions
B,C,... Hence the dynamics responsible for the chiral and $U(1)_A$
symmetry breaking must be strongly suppressed in this hadron. This
should be reflected in the hadron spectrum and  approximate
multiplets of chiral and $U(1)_A$ symmetries should appear.

Unfortunately we cannot solve QCD analytically and hence cannot
clarify all possible important microscopical details of this
physics. What can be  done at present to see this physics at work, 
is to use an
exactly solvable "QCD-like" model that is manifestly chirally
symmetric and confining. This  will be done below in  Sec. 11. 
\footnote{A similar effect is actually well-known in the hydrogen spectrum,
though it is never discussed from the present point of view.
At the quantum-mechanical level the hydrogen spectrum has a symmetry
of the Coulomb potential and this symmetry prescribes a degeneracy
of the 2s-1p and other levels. With the field-theoretical description
this degeneracy is lifted - there appears the Lamb shift. 
The Lamb shift is a result
of the radiative (loop) corrections (which represent
effects of quantum fluctuations of electron and electromagnetic
fields)
 and vanishes  as $1/n^3$,
and much faster with increasing $J$. As a consequence high in the
hydrogen spectrum  the symmetry of the classical Coulomb
potential gets restored. While in QED the loop effects are tiny,
in QCD they are crucial in the low-lying hadrons and it is the loop
effects which shape (via dynamical breaking of chiral symmetry) a
structure of the low-lying spectrum.}

\section{Simple quantum-mechanical example of effective symmetry restoration}

It is instructive  to consider a very simple quantum
mechanical example of symmetry restoration high in
the spectrum \cite{CG2}. Though there are conceptual differences
between a field theory with spontaneous symmetry
breaking and  one-particle quantum mechanics (where
only explicit symmetry breaking is possible), nevertheless
this simple example illustrates how this general phenomenon
comes about.

The example we consider is a two-dimensional harmonic
oscillator. We choose the harmonic oscillator only
for simplicity; the property that will be discussed below
is quite general one and can be seen in other systems.
The Hamiltonian of the system is invariant under 
$U(2) = SU(2)\times U(1)$ transformations. This
symmetry has profound consequences on the spectrum of the system.
 The energy levels of this  system are trivially found and
  are given by
\begin{equation}
E_{N, m} \, = \, ( N \, + \, 1 ); ~ m \, =
\, N, N-2, N-4, \, \cdots \, , -(N-2) , -N \; ,
\label{hoeigen}\end{equation}
where $N$ is the principal quantum number  and m is the
(two dimensional) angular momentum.  As a consequence of the symmetry, 
the levels are $(N+1)$-fold
degenerate.

\begin{figure}
\includegraphics[width=0.5\textwidth,clip=]{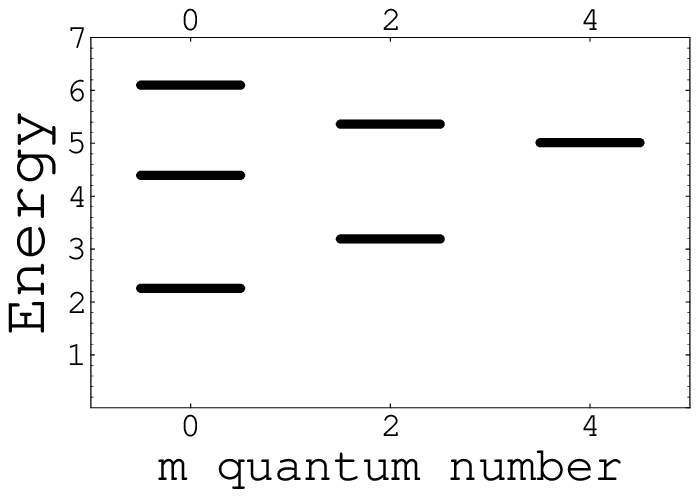}
\includegraphics[width=0.5\textwidth,clip=]{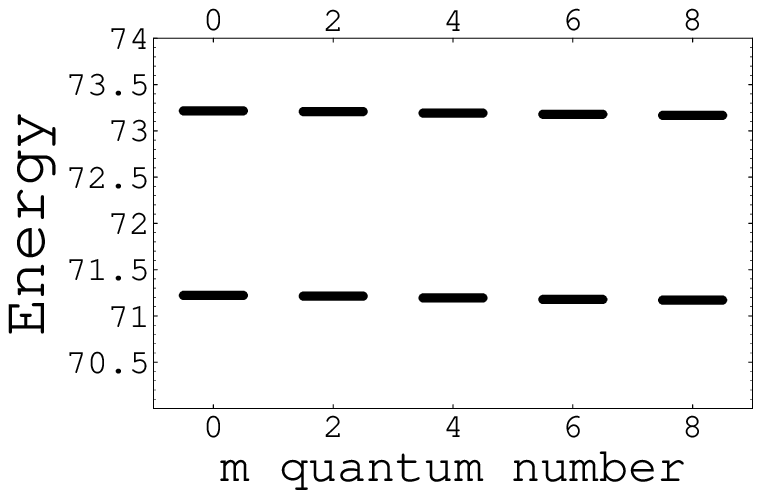}
\caption{The low-lying (left panel) and highly-lying (right panel)
spectra of the two-dimensional harmonic oscillator with the 
$SU(2)$-breaking term.}
\label{to}
\end{figure}

Now suppose we add to the Hamiltonian a $SU(2)$ symmetry breaking
interaction (but which is still $U(1)$ invariant) of the form

\begin{equation}
V_{\rm SB} \, = \, A \, \theta (r - R),
\label{vsb}
 \end{equation}

\noindent
where  $A$ and $R$ are parameters and $\theta$ is the step
function.  Clearly, $V_{\rm SB}$ is not invariant under the
$SU(2)$ transformation. Thus the $SU(2)$ symmetry
is explicitly broken by this additional interaction, that acts
only within a circle of radius $R$.
As a result one would expect that the eigenenergies will not
reflect the degeneracy structure of seen in Eq.~(\ref{hoeigen})
 if the
coefficients $R,A$ are sufficiently large.  Indeed,
 we have solved
numerically for the eigenstates for the case of $A=4$ and $R=1$
in  dimensionless units and one does not see a
multiplet structure in the low-lying spectrum as can be seen in
Fig. \ref{to}.

What is interesting for the present context is the high-lying spectrum.
  In Fig.\ref{to} we have also plotted the energies between 70 and 74 for
   a few of the lower $m$'s.
   A multiplet structure is quite evident---to very good approximation
    the states of different $m$'s form degenerate multiplets and,
    although we have not shown this in the figure these multiplets
     extend in $m$ up to $m=N$.  
     
     How does this happen? The symmetry breaking
      interaction  plays a dominant role in the
       spectroscopy for small energies. Indeed, at small
       excitation energies
       the system is mostly located at distances where the symmetry
       breaking interaction acts and where it is dominant.
Hence  the low-lying spectrum to a very large extent is motivated  
by the symmetry breaking interaction.   However, at high 
excitation energies
the system mostly lives at large distances, where physics is dictated
by the unperturbed harmonic oscillator.  Hence at
 higher energies the spectroscopy
       reveals the $SU(2)$ symmetry of the two-dimensional harmonic
        oscillator.

Note that the high-lying levels are not exactly degenerate. There is always
some splitting between levels and the levels do not transform into each
other under the unbroken symmetry transformation, because the symmetry is
manifestly broken in the Hamiltonian. However, this symmetry breaking
dynamics becomes  inessential in highly excited levels and the splittings
between levels asymptotically vanish. High in the spectrum the symmetry
breaking dynamics can be accounted as a small perturbation.

While this simple quantum-mechanical example nicely illustrates
 effective restoration of the symmetry high in the spectrum, the analogy
 with QCD is not perfect. In QCD chiral symmetry is broken spontaneously,
 while in a few-body quantum mechanical system only explicit symmetry 
 breaking is possible. Hence to consider a model with
 spontaneous breaking of the symmetry we have to consider  more complicated
 field-theoretical models, which will be done in the subsequent sections.

\section{Can the quark-hadron duality and the Operator Product Expansion
predict the rate of the symmetry restoration?}

Initially there was a hope that the chiral symmetry restoration
in highly excited hadrons can be obtained from the asymptotic freedom
and the
Operator Product
Expansion (OPE) of the two-point correlation function at large space-like
momenta and the analyticity of the two-point function \cite{CG1}. 
Indeed, if the spectrum is strictly continuous at $s \rightarrow \infty$
and the function $R$ approaches a constant value at $s \rightarrow \infty$,
then the spectral function at $s \rightarrow \infty$ must be dual to
the leading terms of the perturbation theory at large space-like momenta
(the free loop diagram, etc) and 
hence must be chirally symmetric, as it is observed, e.g. in $e^+e^-$ 
annihilation. While this is definitely true, it is not what we actually
need. We have to consider a (quasi)discrete spectrum where the given
hadron state is isolated. A conjecture of ref. \cite{CG1} was that the
chiral restoration occurs for the discrete spectrum too. 

Then one
can invoke the apparatus of OPE at large space-like momenta and try
to constrain the discrete spectrum of large $N_c$ mesons 
at large excitations \cite{beane,CG2,afon}. However, the Operator Product
Expansion, which is an expansion of the two-point function at 
large space-like
momenta, cannot help us, unfortunately. 
 This is because the OPE is only an asymptotic
expansion. While such a kind of expansion is very useful in the space-like
region, it does not define any unique analytical function which could
be continued to the time-like region. This means that while the real
(correct) spectrum of QCD must be consistent with the OPE, there is an
infinite amount of incorrect spectra that can also be consistent with
the OPE.
For low-lying hadrons the
convergence of the OPE can be improved by means of a Borel transform
and this trick makes the OPE useful in SVZ sum rules \cite{SVZ}. 
However, this kind
of transform cuts a sensitivity to the high-lying spectrum and cannot
be used here. 

Then in order to obtain some information about the
high-lying spectrum one needs to assume something else on  top
of the OPE. Hence the result will be crucially dependent on these assumptions.
Assuming for the chiral partners  asymptotically linear radial Regge 
trajectories with the same intercept and  local duality Shifman has 
obtained
from the OPE a constraint that the rate of the chiral symmetry restoration 
in the large 
$N_c$ mesons
must be $\sim 1/n^{3/2}$ or faster \cite{shifm}. In contrast, the rate
$\sim 1/n^{1/2}$ has been shown in ref. \cite{cata} to be consistent
with the OPE, provided that one
gives up the local duality. The latter rate is actually straightforwardly
 implied by the
linearity of the radial Regge trajectories if the intercepts
for the chiral partners are different. Clearly the asymptotic linearity 
of the Regge trajectories is also a nontrivial dynamical assumption.

These results illustrate that one needs a
microscopic insight and a theory that would incorporate at the same
time chiral symmetry breaking and confinement in order to really
understand effective chiral symmetry restoration.

\section{Chiral multiplets of excited mesons}

In this chapter we classify all possible representations of
$SU(2)_L\times SU(2)_R \times U(1)_A \subset U(2)_L \times U(2)_R$
that are compatible with the Lorentz symmetry and can be
characterized by a definite spin and parity.
Then we confront these representations with the available experimental
data on highly excited mesons \cite{G2,G3}. 
In the following, if it is not specified otherwise, the
 chiral symmetry  will refer specifically to the $SU(2)_L\times SU(2)_R$
symmetry. This group is isomorphic to the group SO(4).

\subsection{Representations of the parity-chiral group}

The irreducible representations of 
the chiral group can be specified by the isospins of the left-handed and
right-handed quarks, $(I_L,I_R)$. The total isospin of the state
can be obtained from the left- and right-handed isospins according to
the standard angular momentum addition rules

\begin{equation}
I= |I_L-I_R|, ... , I_L+I_R.
\label{isospin}
\end{equation}

All hadronic states are characterized by a definite parity. However,
not all  irreducible representations of the chiral group are
invariant under parity. Indeed,  parity transforms  left-handed
quarks into  right-handed ones and vice versa. Hence while representations
with $I_L=I_R$ are invariant under parity (except for the representations (0,0)),
i.e. under a parity
operation every state in the representation transforms into a
state of opposite parity within the same representation, this
is not true for the case $I_L \neq I_R$. In that case 
parity transforms every state in the representation $(I_L,I_R)$
into a state in the representation $(I_R,I_L)$. We can
construct definite parity states only by combining basis vectors from both
these irreducible representations. Hence it is only the direct sum of these
two representations
 
\begin{equation}
(I_L,I_R) \oplus (I_R, I_L), ~~~~~~ I_L \neq I_R,
\label{mult}
\end{equation}

\noindent
 that is invariant under parity. This reducible representation of the
chiral group is an irreducible representation of the larger
group, the parity-chiral group 

\begin{equation}
 SU(2)_L \times SU(2)_R \times C_i,
\label{gr}
\end{equation}

\noindent
where the group $C_i$ consists of two elements: identity
and inversion in 3-dimensional space.
This group is isomorphic to O(4).
This symmetry group is
the symmetry of the QCD Lagrangian (neglecting quark masses),
 however only its
subgroup $SU(2)_I \times C_i$ survives in the broken symmetry mode.
The dimension of the representation (\ref{mult}) is

\begin{equation}
dim_{ (I_a,I_b) \oplus (I_b,I_a)} = 2(2I_a+1)(2I_b+1).
\label{dim}
\end{equation}

 When we consider mesons
of isospin $I=0,1$, only three types of irreducible representations
of the parity-chiral group exist.

{\bf (i)~~~ (0,0) .} Mesons in this representation must have isospin
$I=0$. At the same time $I_R=I_L=0$. This can be achieved when 
either there are no valence quarks in the meson\footnote{This corresponds to the trivial
(identity) representation.
Hence glueballs must
be classified according to this trivial representation \cite{Gglue}; with
no quark content this representation contains one state of only
one parity.}, or both
valence quark and antiquark are right or  left.
If we denote $R$  ($L$) as a column consisting of the right (left) $u$ and $d$
quarks,   then there are two independent (0,0) representations of the chiral
group which are not invariant under parity; they are generated by the
$\bar R R$ and $\bar L L$ quark configurations.

 Then the irreducible representation
of the parity-chiral group can be constructed as a direct sum of these two irreducible
representations of the chiral group.

The basis
states of both parities can  then be written in terms of the right and left
components of two valence quarks as

\begin{equation}
|(0,0); \pm; J \rangle = \frac{1}{\sqrt 2} (\bar R R \pm \bar L L)_J.
\label{00}
\end{equation} 

\noindent
The index $J$ means that a definite spin $J$ can be ascribed
to the given system.
Note that such a system can have spin $J \geq 1$. Indeed, valence
quark and antiquark in the state (\ref{00}) have definite
helicities, because generically helicity = +chirality for quarks and
helicity = -chirality for antiquarks. Hence the total spin projection
of the quark-antiquark system  onto the momentum direction of the quark
is $ \pm 1$. The parity transformation property of the quark-antiquark state
is then regulated by the total spin of the system \cite{LANDAU}

\begin{equation}
\hat P |(0,0); \pm; J \rangle = \pm (-1)^J  |(0,0); \pm; J \rangle.
\label{P00}
\end{equation}

{\bf (ii)~~~ (1/2,1/2).} There are two independent irreducible representations
of the chiral group of this type. We denote them as  $(1/2,1/2)_a$ and $(1/2,1/2)_b$,
respectively. Each of these representations is also an irreducible representation
of the parity-chiral group. If we construct basis vectors with two valence quarks,
then the quark must be right
and the antiquark must be left, and vice versa. These representations combine
states with I=0 and I=1, which must be of opposite parity.
The basis states
within the two distinct representations  
of this type are

\begin{equation}
|(1/2,1/2)_a; +;I=0; J \rangle = \frac{1}{\sqrt 2} (\bar R L + \bar L R)_J,
\label{I0+1}
\end{equation} 

\begin{equation}
|(1/2,1/2)_a; -;I=1; J \rangle = \frac{1}{\sqrt 2} (\bar R \vec \tau L -
\bar L \vec \tau R)_J,
\label{I0+2}
\end{equation} 

\noindent
and

\begin{equation}
|(1/2,1/2)_b; -;I=0; J \rangle = \frac{1}{\sqrt 2} (\bar R L - \bar L R)_J,
\label{I0-1}
\end{equation}

\begin{equation}
|(1/2,1/2)_b; +;I=1; J \rangle = \frac{1}{\sqrt 2} (\bar R \vec \tau L +
\bar L \vec \tau R)_J.
\label{I0-2}
\end{equation}

\noindent
In these expressions $\vec \tau$ are isospin Pauli matrices.
The parity of  every state in these representations is determined as

\begin{equation}
\hat P |(1/2,1/2); \pm; I; J \rangle = \pm (-1)^J |(1/2,1/2); 
\pm; I; J \rangle.
\label{P12}
\end{equation} 

Mesons in the representations of this type can have
any spin.
Note that the sum of the two distinct $(1/2,1/2)_a$ and $(1/2,1/2)_b$
 irreducible representations
of $SU(2)_L \times SU(2)_R$ form an irreducible representation
of $U(2)_L \times U(2)_R$ or $SU(2)_L \times SU(2)_R \times U(1)_A$ groups.

{\bf (iii)~~~ (0,1)$\oplus$(1,0).} The total isospin is 1 and
both the quark and  the antiquark must be right or  left.
 With two valence quarks this representation
is possible only for $J \geq 1$. The basis states are

\begin{equation}
|(0,1)+(1,0); \pm; J \rangle = \frac{1}{\sqrt 2} (\bar R \vec \tau R 
\pm \bar L  \vec \tau L)_J
\label{10}
\end{equation} 

\noindent
with  parities

\begin{equation}
\hat P |(0,1)+(1,0); \pm; J \rangle = \pm (-1)^J 
 |(0,1)+(1,0); \pm; J \rangle.
\label{P10}
\end{equation}

We have to stress that the usual quantum numbers $I,J^{PC}$ are not
enough to specify uniquely
the chiral representation for $J \geq 1$. 
It happens that some of
the physical particles with the given $I,J^{PC}$ belong to
one chiral representation (multiplet), while the other particles with
the same $I,J^{PC}$ belong to the other multiplet. Classification
of the particles according to $I,J^{PC}$ is simply not complete
in the chirally restored regime. This property will have very
important implications as far as the amount of the states
with  the given  $I,J^{PC}$ is
concerned. A detailed discussion of this property is relegated to the
 subsection 7.3.

\subsection{$J=0$ mesons}

 Consider first the mesons of spin $J=0$,
which are the $ \pi (1,0^{-+}), f_0 (0,0^{++}), a_0(1,0^{++})$ and $\eta(0,0^{-+})$ mesons with  $u,d$
quark content only.
Their interpolating fields
are given as

\begin{equation}
 J_\pi(x)  = \bar q(x) \vec \tau \imath \gamma_5 q(x),
\label{ppi}
\end{equation}

\begin{equation}
 J_{f_0}(x)  = \bar q(x)  q(x),
\label{ff0}
\end{equation}

\begin{equation}
 J_{\eta}(x)  = \bar q(x)  \imath \gamma_5 q(x),
\label{eeta}
\end{equation}

\begin{equation}
 J_{a_0}(x)  = \bar q(x) \vec \tau  q(x).
\label{aa0}
\end{equation}

\noindent
These four currents 
transform as scalars or pseudoscalars with respect to
the Lorentz group and belong to an
irreducible representation
of the group
$U(2)_L\times U(2)_R  \supset SU(2)_L\times SU(2)_R \times U(1)_A$. 
It is instructive to see how these currents transform
under different subgroups of the group above.

 The $SU(2)_L\times SU(2)_R$ transformations consist of vectorial
 and axial transformations in the isospin space (\ref{VA}). The axial
transformation  mixes the currents of opposite parity. For instance,

\begin{equation}
  \bar q(x)  q(x) \longrightarrow  \bar q(x) e^{ i \gamma_5 \Theta^a_A \tau^a} q(x)
 = \cos {|\vec \Theta_A|} \bar q(x)  q(x) + \sin {| \vec \Theta_A|}  
 \frac{\vec \Theta_A}{|\vec \Theta_A|} \cdot 
\bar q(x) \vec \tau \imath \gamma_5 q(x),  
\label{f0mix}
\end{equation}

\noindent 
and similar in  other cases. Then, under the axial transformation
the following currents get mixed

\begin{equation}
 J_\pi(x)  \leftrightarrow J_{f_0}(x) 
\label{pif0}
\end{equation} 

\noindent
as well as

\begin{equation}
 J_{a_0}(x)  \leftrightarrow J_{ \eta}(x).
\label{a0eta}
\end{equation} 

\noindent
Hence the currents (\ref{pif0}) form the basis functions of the $(1/2,1/2)_a$
representation of the chiral and parity-chiral groups, while the interpolators
(\ref{a0eta}) transform as $(1/2,1/2)_b$. This can also be seen upon
substituting  of  (\ref{RL})   into (\ref{ppi} - \ref{aa0}).

The $U(1)_A$ transformation  (\ref{VAS}) mixes the currents
of the same isospin but opposite parity:

\begin{equation}
 J_\pi(x)  \leftrightarrow J_{a_0}(x) 
\label{pia0}
\end{equation} 

\noindent
as well as

\begin{equation}
 J_{f_0}(x)  \leftrightarrow J_{ \eta}(x).
\label{f0eta}
\end{equation}

\noindent
All four currents together belong to the  representation
$(1/2,1/2)_a \oplus (1/2,1/2)_b$ which
is an irreducible representation of the group $U(2)_L\times U(2)_R $  and
$SU(2)_L\times SU(2)_R \times U(1)_A $.

If the vacuum were invariant with respect to  $U(2)_L\times U(2)_R $
transformations, then all four mesons, $\pi,f_0,a_0$ and $ \eta$
would be degenerate level-by-level. Once
the $U(1)_A$ symmetry is broken explicitly through
the axial anomaly, but the chiral $SU(2)_L\times SU(2)_R $ 
symmetry is still
intact in the vacuum, then the spectrum would consist of
degenerate $(\pi, f_0)$ and $(a_0,  \eta)$ pairs. If
in addition the chiral  $SU(2)_L\times SU(2)_R $ symmetry is
spontaneously broken 
in the vacuum, the degeneracy is also lifted in  the pairs
above and the pion becomes a (pseudo)Goldstone boson. Indeed,
the masses of the lowest mesons  are \cite{PDG}\footnote{The
$\eta$ meson mass given here was obtained by unmixing the
 $SU(3)$ flavor octet  and singlet states so it represents
the pure $\bar n n = (\bar u u + \bar d d )/\sqrt 2$ state,
see for details ref. \cite{G2}.}

 $$ m_\pi \simeq 140 MeV, ~m_{f_0} \simeq 400 - 1200 MeV,~
m_{a_0} \simeq 985 MeV ,~ m_{\eta } \simeq 782 MeV. $$

\noindent
This immediately shows that both $SU(2)_L\times SU(2)_R $ and
$U(1)_V \times U(1)_A$ are broken in the QCD vacuum to the
vector subgroups $SU(2)_I$ and $U(1)_V$, respectively.

\begin{figure}
\begin{center}
\includegraphics*[width=5cm,angle=-90]{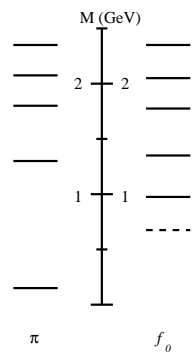}
\end{center}
\caption{Pion and $\bar n n$ $f_0$ spectra.}
\label{pif}
\end{figure}

\begin{figure}
\begin{center}
\includegraphics*[width=7cm]{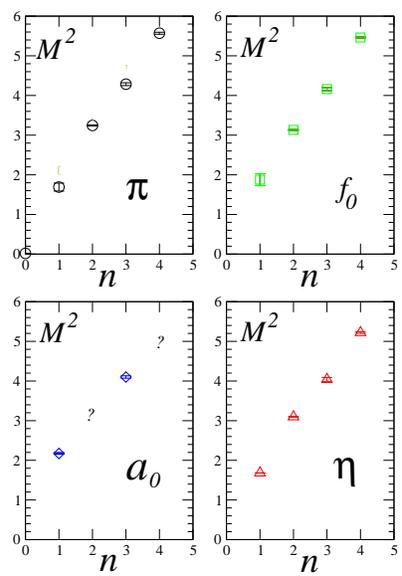}
\end{center}
\caption{Radial Regge trajectories for the four successive
high-lying $J=0$ mesons.}
\label{regg}
\end{figure}

The modern data on the upper part of the $\pi$ and $\bar n n$ $f_0$ 
meson spectra   are summarized in Table 1 and the corresponding
spectra are depicted in Fig.~\ref{pif}. One notices that
the four successive  excited $\pi$ mesons
and the corresponding $\bar n n$ $f_0$ mesons 
form approximate chiral pairs \cite{G2}. 
This pattern is a clear manifestation of the chiral
symmetry restoration.

A similar behavior is observed from a comparison of the
$a_0$ and $\eta$ masses \cite{G2}. However, there are two
missing $a_0$ mesons which must be discovered in order to complete
all chiral multiplets. (Technically the identification of the
spinless states from the partial wave analysis is a rather
difficult task). There is little doubt that these
missing $a_0$ mesons do exist. If one puts the four high-lying 
$\pi$, $\bar n n$ $f_0$, $a_0$ and  $\bar n n$ $\eta$ mesons on the 
{\it radial} Regge trajectories, see Fig. ~\ref{regg}, one clearly notices
that the two missing $a_0$ mesons lie on the approximately linear trajectory with
the same slope as all other mesons \cite{BUGG1,BUGG2}. 
If one reconstructs these
missing $a_0$ mesons according to this slope, then a pattern
of the $a_0 - \eta$ chiral partners appears, similar to the one
for the $\pi$ and $f_0$ mesons.

A one-to-one correspondence between the excited $\pi$ and $\bar n n$ $f_0$
 as well as between  $a_0$ and $\bar n n$ $\eta$ mesons implies that the
 chiral symmetry in this part of the spectrum is realized linearly
 with a small breaking.
\begin{table}
\begin{center}
\caption{Chiral multiplets of $\pi$ and $\bar n n ~~~f_0$  mesons
\cite{G2}.
Comments: (i) $\pi(1300)$ and $f_0(1370)$ are well established states
and can be found in the Meson Summary Table of the Review of Particle 
Physics \cite{PDG}. (ii) $\pi(1812 \pm 14)$ is abbreviated as
$\pi(1800)$ in the Meson Summary Table of the Review of Particle 
Physics \cite{PDG}; $f_0(1770 \pm 12)$ is seen in $\bar p p$
as $\bar n n$ state \cite{BUGG1,BUGG2} and also recently as $15 \sigma$
peak in the $\pi \pi$ channel in
$J/\Psi$ decays at $\sim 1790$ MeV \cite{BES}
(iii) These states are clearly seen in a few different channels
in $\bar p p$ \cite{BUGG1,BUGG2},
though in order to appear in the Meson Summary Table of the Review of Particle 
Physics they must be confirmed by an independent experiment.
 }
 \begin{tabular}{|lllll|} \hline
Chiral multiplet &  Representation & $\chi$ & $So$& Comment\\ \hline
$\pi(1300 \pm 100) - f_0(1370^{+130}_{-170})$ & (1/2,1/2) &
$0.03^{+0.09}_{-0.03}$  & $0.1^{+0.3}_{-0.1}$  & (i) \\

$\pi(1812 \pm 14) - f_0(1770 \pm 12)$ & (1/2,1/2) &
$0.012 \pm 0.007$  & $0.09 \pm 0.06$  & (ii) \\

$\pi(2070 \pm 35) - f_0(2040 \pm 38)$ & (1/2,1/2) &
$0.007^{+0.017}_{-0.007}$  & $0.11^{+0.28}_{-0.11}$  & (iii) \\

$\pi(2360 \pm 25) - f_0(2337 \pm 14)$ & (1/2,1/2) &
$0.005^{+0.009}_{-0.005}$  & $0.08^{+0.13}_{-0.08}$  & (iii) \\

\hline
\end{tabular}
\end{center}
\label{t1}
\end{table}

Both the chiral asymmetries and the spectral overlaps as well as the
assignments of $\pi$ and $\bar n n ~~~f_0$ mesons   where the data
sets are complete enough, are given in Table 1.

It is well seen from  Table 1 that the chiral $(1/2,1/2)_a$  $\pi$ and $\bar n n ~~~f_0$ 
multiplets are very good and the 
chiral symmetry breaking contribution to the hadron mass
is tiny. This means that practically the whole mass of a highly excited
hadron is not related to the chiral symmetry breaking in the vacuum.

A crucial prediction of the approximate linear realization of
chiral symmetry is the existence of 
two missing $a_0$ mesons, see Fig. \ref{regg}.
 The spectrum of the $a_0$ mesons is the worst known
and the observation of the missing $a_0$ mesons is a very important experimental
task. If these $a_0$ mesons do exist and their masses fall into
approximate $a_0 - \eta$ $(1/2,1/2)_b$ chiral multiplets, then it would
also mean existence of approximate $U(1)_A$ parity doublets 
$a_0 - \pi$ along with the existing $f_0 - \eta$ doublets. This would
imply the approximate restoration of the $U(2)_L \times U(2)_R$ symmetry.

\subsection{$J > 0$ mesons}
For the $J \geq 1$ mesons it happens that there is no one-to-one
mapping of the standard quantum numbers $I,J^{PC}$ and representations
of the parity-chiral group. Namely, some of the mesons with the given
quantum numbers $I,J^{PC}$ must belong to one chiral representation while
the other mesons with the same quantum numbers must belong to the
other chiral representation. Hence, the chiral classification of
the $\bar q q$ mesons with $J > 0$ is richer than for $J = 0$ \cite{G3}.

Consider the $\rho(1,1^{--})$ mesons as an example. It was considered natural for
a few decades, starting from the famous Weinberg sum rules \cite{WSR},
that these particles should belong to the (0,1)+(1,0) representation
and their chiral partners must be the axial vector mesons, $a_1(1,1^{++})$.
A rationale for this kind of assumption is as follows.
The rho-mesons can be created from the vacuum by the vector
current, $\bar \psi \gamma^\mu \vec \tau \psi$. Its chiral
partner is the axial vector current, 
$\bar \psi \gamma^\mu \gamma^5 \vec \tau \psi$, which creates
from the vacuum the axial vector mesons, $a_1(1,1^{++})$ . Both
these currents belong to the representation (0,1)+(1,0) and
have the  right-right $\pm$ left-left quark content. 
While this is true, this picture is not complete and leads to 
wrong results once the chiral symmetry is approximately restored
in high-lying states. Indeed, in the chirally restored regime mesons
created by these currents must be degenerate level-by-level and
fill out the (0,1)+(1,0) representations. Hence,
naively  
 the amount of $\rho$ and $a_1$
mesons high in the spectrum should be equal.

This is not correct, however.  The reason is that the
 $\rho$-mesons can be also created
from the vacuum by  other type(s) of  current(s),
$ \bar \psi \sigma^{0i} \vec \tau \psi$ (or by 
$\bar \psi \partial^\mu \vec \tau \psi$). These interpolators belong, however,
to the  $(1/2,1/2)_b$ representation and have the 
left-right $+$ right-left quark content.
In the regime where chiral symmetry is strongly broken (as
in the low-lying states) the physical states are mixtures
of different representations. Hence these low-lying states are
well coupled to both (0,1)+(1,0) and $(1/2,1/2)_b$ interpolators. This
is well confirmed by the lattice calculations, where it is
shown that the $\rho$-meson is strongly coupled to both
$\bar \psi \gamma^\mu \vec \tau \psi$ and 
$ \bar \psi \sigma^{0i} \vec \tau \psi$ currents \cite{BURCH}.
However, when chiral symmetry is  (approximately)
restored, then each physical
state must be strongly dominated by the given representation
and hence will  couple practically only to the interpolator which transforms
according to the same representation. Then in the chirally restored
regime there must be rho-mesons of two kinds, some of them belong to
(0,1)+(1,0) and the other ones transform as  $(1/2,1/2)_b$. The
chiral partners of the latter rho-mesons must be the $h_1(0,1^{+-})$
mesons, as it follows from (\ref{I0-1})-(\ref{I0-2}). Exactly the
same result can be obtained upon performing the chiral transformation
of the $ \bar \psi \sigma^{0i} \vec \tau \psi$ current: it gets mixed
with the $ \varepsilon^{ijk} \bar \psi \sigma^{jk}  \psi$ current. Hence
the latter current is its chiral partner. \footnote{Chiral transformation 
properties of some 
interpolators can be found in ref. \cite{CJ}.} Its quantum numbers coincide
with the quantum numbers of the $h_1$ mesons and it can create from the 
vacuum only mesons of this kind. Similar result can be obtained upon
the chiral transformation of the  interpolator
$\bar \psi \partial^\mu \vec \tau \psi$. Its chiral partner is
$\bar \psi \gamma^5 \partial^\mu   \psi$, which also interpolates
$h_1$ mesons.

All this unambiguously implies that
in the chirally restored regime, some of the
$\rho$-mesons must be degenerate with the $a_1$ mesons ((0,1)+(1,0)
multiplets), but the others
- with the $h_1$ mesons ($(1/2,1/2)_b$ multiplets)\footnote{
Those $\rho(1,1^{--})$ and $\omega(0,1^{--})$ mesons which
belong to $(1/2,1/2)_b$ cannot be seen in $e^+e^- \rightarrow hadrons$.}. 
Consequently, high in the spectra the combined amount
of $a_1$ and $h_1$ mesons must coincide with the amount of
$\rho$-mesons. Then asymptotically there must be two
independent  Regge trajectories for the $\rho$-mesons. The first
one is characterized by the index (0,1)+(1,0) and the second one -
by $(1/2,1/2)_b$.
This is a highly nontrivial prediction of chiral
symmetry.

Actually it is a very typical situation. Consider the
  $f_2(0,2^{++})$ mesons as another example. They can be interpolated
by the tensor field $\bar \psi \gamma^\mu \partial^\nu \psi$
(properly symmetrized, of course), which belongs to the (0,0)
representation. Their chiral partners are the $\omega_2(0,2^{--})$
mesons, which are created by the 
$\bar \psi \gamma^5\gamma^\mu \partial^\nu \psi$ interpolator.
On the other hand  $f_2(0,2^{++})$ mesons can also be created from
the vacuum by the $\bar \psi \partial^\mu \partial^\nu \psi$
type of interpolator, which belongs to the $(1/2,1/2)_a$ representation.
Its chiral partner is 
$\bar  \psi \gamma^5\partial^\mu \partial^\nu \vec \tau \psi$,
which creates the $\pi_2(1,2^{-+})$ mesons. Hence in the chirally
restored regime we have to expect the $\omega_2(0,2^{--})$ mesons
to be degenerate systematically with some of the $f_2(0,2^{++})$ mesons
((0,0) representations) while the $\pi_2(1,2^{-+})$ mesons must
be degenerate with {\it other} $f_2(0,2^{++})$ mesons (forming
$(1/2,1/2)_a$ multiplets). Hence the total number of the $\omega_2(0,2^{--})$
and $\pi_2(1,2^{-+})$ mesons in the chirally restored regime
must coincide with the amount of the $f_2(0,2^{++})$ mesons.
There must be two independent Regge trajectories of $f_2$ mesons;
one of them having the chiral index (0,0) and the other one - $(1/2,1/2)_a$.
These examples can be generalized to mesons of any spin $J \geq 1$ \cite{G3}.

Now we can summarize all possible chiral multiplets of the
$\bar q q$ mesons.

\begin{center}
{\bf J~=~0}

\begin{eqnarray}
(1/2,1/2)_a  &  :  &   1,0^{-+} \longleftrightarrow 0,0^{++}  \label{e0a} \\
(1/2,1/2)_b  &  : &   1,0^{++} \longleftrightarrow 0,0^{-+} \label{e0b} ,
\end{eqnarray}
%
%

 {\bf J~=~2k, ~~~k=1,2,... }

\begin{eqnarray}
 (0,0)  & :  &   0,J^{--} \longleftrightarrow 0,J^{++} \label{e1}  \\
 (1/2,1/2)_a  & : &   1,J^{-+} \longleftrightarrow 0,J^{++} \label{e1a}  \\
 (1/2,1/2)_b  & : &   1,J^{++} \longleftrightarrow 0,J^{-+}  \label{e1b} \\
 (0,1) \oplus (1,0)  & :  &   1,J^{++} \longleftrightarrow 1,J^{--} \label{e11}
\end{eqnarray}

 {\bf J~=~2k-1, ~~~k=1,2,...}

\begin{eqnarray}
 (0,0)  & :  &   0,J^{++} \longleftrightarrow 0,J^{--}  \label{e2} \\
 (1/2,1/2)_a  & : &   1,J^{+-} \longleftrightarrow 0,J^{--} \label{e2a} \\
 (1/2,1/2)_b  & : &   1,J^{--} \longleftrightarrow 0,J^{+-} \label{e2b} \\
 (0,1) \oplus (1,0)  & :  &   1,J^{--} \longleftrightarrow 1,J^{++} \label{e22}
\end{eqnarray}
\end{center}

The available data for the $J=1,2,3$ mesons \cite{BUGG1,BUGG2}
have been systematized in 
Ref. \cite{G3}. Below we show the chiral patterns for the $J=2$
mesons, where the data set seems to be complete (the corresponding
chiral asymmetries and spectral overlaps can be found in Table 2).\\

\begin{center}{\bf (0,0)}\\

{$\omega_2(0,2^{--})~~~~~~~~~~~~~~~~~~~~~~~f_2(0,2^{++})$}\\
\medskip
{$1975 \pm 20   ~~~~~~~~~~~~~~~~~~~~~~~~1934 \pm 20$}\\
{$2195 \pm 30   ~~~~~~~~~~~~~~~~~~~~~~~~2240 \pm 15$}\\

\bigskip
{\bf $(1/2,1/2)_a$}\\

{$\pi_2(1,2^{-+})~~~~~~~~~~~~~~~~~~~~~~~f_2(0,2^{++})$}\\
\medskip
{$2005 \pm 15   ~~~~~~~~~~~~~~~~~~~~~~~~2001 \pm 10$}\\
{$2245 \pm 60   ~~~~~~~~~~~~~~~~~~~~~~~~2293 \pm 13$}\\

\bigskip
{\bf $(1/2,1/2)_b$}\\

{$a_2(1,2^{++})~~~~~~~~~~~~~~~~~~~~~~~ \eta_2(0,2^{-+})$}\\
\medskip
{ $2030 \pm 20  ~~~~~~~~~~~~~~~~~~~~~~~~2030 ~\pm ~?$}\\
{ $2255 \pm 20 ~~~~~~~~~~~~~~~~~~~~~~~~2267 \pm 14$}\\

\bigskip
{\bf (0,1)+(1,0)}\\

{$a_2(1,2^{++})~~~~~~~~~~~~~~~~~~~~~~~\rho_2(1,2^{--})$}\\
\medskip
{ $1950^{+30}_{-70}~~~~~~~~~~~~~~~~~~~~~~~~~~~1940 \pm 40$}\\
{ $2175 \pm 40  ~~~~~~~~~~~~~~~~~~~~~~~~2225 \pm 35$}\\
\end{center}

\begin{table}
\begin{center}
\caption{Chiral multiplets of high-lying $J=2$ mesons with 
u,d valence quark content \cite{G3}. 
Comments: 
(i) These states are clearly seen in a few different channels
in $\bar p p$ \cite{BUGG1,BUGG2},
though in order to appear in the Meson Summary Table of the Review of Particle 
Physics they must be confirmed by an independent experiment.
 }
\begin{tabular}{|lllll|} \hline
Chiral multiplet &  Representation & $\chi$ & Spectral Overlap & Comment\\ \hline
$\omega_2(1975 \pm 20) - f_2(1934 \pm 20)$ & (0,0) &
$0.01 \pm 0.01$  & $0.16 \pm 0.15$  & (i) \\

$\omega_2(2195 \pm 30) - f_2(2240 \pm 15)$ & (0,0) &
$0.01 \pm 0.01$  & $0.17 \pm 0.17$  & (i) \\

$\pi_2(2005 \pm 15) - f_2(2001 \pm 10)$ & $(1/2,1/2)_a$ &
$0.001^{+0.006}_{-0.001}$  & $ 0.02^{+0.09}_{-0.02}$  & (i) \\

$\pi_2(2245 \pm 60) - f_2(2293 \pm 13)$ & $(1/2,1/2)_a$ &
$0.01^{+0.02}_{-0.01}$  & $ 0.18^{+0.27}_{-0.18}$  & (i) \\

$a_2(2030 \pm 20) - \eta_2(2030 \pm ?)$ & $(1/2,1/2)_b$ &
$0.0 \pm ?$  & $0.0 \pm ?$  & (i) \\

$a_2(2255 \pm 20) - \eta_2(2267 \pm 14)$ & $(1/2,1/2)_b$ &
$0.003^{+0.008}_{-0.003}$  & $ 0.05^{+0.015}_{-0.05}$  & (i) \\

$a_2(1950^{+30}_{-70} ) - \rho_2(1940 \pm 40)$ & $(0,1) \oplus (1,0)$  &
$0.003^{+0.018}_{-0.003}$  & $ 0.04^{+0.27}_{-0.04}$  & (i) \\

$a_2(2175  \pm 40 ) - \rho_2(2225 \pm 35)$ & $(0,1)\oplus (1,0)$  &
$0.011^{+0.017}_{-0.011}$  & $ 0.20^{+0.29}_{-0.20}$  & (i) \\

\hline
\end{tabular}
\end{center}
\label{t2}
\end{table}

\noindent
We see systematic patterns of chiral symmetry restoration. In
particular, the amount of the $f_2(0,2^{++})$ mesons coincides with
the combined amount of the $\omega_2(0,2^{--})$ and $\pi_2(1,2^{-+})$
states. Similarly,  the number of the $a_2(1,2^{++})$ states is
the same as the number of the $\eta_2(0,2^{-+})$ and $\rho_2(1,2^{--})$
states together. All chiral multiplets are complete. While the masses of
some of these states can and will be  corrected in  future experiments,
any  new  states that might be discovered in this energy region in other
types of experiments, 
should be either $\bar s s$ states, hybrids or glueballs.
Clearly one sees two independent radial Regge trajectories for both
the $f_2$ and the $a_2$ mesons.

The data sets for the $J=1$ and $J=3$ mesons are less complete
and there are a few missing states to be discovered. Nevertheless,
these spectra also offer  impressive patterns of chiral symmetry:

\begin{center} {\bf (0,0)} \\

{$\omega(0,1^{--})~~~~~~~~~~~~~~~~~~~~~~~f_1(0,1^{++})$}\\
\medskip
{$~~~~~?~~~~~   ~~~~~~~~~~~~~~~~~~~~~~~~1971 \pm 15$}\\
{$~~~~~?~~~~~   ~~~~~~~~~~~~~~~~~~~~~~~~2310 \pm 60$}\\

\bigskip
{\bf (1/2,1/2)}\\

{$\omega (0,1^{--})~~~~~~~~~~~~~~~~~~~~~~~b_1(1,1^{+-})$}\\
\medskip
{$1960 \pm 25   ~~~~~~~~~~~~~~~~~~~~~~~~1960 \pm 35$}\\
{$2205 \pm 30   ~~~~~~~~~~~~~~~~~~~~~~~~2240 \pm 35$}\\

\bigskip
{\bf (1/2,1/2)}\\

{$h_1(0,1^{+-})~~~~~~~~~~~~~~~~~~~~~~~\rho(1,1^{--})$}\\
\medskip
{$1965 \pm 45   ~~~~~~~~~~~~~~~~~~~~~~~~1970 \pm 30$}\\
{$2215 \pm 40   ~~~~~~~~~~~~~~~~~~~~~~~~2150 \pm ~?$}\\

\bigskip
{\bf (0,1)+(1,0)}\\

{$a_1(1,1^{++})~~~~~~~~~~~~~~~~~~~~~~~\rho (1,1^{--})$}\\
\medskip
{$1930^{+30}_{-70} ~~~~~~~~~~~~~~~~~~~~~~~~~1900 ~\pm ~?$}\\
{$ 2270^{+55}_{-40} ~~~~~~~~~~~~~~~~~~~~~~~~~2265 \pm 40$}\\

\end{center}
Here, like for the $J=0,2$ states, we again observe  patterns
of chiral symmetry restoration. Two  $\omega(0,1^{--})$
states are still missing.\\

Below are the multiplets for $J=3$.
\begin{center}{\bf (0,0)}\\

{$\omega_3(0,3^{--})~~~~~~~~~~~~~~~~~~~~~~~f_3(0,3^{++})$}\\
\medskip
{$~~~~~?~~~~~   ~~~~~~~~~~~~~~~~~~~~~~~~~2048 \pm 8$}\\
{$2285 \pm 60   ~~~~~~~~~~~~~~~~~~~~~~~~2303 \pm 15$}\\

\bigskip
{\bf (1/2,1/2)}\\

{$\omega_3 (0,3^{--})~~~~~~~~~~~~~~~~~~~~~~~b_3(1,3^{+-})$}\\
\medskip
{$1945 \pm 20   ~~~~~~~~~~~~~~~~~~~~~~~~2032 \pm 12$}\\
{$2255 \pm 15   ~~~~~~~~~~~~~~~~~~~~~~~~2245 \pm ~?$}\\

\bigskip
{\bf (1/2,1/2)}\\

{$h_3(0,3^{+-})~~~~~~~~~~~~~~~~~~~~~~~\rho_3(1,3^{--})$}\\
\medskip
{$2025 \pm 20   ~~~~~~~~~~~~~~~~~~~~~~~~1982 \pm 14$}\\
{$2275 \pm 25   ~~~~~~~~~~~~~~~~~~~~~~~~2260 \pm 20$}\\

\bigskip
{\bf (0,1)+(1,0)}\\

{$a_3(1,3^{++})~~~~~~~~~~~~~~~~~~~~~~~\rho_3 (1,3^{--})$}\\
\medskip
{$2031 \pm 12  ~~~~~~~~~~~~~~~~~~~~~~~~ 2013 \pm 30$}\\
{$2275 \pm 35 ~~~~~~~~~~~~~~~~~~~~~~~~~2300^{~+~50}_{~-~80}$}\\

\end{center}

Data on $J \geq 4$ is scarce.  While some of the multiplets
are well seen, it is not yet possible to provide any
systematic analysis. The prediction is that for $J=4$ the
pattern should be the same as for $J=2$, while for $J=5$ it should
be similar to the $J=1,3$ cases.

It is important to see whether there are also signatures
of the $U(1)_A$ restoration. This 
can happen if two conditions are fulfilled \cite{CG1}: (i) unimportance
of the axial anomaly in excited states, (ii) chiral
$SU(2)_L \times SU(2)_R$ restoration (i.e. unimportance of the
quark condensates which break simultaneously both types
of symmetries in the vacuum state). 
Some evidence for the $U(1)_A$ restoration
is seen from the $J=0$
data. Yet  missing $a_0$  states have
to be discovered to complete the $U(1)_A$ multiplets in the
$J=0$ spectra. Below we  demonstrate that the
data on, e.g. the $J=2$ mesons present convincing evidence on $U(1)_A$
restoration.

First, we have to consider which mesonic states can be 
expected to be $U(1)_A$ partners. The $U(1)_A$ transformation
connects interpolators of the same isospin but opposite parity.
But not all such interpolators can be connected by the $U(1)_A$
transformation. For instance, the vector currents 
$\bar \psi \gamma^\mu \psi$
and $\bar \psi \vec \tau\gamma^\mu \psi$ are invariant under
$U(1)_A$. Similarly, the axial vector interpolators
 $\bar \psi \gamma^5 \gamma^\mu \psi$
and $\bar \psi \vec \tau \gamma^5 \gamma^\mu \psi$ are also invariant under
$U(1)_A$. Hence those interpolators (states) that are members
of the $(0,0)$ and $(0,1)+(1,0)$ representations of
$SU(2)_L \times SU(2)_R$ are invariant with respect to
$U(1)_A$. However,  interpolators (states) from the {\it distinct}
(1/2,1/2) representations which have the same isospin but
opposite parity  transform into each other under $U(1)_A$.
For example, $\bar \psi \psi \leftrightarrow \bar \psi \gamma^5 \psi$,
 $\bar \psi \vec \tau \psi \leftrightarrow \bar \psi \vec \tau \gamma^5 \psi$,
and those with derivatives: 
$\bar \psi \partial^\mu \psi \leftrightarrow \bar \psi \gamma^5  
\partial^\mu \psi$, $\bar \psi \vec \tau \partial^\mu \psi 
\leftrightarrow \bar \psi \vec \tau \gamma^5  \partial^\mu \psi$, etc.
If the corresponding states are systematically degenerate, then
it is a signal that $U(1)_A$ is restored. In what follows we show
that it is indeed the case.

\begin{center}{\bf J=1}\\

{$\omega(0,1^{--})~~~~~~~~~~~~~~~~~~~~~~~h_1(0,1^{+-})$}\\
\medskip
{$1960 \pm 25   ~~~~~~~~~~~~~~~~~~~~~~~~1965 \pm 45$}\\
{$2205 \pm 30   ~~~~~~~~~~~~~~~~~~~~~~~~2215 \pm 40$}\\

\bigskip

{$b_1 (1,1^{+-})~~~~~~~~~~~~~~~~~~~~~~~\rho(1,1^{--})$}\\
\medskip
{$1960 \pm 35   ~~~~~~~~~~~~~~~~~~~~~~~~1970 \pm 30$}\\
{$2240 \pm 35   ~~~~~~~~~~~~~~~~~~~~~~~~2150 \pm ~?$}\\

\bigskip
{\bf J=2}\\

{$f_2(0,2^{++})~~~~~~~~~~~~~~~~~~~~~~~\eta_2(0,2^{-+})$}\\
\medskip
{$2001 \pm 10  ~~~~~~~~~~~~~~~~~~~~~~~~2030 ~\pm ~?$}\\
{$2293 \pm 13   ~~~~~~~~~~~~~~~~~~~~~~~~2267 \pm 14$}\\

\bigskip

{$\pi_2(1,2^{-+})~~~~~~~~~~~~~~~~~~~~~~~a_2 (1,2^{++})$}\\
\medskip
{$2005 \pm 15  ~~~~~~~~~~~~~~~~~~~~~~~~ 2030 \pm 20$}\\
{$2245 \pm 60 ~~~~~~~~~~~~~~~~~~~~~~~~ 2255 \pm 20 $}\\

\bigskip

{\bf J=3}\\

{$\omega_3(0,3^{--})~~~~~~~~~~~~~~~~~~~~~~~h_3(0,3^{+-})$}\\
\medskip
{$1945 \pm 20   ~~~~~~~~~~~~~~~~~~~~~~~~2025 \pm 20$}\\
{$2255 \pm 15   ~~~~~~~~~~~~~~~~~~~~~~~~2275 \pm 25$}\\

\bigskip

{$b_3 (1,3^{+-})~~~~~~~~~~~~~~~~~~~~~~~\rho_3(1,3^{--})$}\\
\medskip
{$2032 \pm 12   ~~~~~~~~~~~~~~~~~~~~~~~~1982 \pm 14$}\\
{$2245 ~\pm ~?   ~~~~~~~~~~~~~~~~~~~~~~~~2260 \pm 20$}\\

\end{center}

 We see clear approximate doublets of $U(1)_A$ restoration. Hence
the two distinct (1/2,1/2) multiplets of $SU(2)_L \times SU(2)_R$
can be combined into one multiplet of $U(2)_L \times U(2)_R$.
So we conclude that the whole chiral symmetry  of the
QCD Lagrangian $U(2)_L \times U(2)_R$ gets approximately 
restored high in the hadron spectrum.

\section{Chiral multiplets of excited baryons}

Now we will consider chiral multiplets of excited baryons \cite{CG1,CG2}.
The nucleon or delta states have a half integral isospin. Then
such a multiplet cannot be an irreducible representation of the
chiral group $(I_L,I_R)$ with $I_L = I_R$, because in this case
the total isospin can only be integral. Hence the minimal possible
representation that is invariant under parity transformation is
the one of (\ref{mult}).
Empirically, there are no known baryon resonances within the two
light flavors sector which have an isospin greater than 3/2.
Thus we have a constraint from the data that
if chiral symmetry is effectively restored for very highly
excited baryons, the only possible representations for the
 observed baryons have $I_L + I_R \le 3/2$,  i.e. the
 only possible representations are

\begin{equation}
\left(\frac{1}{2},0 \right) \oplus \left(0,\frac{1}{2} \right);~
\left(\frac{3}{2},0 \right) \oplus \left(0,\frac{3}{2} \right);~
\left(\frac{1}{2},1 \right) \oplus \left(1,\frac{1}{2} \right).
\label{list}
\end{equation}

\noindent
These multiplets can correspond to states of any fixed spin.

The same classification can actually be obtained assuming that the
chiral properties of excited baryons are determined by three
massless valence quarks which have a definite chirality. Indeed
 one quark field transforms as

\begin{equation}
q \sim \left (\frac{1}{2},0 \right ) \oplus \left (0,\frac{1}{2} \right).
\label{qq}
\end{equation}
Then all possible representations for the three-quark baryons
in the chirally restored phase can
be obtained as a direct product of three "fundamental" representations
(\ref{qq}). Using the standard isospin coupling rules
separately for the left and right quark components, one
easily obtains a decomposition of this direct product \cite{CJ}
and arrives at the result shown in (\ref{list}).

The $(1/2,0) \oplus (0,1/2)$ multiplets contain only isospin
1/2 states and hence correspond to parity doublets of nucleon
states (of any fixed spin).\footnote{If one distinguishes
nucleon states with different electric charge, i.e. different
isospin projection, then this ``doublet'' is actually a quartet.}
Similarly, the $(3/2,0) \oplus (0,3/2)$
 multiplets contain only isospin 3/2 states and hence correspond
to parity doublets of $\Delta$ states (of any fixed spin).\footnote{
Again, keeping in mind different charge states of a delta
resonance it is actually an octet.}
However, the $(1/2,1) \oplus (1,1/2)$ multiplets contain both
isospin 1/2 and isospin 3/2 states and hence correspond to
 multiplets containing both nucleon and $\Delta$ states of
both parities and any fixed spin.\footnote{This representation
is a 12-plet once we distinguish between different charge states.}

Summarizing, the phenomenological consequence of the effective
restoration of chiral symmetry
high in the $N$ and $\Delta$ spectra is that the baryon states
will fill out  the  approximate irreducible
representations of the parity-chiral group (\ref{gr}).
If $(1/2,0) \oplus (0,1/2)$ and $(3/2,0) \oplus (0,3/2)$
multiplets were realized in nature, then the spectra of highly excited
nucleons and deltas would consist of parity doublets. However,
the energy of the parity doublet with  a given spin in
the nucleon spectrum {\it a-priori} would not be degenerate with the
 doublet with the same spin in the delta spectrum;
these doublets would belong to different
representations,  i.e. to distinct
multiplets and their energies
are not related.   On the other hand,
if $(1/2,1) \oplus (1,1/2)$ were realized, then the highly
lying states in the $N$ and $\Delta$ spectra
would have a $N$ parity doublet and a $\Delta$
parity doublet with the same spin and which are degenerate in mass.
In either of cases the highly lying spectrum  must systematically
consist of parity doublets.

If one looks carefully at the nucleon spectrum, see Fig. \ref{ns}, and the 
delta spectrum
one notices that a systematic parity doubling in the nucleon
spectrum appears at  masses of 1.7 GeV and above, while  
parity doublets in the delta spectrum insist at  masses of 
1.9 GeV and higher. This means that a parity doubling in both cases
is seen at approximately the same excitation energy with respect
to the corresponding ground state. There are no approximately
degenerate doublets in the Delta spectrum in the 1.7 GeV region.
This fact implies that at
least those nucleon doublets that are seen at $\sim 1.7$ GeV belong
to a $(1/2,0) \oplus (0,1/2)$ representation. Chiral multiplets of
excited {\it well established} nucleons are presented in Table 3.

\begin{table}
\begin{center}
\caption{Chiral multiplets of excited nucleons.
Comments: (i) All these states are well established and
can be found in the Baryon Summary Table of the Review of Particle 
Physics. 200 MeV is taken as an interval between the consequent 
multiplets in order to evaluate the spectral overlap. (ii) There
are two possibilities to assign the chiral representation:
$(1/2,0) \oplus (0,1/2)$ or $(1/2,1) \oplus (1,1/2)$ because
there is a possible chiral pair in the $\Delta$ spectrum
with the same spin with similar mass.
 }
\begin{tabular}{|llllll|} \hline
Spin & Chiral multiplet &  Representation & $\chi$ & Spectral Overlap & Comment\\ \hline
1/2& $N_+(1710 \pm 30) - N_-(1650^{+30}_{-10})$ & $(1/2,0) \oplus (0,1/2)$ &
$0.02 \pm 0.02$  & $0.3 \pm 0.3$  & (i) \\

3/2& $N_+(1720^{+30}_{-70}) - N_-(1700^{+50}_{-50})$ & $(1/2,0) \oplus (0,1/2)$ &
$0.01^{+0.03}_{-0.01}$   &  $0.1^{+0.4}_{-0.1}$ & (i) \\

5/2&$N_+(1680^{+10}_{-5}) - N_-(1675^{+10}_{-5})$ & $(1/2,0) \oplus (0,1/2)$ &
$0.002^{+0.006}_{-0.002}$   &  $0.025^{+0.1}_{-0.025}$ & (i) \\

9/2&$N_+(2220^{+90}_{-40}) - N_-(2250^{+60}_{-80})$ &
 see comment (ii) &
$0.01^{+0.03}_{-0.01}$   &  $0.15^{+0.75}_{-0.15}$ & (i),(ii) \\

\hline
\end{tabular}
\end{center}
\label{t3}
\end{table}

Below we show $N$ and $\Delta$ high-lying states 
 in the energy range of 1.9 GeV and higher:

$${\bf J=\frac{1}{2}:}
 ~N^+(2100)~(*),~N^-(2090)~(*),~\Delta^+(1910)~~~~,~\Delta^-(1900) (**);$$

$$ {\bf J=\frac{3}{2}:}
 ~N^+(1900) (**),~N^-(2080) (**),~\Delta^+(1920)~~~~,~\Delta^-(1940)~(*);$$

$${\bf J=\frac{5}{2}:}
 ~N^+(2000) (**),~~N^-(2200) (**),~ \Delta^+(1905)~~~~,~\Delta^-(1930)~~~~;$$

$${\bf J=\frac{7}{2}:}
 ~N^+(1990) (**),~ N^-(2190)~~~~,~\Delta^+(1950)~~~~,~\Delta^-(2200)~(*);$$

$${\bf J=\frac{9}{2}:}
 ~N^+(2220)~~~,~N^-(2250)~~~~,~\Delta^+(2300) (**),~\Delta^-(2400)(**);$$

$${\bf J=\frac{11}{2}:}
 ~~~~~~~~~?~~~~~~~~,~N^-(2600)~~~~,~\Delta^+(2420)~~~~,~~~~~~~~?~~~~~~~~;$$

$${\bf J=\frac{13}{2}:}
 ~N^+(2700) (**), ~~~~~~~?~~~~~~~~,~~~~~~~~?~~~~~~~~,~\Delta^-(2750)(**);$$

$${\bf J=\frac{15}{2}:}
 ~~~~~~~~?~~~~~~~~,  ~~~~~~~~?~~~~~~~~,~\Delta^+(2950) (**),~~~~~~~~?~~~~~~~~.$$

Obviously, the 1- and some of the 2-star states according to the
Particle Data Group classification  should not be taken
very seriously. So, the most reliable prediction of the chiral symmetry
restoration is the existence of the approximately degenerate chiral
partners of the high-lying well established states. There are two states
of this kind in the nucleon spectrum: $11/2^-, N(2600)$ and
$7/2^-, N(2190)$. So an experimental discovery of the lowest  
$11/2^+$ state is of crucial importance. A confirmation as well as
a refinement of the mass of the lowest $7/2^+$ state is also important.

It is difficult to say whether one sees or not 
an approximate degeneracy between
the  $N$ and $\Delta$ doublets. The reason is that there are no "quartets"
where at least three resonances are well established. Then  it 
is an open question
which particular representations are realized
in the spectrum above 1.9 GeV.
If an approximate mass degeneracy between some $N$ and $\Delta$ doublets 
at $M \geq 1.9$ GeV is accidental, then the baryons in this mass
region are
organized according to 
$(1/2,0) \oplus (0,1/2)$ for $N$ and
$(3/2,0) \oplus (0,3/2)$ for $\Delta$ parity-chiral doublets. If not,
then the high lying spectrum forms $(1/2,1) \oplus (1,1/2)$ multiplets.
It can also be possible that in the narrow energy interval
more than one parity doublet in the nucleon and delta spectra
is found for a given spin. This would then mean that different
doublets  belong to different parity-chiral multiplets.
Systematic experimental exploration of the high-lying states
is required in order to assign unambiguously baryons to
the multiplets.

Even though we cannot assign unambiguously the high-lying states
to specific chiral multiplets, a direct implication of the chiral symmetry
restoration is, in any case, that there must be approximately degenerate chiral
partners for  well-established states. Hence, like in the nucleon
spectrum, there are at least a few definitely missing states in the
Delta spectrum. A verification of the 1-star
$3/2^-, \Delta(1940)$ state as well as a discovery of the lowest 
$7/2^-$ and $11/2^-$ states is of extraordinary importance.

\section{Generalized linear sigma-model. Mesons.}

How should the effective chiral Lagrangian for the approximate
chiral multiplets look like?
There is only one chiral multiplet of fields $(\sigma ,\vec \pi)$ 
in the standard  $SU(2)_L \times SU(2)_R$ invariant
linear sigma-model\footnote{Here $\vec \pi$ means a vector in the isospin
space.}
of Gell-Mann and Levy \cite{GL}. In the Wigner-Weyl mode
the pion and sigma mesons are strictly degenerate, what is constrained
by chiral symmetry, while in the Nambu-Goldstone mode the pion becomes
massless. This kind of model cannot illustrate  physics
relevant to highly-excited states.
Let us extend the sigma-model and introduce an infinite
amount of excited chiral pairs $( \sigma_j, {\vec \pi}_j)$ \cite{CG3}. 
In this respect
the model mimics large $N_c$ QCD.  These fields enter the
Lagrangian in a chirally invariant way as members of $\left (
\frac{1}{2},\frac{1}{2} \right )_a $ chiral multiplets:
\begin{eqnarray}\left [V^a,\pi^b_j \right]  =  
i \,\epsilon^{a b c} \, \pi^{c}_j
&{}& \left [ V^a,\sigma_j \right ]  =  0 \nonumber\\
\left[A^a,\pi^b_j \right] = i \delta_{a b} \, \sigma_j \; &{}&
\left[A^a,\sigma_j \right] = i \pi^a_j
\label{sigtr}
\end{eqnarray}
where $V^a$  ($A^a$) represent the generators of vector (axial)
rotations and $a,b$ are isospin indices.  The key point of the
chiral invariance is that the axial rotation transforms
the chiral partners of opposite parity into each other.

The chiral group $SU(2)_L \times SU(2)_R$ is isomorphic to the $SO(4)$
and hence describes rotations  of 
$( \sigma_j, {\vec \pi}_j) \equiv (\sigma_j, \pi^1_j, \pi^2_j, \pi^3_j)$
in the 4-dimensional space. The chiral-invariant Lagrangian must consist
of only scalars with respect to the chiral rotations and hence can contain
any possible scalar products  
$( \sigma_i, {\vec \pi}_i) \cdot ( \sigma_j, {\vec \pi}_j)$. To this
type belong the kinetic energy terms,
$ \partial^\mu  \sigma_j \,
\partial_\mu \sigma_j \, +
\partial^\mu  \vec{\pi}_j \cdot \partial_\mu \vec{\pi}_j $,
the mass terms, $ (\sigma_j^2 \, + \vec{\pi_j} \cdot
\vec{\pi_j})$, as well as all possible interaction terms containing
invariants $\left (  \sigma_i \sigma_j  +
 \vec{\pi}_i \cdot \vec{\pi}_j  \right )$. The coupling constants are
 input parameters. We want to ensure the effective chiral symmetry 
 restoration in the high-lying mesons. Physically the effective chiral
 restoration in QCD means that excited hadrons decouple from the quark 
 condensates of the vacuum. In the sigma-model the corresponding order
 parameter is the vacuum expectation of the $\sigma$-field in the 
 Nambu-Goldstone phase. We construct the model in such a way that
 only the ground state field $\sigma_1$ acquires the nonzero vacuum
 expectation value. Then  the  chiral restoration
 in a given pair $( \sigma_k, {\vec \pi}_k)$ is provided when the coupling
 constant of this chiral pair to the ground state pair 
 $( \sigma_1, {\vec \pi}_1)$ is small and asymptotically vanishes.

Then the Lagrangian is given by

\begin{eqnarray}
{\cal L} &=& \sum_j \frac{1}{2} \left ( \partial^\mu  \sigma_j \,
\partial_\mu \sigma_j \, +
\partial^\mu  \vec{\pi}_j \cdot \partial_\mu \vec{\pi}_j  \right )
 - \frac{m_o^2}{2} \left ( \alpha (\sigma_1^2 + \vec{\pi}_1 \cdot
\vec{\pi}_1 )  +  \frac{g}{2 m_o^2} (\sigma_1^2 + \vec{\pi}_1
\cdot \vec{\pi}_1 )^2 \right ) \nonumber \\ & - &  \frac{m_o^2}{2}
\sum_{j=2}^\infty \, \left( j^2 (\sigma_j^2 \, + \vec{\pi_j} \cdot
\vec{\pi_j}) \, + \, \frac{g}{j \, m_o^2 \,} \left ( (\sigma_1
\sigma_j + \vec{\pi}_1\cdot \vec{\pi}_j)^2 \, + \, (\sigma_1^2  +
\vec{\pi}_1\cdot \vec{\pi}_1) \, ( \sigma_j^2 + \vec{\pi}_j\cdot
\vec{\pi}_j)  \right )\right) \label{L}
\end{eqnarray}

\noindent
where $m_o$ has the dimension of a mass and $\alpha$ and $g$ are
dimensionless constants.

 In the
chirally broken phase $\sigma_1$ (and no other fields) acquires a
vacuum expectation value and the excitation associated with
$\pi_1$ becomes massless Goldstone boson. The parameter $\alpha$ controls
the spontaneous symmetry braking; $\alpha>0$ yields the Wigner-Weyl
mode while $\alpha<0$ yields the Nambu-Goldstone mode. 

In the weak coupling limit, $g \ll 1$, the theory 
can be treated classically and  loop contributions can
be neglected. However,  even
in the weakly coupled limit the interaction terms play an essential
role when $\alpha < 0$ since it determines the amount of chiral
symmetry breaking.

It is easy to see that the
minimum of the potential is given by conditions like in the standard
sigma-model:
\begin{eqnarray}
\langle \sigma_j \rangle & =&  0 \; \; \; \; {\rm for} \, \, \alpha > 0 \nonumber \\
\langle \sigma_j \rangle & = & \pm \delta_{j 1} \, m_o \,
\sqrt{\frac{-\alpha}{g}} \; \; \; \; {\rm for} \, \, \alpha \le 0
\; .\end{eqnarray} 

\noindent
The mass spectrum in the Wigner-Weyl and Nambu-Goldstone modes is:
\begin{equation}
 {\rm for}\, \,  \alpha > 0   \; \left \{ \begin{array}{l}
m^2_{\pi_1} =  \alpha m_o^2  \\ \\
m^2_{\sigma_1}  = \alpha m_o^2  \\ \\
m^2_{\pi_j}  = j^2 m_o^2 \; \; (j \ge 2) \\ \\
m^2_{\sigma_j}  =  j^2 m_o^2\; \; (j \ge 2)
\end{array} \right . \nonumber
\end{equation}
\begin{equation}
{\rm for} \, \,  \alpha \le 0 \; \left \{ \begin{array}{l}
m^2_{\pi_1} =  0 \\ \\
m^2_{\sigma_1}  = - 2 \alpha m_o^2  \\ \\
m^2_{\pi_j}  =  j^2  m_o^2+ \frac{2 g\langle \sigma_1 \rangle^2
}{j}
\\\; \; \; \; \; =\left( j^2 + \frac{2 \alpha}{j} \right ) m_o^2\; \; (j \ge 2) \\ \\
m^2_{\sigma_j}  =   j^2 m_o^2 + \frac{4  g \langle \sigma_1
\rangle^2 }{j}
\\\; \; \; \; \; =  \left ( j^2 + \frac{4 \alpha}{j} \right)  m_o^2 \; \; (j \ge 2)
\end{array} \right .
\end{equation}

\begin{figure}
\begin{center}
\includegraphics*[width=7cm]{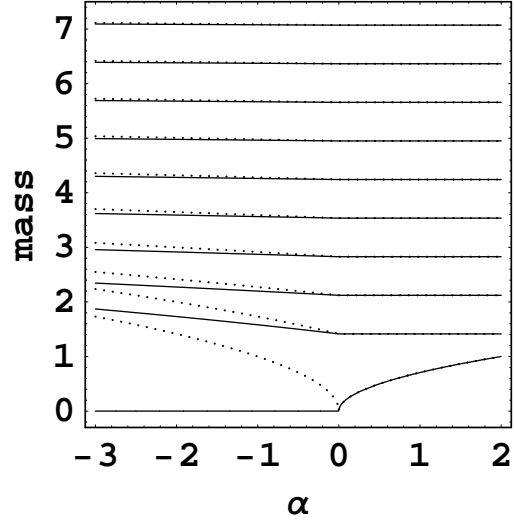}
\end{center}
\caption{The mass spectrum of the model of Eq.~(\ref{L}) in units
of the mass parameter $m_o$.  The solid lines correspond to pions
while the dotted lines correspond to $\sigma$-mesons. \label{toy}}
 \end{figure}

  The
spectrum is shown in Fig.~\ref{toy}.

In the Wigner-Weyl mode all $(\sigma_j , \vec \pi_j)$
states occur in chiral pairs and each pair has a  non-zero chiral-invariant
mass. In the Nambu-Goldstone mode the chiral symmetry is broken. In this
case the ground state pion turns into the Goldstone mode with zero mass. A
degeneracy in all otherwise chiral pairs is lifted, because the mass of each
excited meson consists of the chiral-invariant part, $jm_0$, and the
chiral-non-invariant part which is due to the coupling of the given state
with the chiral order parameter, the vacuum expectation value of the
ground state sigma-field. The smaller  the coupling, which is regulated
by $g/j$, the smaller is the chiral-non-invariant piece of the meson
mass.

In the
spontaneously broken phase this model
exhibits the phenomenon of effective
chiral restoration.   While the lowest-lying states have no hint of a
chiral multiplet structure, as one goes higher in the spectrum the
states fall into nearly degenerate multiplets which to
increasingly good approximation look like  pions and
$\sigma$-mesons in linearly realized $\left (
\frac{1}{2},\frac{1}{2} \right )_a$ representations.
Indeed, the mass splitting between the excited sigma and pi-mesons,
$\Delta m_j = |m_{\pi_j} - m_{\sigma_j}|$,
scales like $1/j^2$ and vanishes asymptotically high. 
At large $j$ the masses become increasingly insensitive
to the chiral order parameter $\sqrt{g} \langle \sigma_1 \rangle$
and the spectrum approximates a
Wigner-Weyl mode spectrum increasingly accurately.
 Obviously, the near degeneracy of the  parity
doublets reflects the underlying chiral symmetry of the model. 
It happens because the high-lying states gradually decouple from
 the
chiral symmetry breaking order parameter and are insensitive to its
magnitude:

\begin{eqnarray}
\frac{\partial m_{\sigma_j}}{\partial  \left(\sqrt{g} \langle
\sigma_1 \rangle \right) } & = & \frac{4 m_o \sqrt{-\alpha}}{ j
m_{\sigma_j}}
\rightarrow \frac{4 \sqrt{-\alpha} }{ j^2 }, \nonumber \\
\frac{\partial m_{\pi_j}}{\partial \left ( \sqrt{g} \langle
\sigma_1 \rangle \right ) } & = & \frac{2 m_o \sqrt{-\alpha}}{ j
m_{\pi_j}} \rightarrow \frac{2 \sqrt{-\alpha}}{ j^2 },
\end{eqnarray}
where the arrow indicates the asymptotic behavior at large $j$.

This extended linear $\sigma$-model explicitly demonstrates
chiral restoration high in the spectrum.
Then a question arises
whether this model illustrates some generic behavior or it is
only specific to the linear realization of the chiral symmetry?
Indeed,
in the Nambu-Goldstone mode one can always  make a field
redefinition to the standard nonlinear realization \cite{WE,C,C1}
 in which fields
of opposite parity are decoupled,
i.e. do not transform into each other under chiral
transformation. In this case the act of making an axial rotation
does not transform a field into its chiral partner but instead
creates a massless Goldstone boson (pion) from the vacuum.
Can it prevent \cite{jaffe} the chiral restoration high in the spectrum?
The answer is obviously negative, because
 the Lagrangian of Eq.~(\ref{L}) rewritten in terms
of these new fields cannot alter the spectrum.
This reflects the general situation that  field redefinitions
themselves cannot modify any physical content of a theory. In
this context it is useful to recall that the physics is not in the
fields, but in the states, which appear once one applies fields on
the vacuum. The physics of these states is controlled only by the
microscopical theory.

The model  illustrates many of the salient
points relevant to the issue.
Firstly, it illustrates the most important point: while the physics of the
low-lying states is crucially determined by the spontaneous breaking
of chiral symmetry, in the high-lying states the effects of
chiral symmetry breaking represent only a small correction.
Secondly, it shows the
gradual nature of the conjectured effect.  The effect is never
absolute but always approximate; for any given strength of the
coupling $\alpha$ it becomes increasingly accurate as one goes up
in the spectrum. Thirdly, it makes very clear that the key issue
is the coupling of the state to the dynamics responsible for
spontaneous chiral symmetry breaking---in this case the coupling
to $\sqrt{g} \langle \sigma_1 \rangle$ which plays the role of the 
chiral order parameter. Once the coupling of a state to the chiral
order parameter becomes small, then the mass of the state almost
entirely has a chiral-invariant nature. Hence the splitting between
chiral partners becomes small and the state decouples from the
Goldstone bosons.

It is instructive to clarify the generic reason for the decoupling of
the states with approximate chiral symmetry from the Goldstone
bosons. The structure of the coupling of the given pair 
$(\sigma_j, \vec \pi_j)$ to the ground state pair $(\sigma_1, \vec \pi_1)$
is prescribed by chiral symmetry: The interaction terms must be
chiral-invariant and hence must depend on the scalar products
$(\sigma_1 \sigma_j + \vec \pi_1 \cdot \vec \pi_j)$. 
Then the coupling of a given state
to the chiral order parameter, $\sqrt{g}\langle \sigma_1 \rangle$,
 proceeds necessarily in parallel with the
coupling of the same state to the $\pi_1$ field (which is the Goldstone
boson field in the Nambu-Goldstone phase), with exactly the
same coupling strength. Then if the state in question decouples
from the order parameter, it necessarily decouples from the Goldstone
bosons.

It is rather clear that the chiral restoration has been built in  
the model. The first assumption which is crucial is that all states
are in chiral pairs. This means that this kind of model can illustrate
physics only for that part of the real spectrum where the states
can be systematically organized into chiral multiplets. For that
a one-to-one mapping of the corresponding opposite parity states
is required. Certainly there is no such a mapping for the low-lying
states in the real hadron spectrum. This implies that the chiral
symmetry breaking effects are so strong here that the linear
realization (which is broken by the vacuum) is not adequate and 
instead only the nonlinear
realization of chiral symmetry can be applied in this part of the
spectrum. Within the nonlinear 
realization of chiral symmetry the one-to-one mapping is not required
since the axial transformation of a hadron creates the same hadron
and pions, rather than a hadron of opposite parity. Clearly the
present kind of models cannot explain a transition from the low-lying
part of the spectrum with the dominant  nonlinear effects to the high-lying
part of the spectrum with the approximate linear realization, where 
chiral symmetry breaking can be considered only as a small perturbation.
A microscopical picture is required to interpolate both parts of the
spectra. 

The second crucial assumption is that the coupling of the states to the
chiral order parameter scales like $1/j$ and hence for large $j$ the
splittings between the chiral partners disappear. Actually we could choose
any other scaling laws, like $1/j^2$, etc. 
All parameters of this effective model are  input parameters
and have been chosen in such a way that guarantees the effective chiral
restoration. No justification  why these parameters
should behave in a prescribed way can be given within the model.
This requires a microscopical approach.

What else distinguishes this model from QCD is that here chiral symmetry
is broken classically, at the tree level,
while in QCD chiral symmetry breaking is a
quantum phenomenon. This means that within a microscopical QCD
picture some tree parameters of this effective model have in fact
a quantum origin.

\section{Generalized linear sigma-model. Baryons.}

Here we outline a   model that exhibits effective
chiral restoration in excited baryons, which is similar to the
meson  model of the previous section. All critical remarks at
the end of the previous section equally apply in the present case.

To this end we need an effective Lagrangian that allows  baryons 
of opposite parity to be in chiral multiplets. It is an important point,
because a usual chiral-invariant
massless fermion field, for example a quark, does not have an
independent chiral partner of opposite parity - there are no quarks
of negative parity. In this case its chiral partner is simply $\gamma_5 q$
and a pair of fields $(q, \gamma_5 q)$ transform into each other under
a chiral transformation according to $(0,1/2) \oplus (1/2,0)$.

A framework how to construct a chiral-invariant
Lagrangian with two massive baryon
fields of opposite parity that are chiral partners, has been given
by Lee in his text long ago \cite{LEE}.
The most important element of the Lee
model is that baryons of opposite parity have a 
nonzero {\it chiral invariant} mass and transform into each other
under an axial rotation. This is similar
to the $(\sigma, \vec \pi)$ chiral pairs of the previous section.
It is this feature which distinguishes Lee's model from the
Gell-Mann - Levy sigma model, where a baryon of only one parity participates
and hence its chiral-invariant mass must be necessarily zero.

Lee's model required that in the
Nambu-Goldstone mode pions decouple from baryons. Clearly it is
not the case in the low-lying baryons and consequently he dismissed this
model as "physically uninteresting". 
  DeTar and
Kunihiro  have added into Lee model one more chiral-invariant meson-nucleon
coupling term and adjusted it
to the low-lying
baryon phenomenology of N and N(1535) so that  $g_{\pi N_-N_-}$
and $g_{\pi N_+N_+}$ become not zero \cite{DETAR}.
 This model has intensively 
been used for the description of the lowest baryons of positive
and negative parity N and N(1535)
and reviewed in great detail in \cite{TIT}. There was
also an attempt to extend the model to some other low-lying
(and not so low-lying) baryons in ref. \cite{JIDO}. 

However, in the low-lying baryons  the chiral symmetry 
breaking effects
are very strong and consequently there is not a one-to-one mapping of
baryons of opposite parity \cite{CG1,CG2}. In this domain
the nonlinear realization of chiral symmetry \cite{WE,C,C1}
should be considered,
which does not transform the baryons of opposite parity into each other
and hence does not require the one-to-one correspondence of positive
and negative parity states \cite{jaffe,JPS}. Nevertheless, this model can be 
used as an effective  model for the high-lying baryons,
where chiral symmetry breaking effects are expected to be
only a small perturbation. In this part of the baryon
spectrum it is natural to expect
an approximate parity doubling \cite{G1,CG1,CG2,G}. 
This particular domain of application is
discussed below.

Consider a pair of the isodoublet fermion fields 

\begin{equation}
\Psi = \left(\begin{array}{c}
\Psi_+\\
\Psi_-
\end{array} \right),
\label{doub}
\end{equation}

\noindent
where the bispinors
$\Psi_+$ and $\Psi_-$ have positive and negative parity, respectively,
because the parity  on the doublet space is defined to be

\begin{equation}
P:~~~~ \Psi(\vec x, t) = \sigma_3 \gamma_0 \Psi(-\vec x, t).
\end{equation}

\noindent
The chiral transformation law
 under the $(0,1/2) \oplus (1/2,0)$ representation 
of $SU(2)_L \times SU(2)_R$ is defined as

\begin{equation}
\Psi \rightarrow 
\exp \left( \imath \frac{\theta^a_V \tau^a}{2}\right)\Psi; ~~
\Psi \rightarrow 
\exp \left(  \imath \frac{\theta^a_A\tau^a}{2} \sigma_1
\right)\Psi.
\label{VAD}
\end{equation}

\noindent
Here $\sigma_i$ is a Pauli matrix that acts in the  $2 \times 2$
space of the parity doublet. While in the  chiral transformation
law (\ref{VA}) the axial rotation mixes the massless Dirac spinor $q$
with $\gamma_5 q$ , in the present  transformation
 a mixing of two fields $\Psi_+$ and $\Psi_-$ is provided.
Then  the chiral-invariant Lagrangian of the free parity doublet is given as

\begin{eqnarray}
\mathcal{L}_0 & = & i \bar{\Psi} \gamma^\mu \partial_\mu \Psi - m_0 \bar{\Psi}
\Psi  \nonumber \\
& = &  i \bar{\Psi}_+ \gamma^\mu \partial_\mu \Psi_+ + 
i \bar{\Psi}_- \gamma^\mu \partial_\mu \Psi_-
- m_0 \bar{\Psi}_+ \Psi_+ - m_0 \bar{\Psi}_- \Psi_- 
\label{lag}
\end{eqnarray}

\noindent
 A crucial
property of the Lee Lagrangian (\ref{lag}) is that  the fermions
$\Psi_+$ and $\Psi_-$ are exactly degenerate and
have a nonzero chiral-invariant mass $m_0$. In contrast, for
usual  "naive" fermions chiral symmetry restricts particles to be massless.

From the axial transformation law (\ref{VAD}) one can read off the
axial charge matrix, which is $\gamma_5 \sigma_1$. Hence the diagonal axial
charges of the opposite parity baryons are exactly 0, $g_+^A = g_-^A = 0$,
while the off-diagonal axial charge is 1,  $ |g_{+-}^A| = |g_{-+}^A| = 1$.
This is another crucial property that distinguishes the parity
doublets from the "naive" fermions where $g^A = 1$.

Note that the same chiral-invariant Lagrangian can be written in different
forms. This can be achieved by a redefinition of the baryon fields. For
instance, defining new baryon fields as

\begin{equation}
N_+ = \frac{1}{\sqrt 2}(\Psi_+ + \gamma_5 \Psi_-) ~~~~~~~~~
N_- = \frac{1}{\sqrt 2}(\gamma_5 \Psi_+ - \Psi_-), 
\label{nva}
\end{equation}

\noindent
one arrives at the Lagrangian with the free mass term

\begin{equation}
-m_0(\bar N_+ \gamma_5 N_- - \bar N_- \gamma_5 N_+),
\label{nm}
\end{equation}

\noindent
used in \cite{TIT}.

One can add the chiral-invariant interaction of the baryon fields 
with the chiral doublet
 $( \sigma, \vec \pi)$ fields that satisfy the meson part of the
Gell-Mann - Levy $\sigma$-model. Then there are two possible
independent interaction terms and the Lagrangian takes the form:

\begin{eqnarray}
\mathcal{L} & = & i \bar{N}_+ \gamma^\mu \partial_\mu N_+ + 
i \bar{N}_- \gamma^\mu \partial_\mu N_-
-m_0(\bar N_+ \gamma_5 N_- - \bar N_- \gamma_5 N_+)  \nonumber \\
& & -g_+\bar{N}_+(\sigma + i \gamma_5 \vec \tau \cdot \vec \pi)N_+
-g_-\bar{N}_-(\sigma - i \gamma_5 \vec \tau \cdot \vec \pi)N_-
+ \mathcal{L}(\sigma,\vec \pi).
\label{lag2}
\end{eqnarray}

\noindent
All tree-level parameters $m_0,g_+,g_-$ of this effective Lagrangian
are input parameters. The coupling constants $g_+,g_-$ induce
the coupling of the baryons with the $\sigma$ field and hence their
values will determine the chiral-non-invariant part of the baryon masses
in the Nambu-Goldstone mode. The chiral symmetry structure of the
couplings requires, at the same time, that the baryon fields in the
Nambu-Goldstone mode will couple with the Goldstone bosons. 

In the Nambu-Goldstone mode,
one chooses a chiral-non-invariant vacuum in the standard way
with the vacuum expectation value 
$\langle 0 | \sigma | 0 \rangle = \sigma_0 = f_\pi$.
Then one can  diagonalize the tree-level Lagrangian
and we arrive at the following mass eigenvalues of the
positive and negative parity baryons \cite{DETAR,TIT}

\begin{equation}
m_\pm = \frac{1}{2}\left(\sqrt{(g_+ + g_-)^2\sigma_0^2 + 4m_0^2}
\pm (g_+ - g_-)\sigma_0 \right).
\label{masses}
\end{equation}

Hence the chiral symmetry gets broken in the particle
spectrum: $m_+ \neq m_- \neq m_0$ ; $m_\pi = 0$; $m_\sigma \neq 0$;
$ g_{\pi N_\pm N_\pm} \neq 0$; $ g_{\pi N_+ N_-} \neq 0$. 
 This is because in the Nambu-Goldstone mode in addition to  the 
chiral-invariant mass $m_0$
baryons acquire a chiral-non-invariant piece of mass according
to the standard Gell-Mann - Levy scenario. As was mentioned above this model
has intensively been used  for  low-lying baryons
{\it in the strong coupling regime}.

However, it cannot be used as effective field theory of 
low-lying baryons.
Firstly, in the strong coupling regime, $g_\pm \gg 0$, quantum
loop effects are crucially important and the tree-level masses 
(\ref{masses}) should be rather far from the actual dressed masses.
Secondly, the chiral symmetry breaking is so strong in the low-lying
baryons that there are no actual chiral partners, because there is
not a one-to-one mapping of the positive and negative parity states in
this part of the spectrum. Only the nonlinear realization applies here,
which does not transform baryons of opposite parity into each other and
hence does not require such a mapping.

 We have already discussed in Sec. 8, that all excited nucleons in
the 1.7 GeV mass region systematically fall into 
$(0,1/2) \oplus (1/2,0)$ representations and are nearly degenerate. 
What will be the effective field theory for these and other 
higher-lying baryons \cite{CGHJO}?
The effective restoration of chiral symmetry means that the hadrons
decouple from the quark condensate and are in approximate chiral
multiplets. In the present case it is the
vacuum expectation value of the sigma field, $\sigma_0$,  which plays 
the role of the
quark condensate in QCD. A coupling to this chiral order parameter
is regulated by the coupling constants $g_+, ~g_-$, which are input parameters. 
Thence, the effective restoration of chiral symmetry requires these
couplings to be small. If they are small,
i.e. we are in the  weak coupling regime, then it {\it is} legitimate to use
the Lagrangian above at the tree level and to consider  loop corrections
to it as a perturbation.

Hence in the weak coupling regime, 
$ g_\pm \rightarrow 0; ~~~  g_\pm \sigma_0 \ll m_0$,
in the Nambu-Goldstone mode with $\sigma_0$ being fixed, 
we arrive at

\begin{equation}
m_+ \approx m_- \approx m_0;~~ g^A_+ \approx 0;~~ g^A_- \approx 0;
~~ g_{\pi N_+ N_+} \approx 0;~~ g_{\pi N_- N_-} \approx 0; ~~
g^A_{+-} \approx 1; ~~ g_{\pi N_+ N_-} \approx 0.
\label{constants}
\end{equation}

\noindent
A fundamental reason for these relations is that in this regime
baryons decouple from the chiral order parameter $\sigma_0$.\footnote{This
regime of effective symmetry restoration should not be confused with the
symmetry restoration at high temperatures or densities, $\sigma_0 \rightarrow 0$,
studied in \cite{DETAR,TIT}.}
Consequently
the chiral-non-invariant part of their mass becomes small.
If the chiral-non-invariant part of the baryon mass becomes small
and asymptotically vanishes, then this baryon necessarily decouples
from the Goldstone bosons, because the coupling of the baryon to
the chiral order parameter and to the Goldstone boson is regulated
by the same parameters $g_\pm$. This latter feature is one of the
most important implications of the chiral invariance.
So, like in the
previous section, one concludes: if the chiral-non-invariant part
of the baryon masses becomes small and asymptotically vanishes,
the chiral partners get approximately degenerate and
 pions decouple from such baryons. This is a highly
nontrivial prediction of effective chiral restoration in excited hadrons.
Since the coupling of  baryons with the pion becomes small for
approximate chiral multiplets, a systematic
loop expansion can be organized with the tree-level masses and
other constants being the
first approximation. The same equally applies to the meson model of the
previous section.\footnote{Actually, there is a regime $g_+ = g_- \gg 0$
within this model, where the tree-level masses are also
degenerate. In this case a large part of the baryon mass is not
chiral invariant and baryons do not decouple from pions. We refer
this regime to as "accidental degeneracy" and it should not be mixed up
with the effective chiral restoration regime.}

Now we want to address questions raised in ref. \cite{jaffe,JPS}. Will
physics look differently if we use instead a nonlinear realization
of chiral symmetry? Certainly, in the Nambu-Goldstone mode
we can rewrite the chiral Lagrangian in terms of the new baryon
fields that transform nonlinearly. In the latter case the axial
rotation does not connect the chiral partners with each other, like in
(\ref{VAD}), but instead it  transforms each baryon into itself, plus
a number of pions. A coupling of the baryons to the Goldstone bosons
splits "would be" degenerate baryons of opposite parity. Then the
effective chiral restoration requires all the coefficients of the
chirally allowed operators ($m_1,c_{2-4}$ in notations of \cite{jaffe,JPS})
of the effective chiral Lagrangian be suppressed, which looks surprisingly,
at least on the first value.

All this is correct but somewhat misleading. The reason is that in
the nonlinear realization the spontaneous breaking of chiral
symmetry is built in from the outset; the underlying dynamics
is not explicit and encoded in the coefficients of various
tree-level operators. The effective chiral restoration indeed
requires a conspiracy among the coefficients. However, if the states
in question decouple from the quark condensate, then all numerous
unconnected coefficients of the effective nonlinear chiral Lagrangian
which break chiral symmetry are suppressed automatically, without any
fine tuning. The conspiracy turns out to be a simple consequence
of effective chiral restoration.
The reader can check it straightforwardly: Take the Lagrangian
(\ref{lag2}) in the effective symmetry restoration regime ($\sigma_0 = const$,
$g_+,g_- \rightarrow 0$) and rewrite it in terms of the new nonlinear
fields. Then it will turn out that all  coefficients  
$m_1,c_{2-4} \rightarrow 0$. The reason for such a suppression has actually 
been clarified above: The coupling of the states in question
to the Goldstone bosons and to the quark condensate proceeds
simultaneously. It is a fundamental requirement of chiral symmetry. If
the states decouple from the quark condensate, and their
mass becomes predominantly chiral-invariant, then they necessarily
decouple from the Goldstone bosons.

\section{Solvable confining and chirally symmetric models}

\subsection{'t Hooft model}

\begin{figure}
\begin{center}
\includegraphics*[width=7cm]{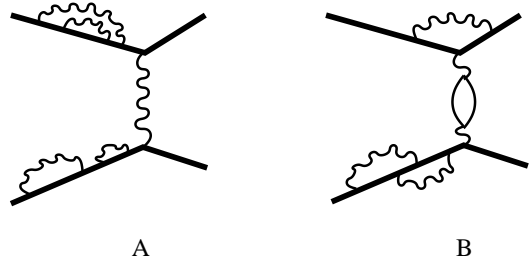}
\end{center}
\caption{ Allowed in $N_c=\infty$ planar  (A) 
and suppressed non-planar (B) diagrams.}
\label{planar}
\end{figure}

It is natural to ask a question about chiral restoration in excited
hadrons within the exactly solvable 't Hooft model 
\cite{HOOFT1,CCG,BG,LI,LENZ,ZHI,NK}. The 't Hooft model is QCD in the
large $N_c$ limit in 1+1 dimensions. In the large $N_c$ limit only
the planar diagrams survive like those shown in Fig. \ref{planar} and
there are no vacuum fermion loops as well as vertex corrections. Once
a gauge is appropriately chosen a complicated nonlinear gluonic
field is reduced to an instantaneous linearly rising Coulomb potential.
In the weakly coupled regime $N_c \rightarrow \infty$,
$m_q \gg g \sim 1/\sqrt N_c$ the theory can be solved exactly and
all  interesting quantities like the spectrum of hadrons, the quark condensate,
etc. can be calculated.

\begin{figure}
\begin{center}
\includegraphics*[width=15cm]{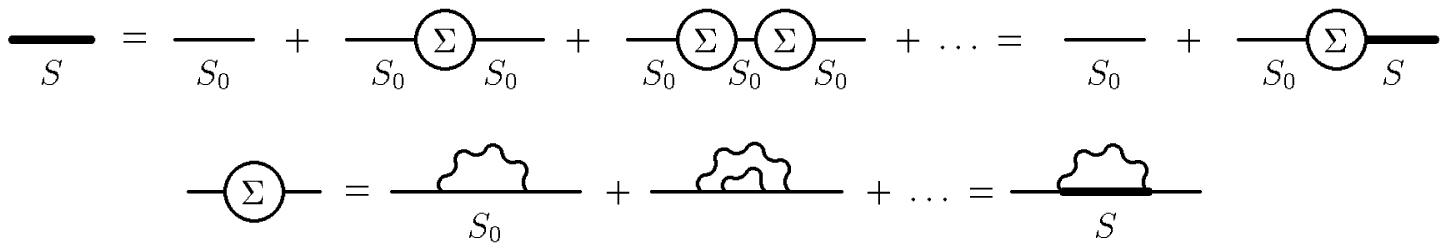}
\end{center}
\caption{ Dressed quark Green function and Schwinger-Dyson equations.}
\label{sigma}
\end{figure}

It is instructive to outline how this field theory is solved. We shall
not present here any formulae, because similar expressions will
appear in the next section where we will be discussing a 3+1 dimensional
generalization of the 't Hooft model. If we are interested in the
quark-antiquark bound states, then the first step is to find the dressed
quark Green function. To this end one must solve the Dyson-Schwinger
equation or related gap equation (this theory is very similar to the
BCS theory of superconductivity), see Fig. \ref{sigma}.

\begin{figure}
\begin{center}
\includegraphics*[width=7cm]{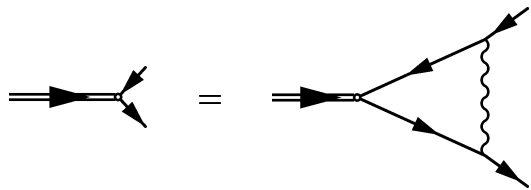}
\end{center}
\caption{ Homogeneous Bethe-Salpeter equation for the quark-antiquark
bound states.}
\label{bs}
\end{figure}

\noindent
 As an output 
the chiral symmetry
gets broken and we find a dynamical mass of quarks as well as a quark
condensate. Given the dressed quark Green function one is to solve
as the second step the homogeneous
Bethe-Salpeter equation for the quark-antiquark
bound states, see Fig. \ref{bs}. 
With the instantaneous linearly rising potential the
loop integrals are infrared divergent. Consequently the single
quark Green function is divergent. However, the same singularity appears
in the kernel of the Bethe-Salpeter equation and cancels out exactly
the singularity of the quark Green function. As a result the physically
observable color-singlet mass of a meson is a finite and well defined
quantity. Hence to perform this program one has to choose any possible
infrared regularization and in the final result for meson mass
take the infrared limit. The final result does not depend on the type
of the infrared regulator. The same is true with respect to the chosen
gauge. The quark self-energy is a gauge-variant quantity, while the
color-singlet meson mass is a gauge-invariant one. The Lorentz
covariance is also manifest only for  color-singlet quantities,
while it is explicitly broken for  gauge-variant quantities like
the dressed quark propagator in the Coulomb gauge \cite{BG}.

There is no rotation and spin in 1+1 dimensions. Hence the meson states
 are characterized only by  the radial
 quantum number $n$ and parity. The ground state, the "pion", has
 zero mass in the chiral limit. Asymptotically the states lie
 on the linear radial Regge trajectory
 
\begin{equation}
 M_n^2 = N_c g^2 \pi n, 
\label{hr}
\end{equation}

\noindent
with the correction term being $\sim \ln n$.  
States with  subsequent values of $n$ have
 opposite parities. Then it is not possible to identify chiral
multiplets. Effective chiral restoration does not occur in 1+1 dimensions
and the chiral symmetry is always strongly broken in all states, irrespective
how high in the spectrum we are.

However, this is specific to 1+1 dimensions where rotational motion of
quarks is impossible. The valence quarks can perform only an
oscillatory motion. Then at the turning points, where the quarks
get slow, the chiral symmetry is maximally violated.
To observe the effective chiral restoration we
have to consider a similar model in 3+1 dimensions.
In the latter case the valence quarks can rotate and hence
can be always ultra-relativistic. This model is discussed in what 
follows.

\subsection{ Generalized Nambu and Jona-Lasinio model. Introduction to the model 
and general remarks }

Only solution of QCD
can give answers about actual masses and wave functions of
all excited states. However, there is no hope that QCD can be solved
analytically. Lattice studies cannot help us either, because it
is an intrinsic difficulty to extract on the lattice highly
excited states from the Euclidean correlation functions. Hence at
the present stage the only useful tool to address the problem is
modeling.

The model must contain all principal elements of QCD that are relevant
to the present problem. It must be (i) 3+1 dimensional, (ii) exactly
solvable,
(iii) relativistic and field-theoretical in nature, (iv) chirally
symmetric, (v) confining, (vi) chiral symmetry must be dynamically
broken, (vii) the axial vector current must be conserved. 
Such a model can be used as a test to verify the principal
question of whether effective chiral restoration occurs or not
and if it does occur it can be used as a laboratory to understand
the microscopical mechanism of the restoration.
It is highly nontrivial to meet all these requirements within
one and the same model. For instance, the Nambu and Jona-Lasinio
model \cite{NJL,VWNJL} or a specific model like the instanton-liquid model
\cite{SD} cannot be applied to excited hadrons, because these
models are not confining. On the contrary, potential constituent quark
models  do not satisfy the conditions (iii) and (iv).

Such a model does exist, however. It is a generalized Nambu and Jona-Lasinio
model (GNJL) \cite{orsay}. This model can be considered as 
a generalization
of the large $N_c$ 1+1 dimensional 't Hooft model to 3+1 dimensions. It
is postulated within this model that there exists an 
instantaneous Lorentz-vector Coulomb-like linear potential between quarks.
Then chiral symmetry breaking
is described by the standard   summation of the valence quarks
self-interaction loops in the rainbow approximation (the Schwinger-Dyson or gap
equations), while mesons are obtained from the Bethe-Salpeter equation for the
quark-antiquark bound states, which is well justified and consistent
in the large $N_c$ limit.

In the 1+1 dimensional 't Hooft model
the linear instantaneous Lorentz-vector confining 
potential
appears automatically as the Coulomb interaction in 1+1 dimensions, once
a proper gauge is chosen. It is clear that in QCD in 3+1 dimensions
there are different kinds of gluonic interactions between quarks. It
is hopeless to solve even large $N_c$ QCD with full gluodynamics. However,
for our present purpose it is not required. In order to answer a principal
question about chiral restoration one needs a model
satisfying all the conditions (i)-(vii). The GNJL model is the only known
such a model. Note that the instantaneous Lorentz-vector linear confining
potential is a principal ingredient of the Gribov-Zwanziger
 scenario for confinement
in Coulomb gauge \cite{Gribov}. It has become a very popular subject
of investigations in recent years, see e.g. 
\cite{Zwanziger:2003de,Greensite:2004ke,SS,R,Nakamura:2005ux,AL}.

The  dynamical chiral symmetry breaking as well as the properties
of the lowest meson excitations have been studied within the present
model  in great detail long ago 
\cite{orsay,Adler:1984ri,Alkofer:1988tc,BR,COT,BN}.
The chiral symmetry restoration in excited hadrons has been addressed
only very recently. The spectrum of the excited heavy-light mesons as well
as  the Dirac structure of the effective single quark potential leading
to the effective chiral restoration have been studied in ref. \cite{KNR}.
In this work the harmonic confining inter-quark potential has been used. 
The quantum nature of chiral symmetry breaking as well as  the transition to the
semiclassical regime in the highly excited mesons, where 
quantum fluctuations of the quark fields represent only a small
correction to the chiral-invariant classical contributions, have
been discussed in ref. \cite{GNR}. This work illustrates and
clarifies the most
fundamental physical origin of effective chiral restoration \cite{G}.
In the subsequent paper \cite{GN}
a decoupling of the highly-excited hadrons from the Goldstone
bosons has been addressed. A complete spectrum of the light-light
mesons with the linear confining potential 
with all the chiral multiplets required by the chiral symmetry
have been obtained
in refs. \cite{WG1,WG2}. This phenomenon has also been shortly mentioned
within a slightly different model in ref. \cite{swanson}.
These works represent an exhaustive proof of
the effective chiral symmetry restoration, at least within the given
model.

\subsection{Chiral symmetry breaking}
The GNJL model is described by the Hamiltonian \cite{orsay}
\begin{eqnarray} 
\hat{H} & = & \int d^3x\bar{\psi}(\vec{x},t)\left(-i\vec{\gamma}\cdot
\vec{\bigtriangledown}+m\right)\psi(\vec{x},t) \nonumber \\
 &+& \frac12\int d^3
xd^3y\;J^a_\mu(\vec{x},t)K^{ab}_{\mu\nu}(\vec{x}-\vec{y})J^b_\nu(\vec{y},t),
\label{H} 
\end{eqnarray} 
with the quark current--current 
($J_{\mu}^a(\vec{x},t)=\bar{\psi}(\vec{x},t)\gamma_\mu\frac{\lambda^a}{2}
\psi(\vec{x},t)$) interaction parametrized by an instantaneous 
confining kernel $K^{ab}_{\mu\nu}(\vec{x}-\vec{y})$ of a generic
form.

The simplest confining interaction  is the instantaneous
Coulomb-like potential which is taken to
have the linear form

\begin{equation} 
K^{ab}_{\mu\nu}(\vec{x}-\vec{y})=g_{\mu 0}g_{\nu 0}
\delta^{ab} V (|\vec{x}-\vec{y}|); ~~~~~
\frac{\lambda^a \lambda^a}{4}V(r) = \sigma r.
\label{KK}
\end{equation}

The Fourier transform of this potential does not exist
and any loop integral is infrared divergent. 
 Hence it is required to perform a infrared
regularization, i.e. to suppress the contributions
around $p=0$.  One of the possibilities, used in ref. \cite{orsay},
 is to substitute the strictly
linear potential by

\begin{equation} 
V(r) = \sigma r  ~~~~~ \Longrightarrow ~~~~~ \sigma r e^{-\mu_{IR} r}.
\label{reg}
\end{equation}

\noindent
Then the potential in the momentum space is well defined,

\begin{equation}
V(p)=-\int d^3 r e^{i{\vpp\vec r}} V(r)
 =\frac{8\pi\sigma}{p} Im \frac{1}{(\mu_{IR} -i p)^3},
\label{FV} 
\end{equation}

\noindent 
and all loop integrals are finite. The physical results should
not be dependent on the infrared regulator $\mu_{IR}$ in the
infrared limit $\mu_{IR} \rightarrow 0$. 

The other possibility is to use the following regularized potential
in the momentum space \cite{AL,WG1,WG2}

\begin{equation}
V(p)= \frac{8\pi\sigma}{(p^2 + \mu_{IR}^2)^2}.
\label{FV} 
\end{equation}

\noindent
Then, upon  transformation back into the configurational
space, 

\begin{equation}
- \frac{1}{(2\pi)^3}\int d^3 p e^{i\vec p \vec r} V(p) = 
-\frac{\sigma exp(-\mu_{IR} r)}{\mu_{IR}} =
\sigma r - \frac{\sigma}{\mu_{IR}} + {\cal{O}}(\mu_{IR}),
\label{div}
\end{equation}

\noindent
one recovers that the potential contains the required linear
potential plus a constant term that diverges in the infrared limit 
$\mu_{IR} \rightarrow 0$ and  is irrelevant to observables.
Indeed,  while the single quark Green function is divergent 
in the infrared limit (i.e.
it cannot be observed), the observable color-singlet meson mass is finite. 
 This is because the infrared divergence of the
single quark Green function exactly cancels out with the infrared
divergence of the kernel in the Bethe-Salpeter equation \cite{WG2}.
Note that there are no ultraviolet divergences with the linear
potential, which would persist if
there have been in addition the Coulomb potential.

The Dirac operator for the dressed quark is
\begin{equation}
D(p_0,\vec{p})= i S^{-1}(p_0,\vec{p}) = D_0(p_0,\vec{p})-\Sigma(p_0,\vec{p}),
\label{SAB}
\end{equation}

\noindent
where  $D_0$ is the bare Dirac operator with the bare quark mass $m$, 

\begin{equation}
D_0(p_0,\vec{p})=i S_0^{-1}(p_0,\vec{p})=p_0\gamma_0-\vec{p}\cdot\vec{\gamma}-m.
\label{bare}
\end{equation}

\noindent
Paremetrising the self-energy operator in the form

\begin{equation}
\Sigma(\vec p) =[A_p-m]+(\vec{\gamma}\hat{\vec{p}})[B_p-p],
\label{SE} 
\end{equation}

\noindent
where functions $A_p$ and $B_p$ are yet to be found, the
Schwinger-Dyson equation for the self-energy operator in the
ladder approximation,which is valid in the large $N_c$ limit
for the instantaneous interaction, see
 Fig. \ref{sigma}, takes the form

\begin{equation}
i\Sigma(\vec{p})=\int\frac{d^4k}{(2\pi)^4}V(\vec{p}-\vec{k})
\gamma_0\frac{1}{S_0^{-1}(k_0,\vk)-\Sigma(\vk)}\gamma_0. 
\label{Sigma03} 
\end{equation}

 There are poles  in eq. (\ref{Sigma03}) at $k_0^2 = \omega_k^2$,
 where $\omega_k = \sqrt { A_k^2 + B_k^2}$. The integration over
 $k_0$ can be trivially done with the help of the standard  $+i\epsilon$
 rule for the retarded Green functions. 
 Note that the $p_0\gamma_0$ component of the Dirac operator is
 not dressed because of the instantaneous nature of the interaction.
 Then the Schwinger-Dyson
 equation is reduced to the nonlinear gap equation for the chiral angle
 $\varphi_p$
 
 \begin{equation}
 A_p \cos \varphi_p - B_p \sin \varphi_p = 0,
 \label{gap}
 \end{equation}
 
\noindent
where
 
\begin{eqnarray}
A_p & = & m+\frac{1}{2}\int\frac{d^3k}{(2\pi)^3}V
(\vec{p}-\vec{k})\sin\vp_k,\quad  \\
B_p & = & p+\frac{1}{2}\int \frac{d^3k}{(2\pi)^3}\;(\hat{\vec{p}}
\hat{\vec{k}})V(\vec{p}-\vec{k})\cos\vp_k, 
\label{AB} 
\end{eqnarray}  
 
The functions $A_p,B_p,$  i.e. the quark self-energy
  are divergent in the
 infrared limit. However, there is no infrared divergence in the gap equation 
so this equation can be solved directly in the infrared limit.
  This gap equation for the linear potential has been solved
 numerically in ref. \cite{Adler:1984ri} and the solution has
 been repeatedly reconfirmed in many subsequent works on the model.
  Alternatively, it
can be solved for small but finite values of the infrared regulator.
This solution  in the chiral limit $m=0$
is shown in Fig. \ref{dm} where in addition
convergence with respect
to the mass of the infrared regulator in eq. (\ref{FV}) is demonstrated
\cite{AL,WG2}. 

\begin{figure}[t]
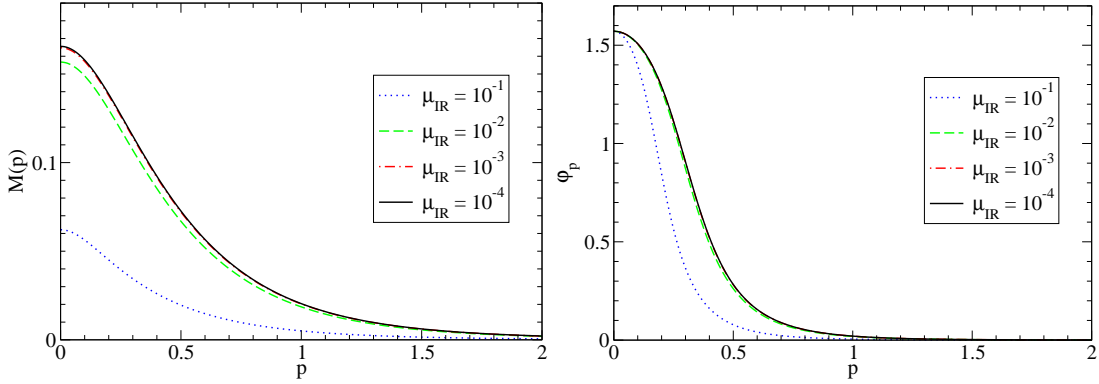

\includegraphics[width=0.45\hsize,clip=]{m.eps}
\includegraphics[width=0.45\hsize,clip=]{chiralangle.eps}
\caption{Dynamical mass and chiral angle in the chiral limit
for different values of the infrared regulator $\mu_{\rm IR}$.
All quantities are given in appropriate units of $\sqrt{\sigma}$.}
\label{dm}
\end{figure}

For free quarks the chiral angle, $\tan \varphi_p^0 = m/p$, reduces to
the Foldy angle that diagonalizes the Dirac Hamiltonian,
$H = \vec \alpha \cdot \vec p + \beta m.$ Then one can introduce the
effective dynamical mass of quarks

\begin{equation}
M(p) = p \tan \varphi_p,
\label{dyna}
\end{equation} 

\noindent
which is also shown in the chiral limit in Fig. \ref{dm}.
The nontrivial
solution of the gap equation  as well as the nonzero value of
the dynamical mass signal dynamical breaking of chiral symmetry in the
vacuum.
Both the chiral angle and the dynamical mass of quarks vanish
fast at larger quark momenta. This property is crucial for a proper
understanding of chiral symmetry restoration in excited hadrons.

The
quark condensate is given as

\begin{equation}
\langle\bar{q}q\rangle=-\frac{N_C}{\pi^2}\int^{\infty}_0 dp\;p^2\sin\vp_p.
\label{Sigma1}
\end{equation} 

\noindent
The numerical value of the quark condensate is 
$\langle\bar{q}q\rangle=(-0.231\sqrt{\sigma})^3$. 
If we fix the string tension from the phenomenological
angular Regge trajectories, then $\sqrt{\sigma}\approx 300$ -- $400$ MeV 
and hence the quark
condensate is between $(-70\ \mbox{MeV})^3$ and $(-90\ \mbox{MeV})^3$ which
obviously underestimates the phenomenological value.
Probably this indicates
that other gluonic interactions also contribute to chiral symmetry
breaking.
 
\subsection{Bethe-Salpeter equation for mesons and chiral
symmetry properties of the spectrum}

The homogeneous Bethe-Salpeter equation for a quark-antiquark
bound state in the rest frame with the instantaneous interaction
is  

\begin{eqnarray}
\chi(\mu,\vpp)&= &- i\int\frac{d^4q}{(2\pi)^4}V(|\vpp-\vq|)\;
\gamma_0 S(q_0+\mu/2,\vpp-\vq) \nonumber \\
& \times & \chi(\mu,\vq)S(q_0-\mu/2,\vpp-\vq)\gamma_0.
\label{GenericSal}
\end{eqnarray}

\noindent
Here $\mu$ is meson mass and $\vec p$ is relative momentum. The Bethe-Salpeter
equation can be solved by means of expansion of the vertex function
$\chi(\mu,\vpp)$  into a set of all possible independent Poincar\'{e}-invariant
amplitudes consistent with $I,J^{PC}$. Then the Bethe-Salpeter equation
transforms into a system of coupled equations \cite{WG2}. The infrared
divergence cancels exactly in these equations and they can be solved
either in the infrared limit $\mu_{IR} = 0$ or for very small $\mu_{IR}$.

It is instructive to study the symmetry properties of these equations
in the limit $M(p)=0$ \cite{WG2}. In this limit the chiral symmetry is
unbroken and the Bethe-Salpeter amplitudes exactly fall into chiral
representations (\ref{e0a})-(\ref{e22}) and multiplets of $U(1)_A$. 
While it has been demonstrated
analytically, this property is expected a-priory, because in this case
the quarks must have a definite chirality and the representations 
(\ref{e0a})-(\ref{e22}) exhaust all possible chiral representations
for the quark-antiquark system that are consistent with the Poincar\'{e}
invariance.
In the highly excited mesons a typical momentum of valence quarks
is large. Hence, since the dynamical mass of quarks $M(p)$
vanishes at large momenta, one expects that the highly-excited mesons
fall into approximate multiplets of chiral and $U(1)_A$ groups.

\begin{figure}
\begin{center}
\includegraphics*[height=5cm]{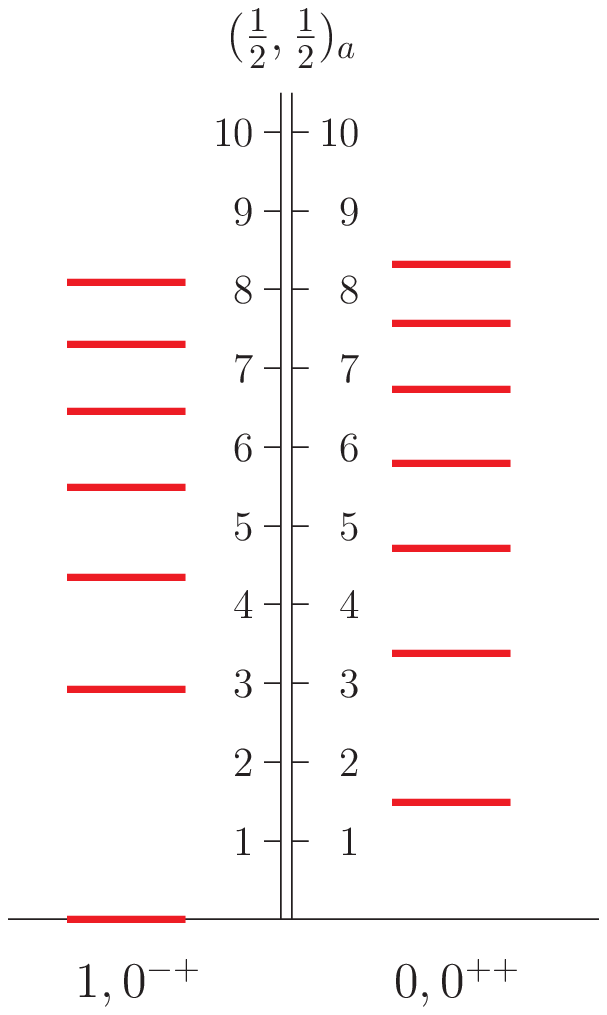}
\hspace*{1cm}
\includegraphics[height=5cm]{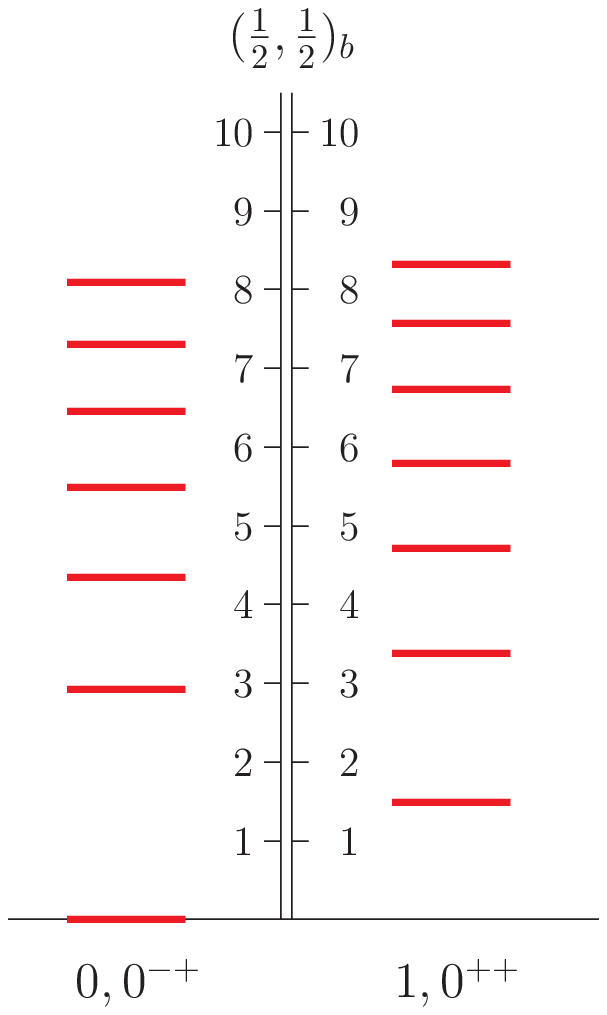}
\end{center}
\caption{ Spectra of $J=0$ mesons. Masses are in units of $\sqrt \sigma$.}
\label{s0}
\end{figure}

\begin{figure}
\begin{center}
\includegraphics[height=5cm,clip=]{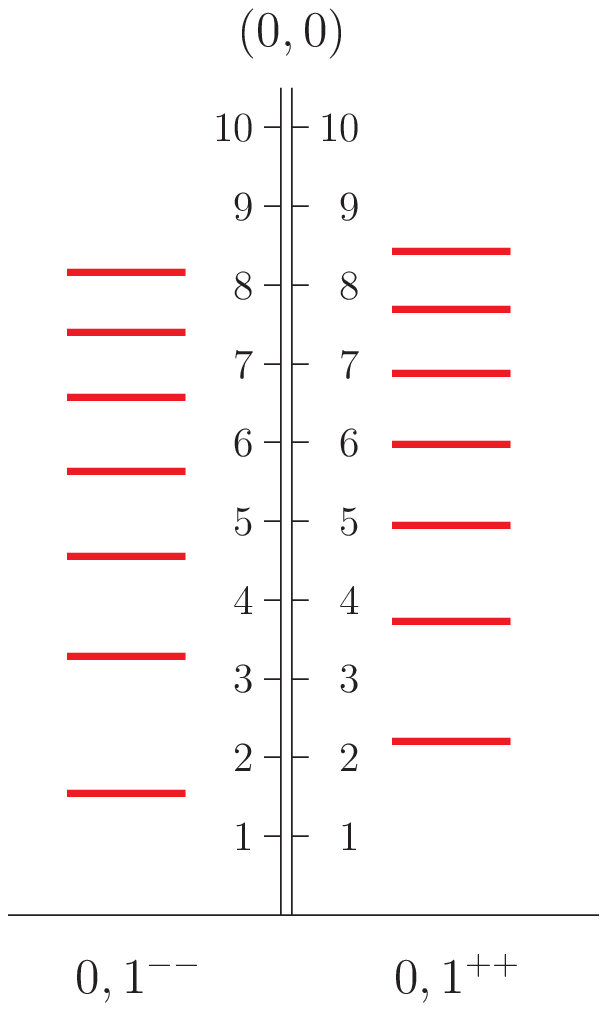}
\hspace*{0.5cm}
\includegraphics[height=5cm,clip=]{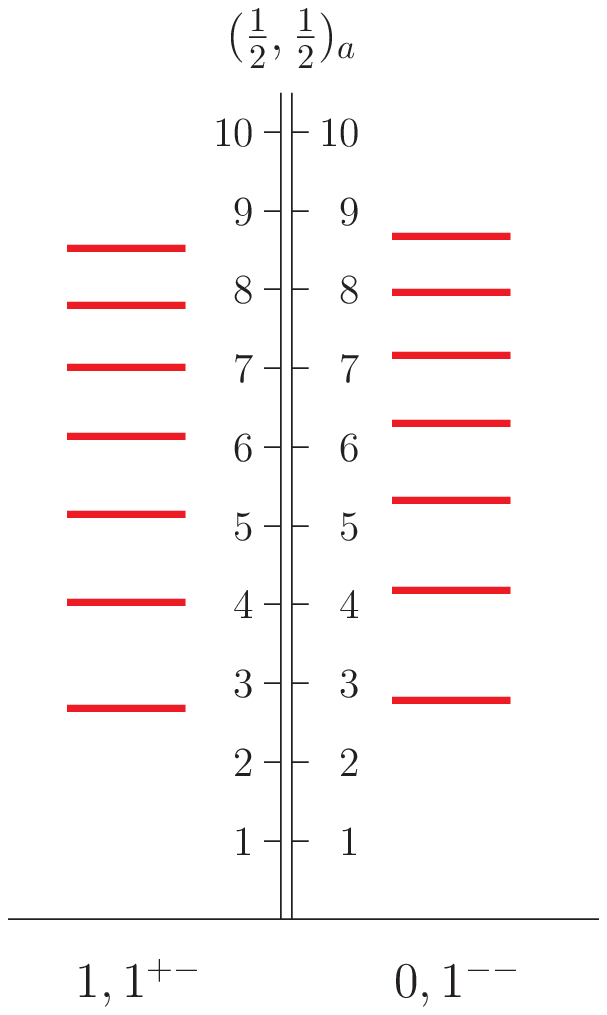}
\hspace*{0.5cm}
\includegraphics[height=5cm,clip=]{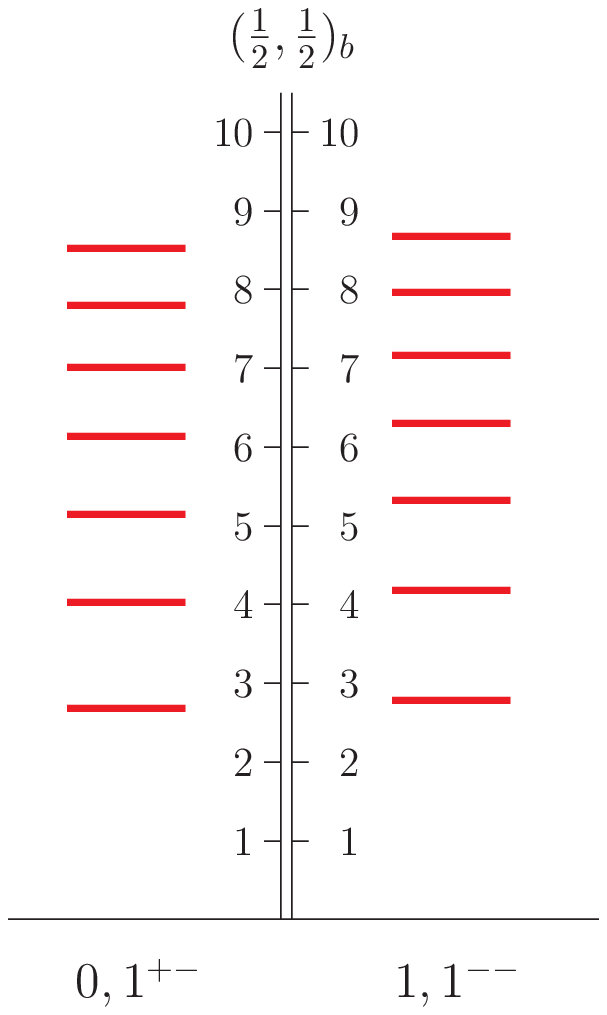}
\hspace*{0.5cm}
\includegraphics[height=5cm,clip=]{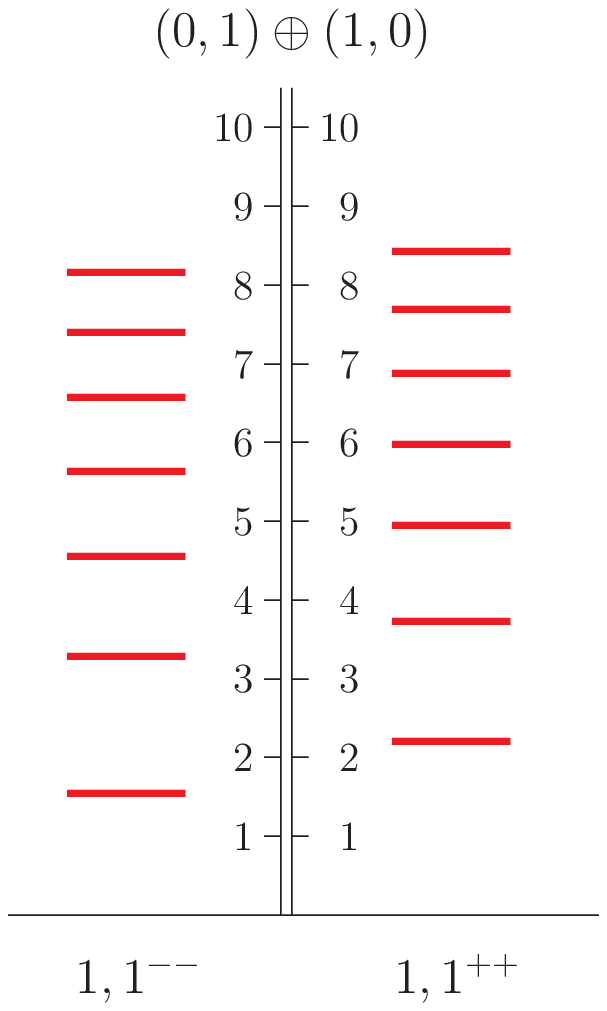}
\end{center}
\caption{ Spectra of $J=1$ mesons. Masses are in units of $\sqrt \sigma$.}
\label{s1}
\end{figure}

\begin{figure}
\begin{center}
\includegraphics[height=5cm,clip=]{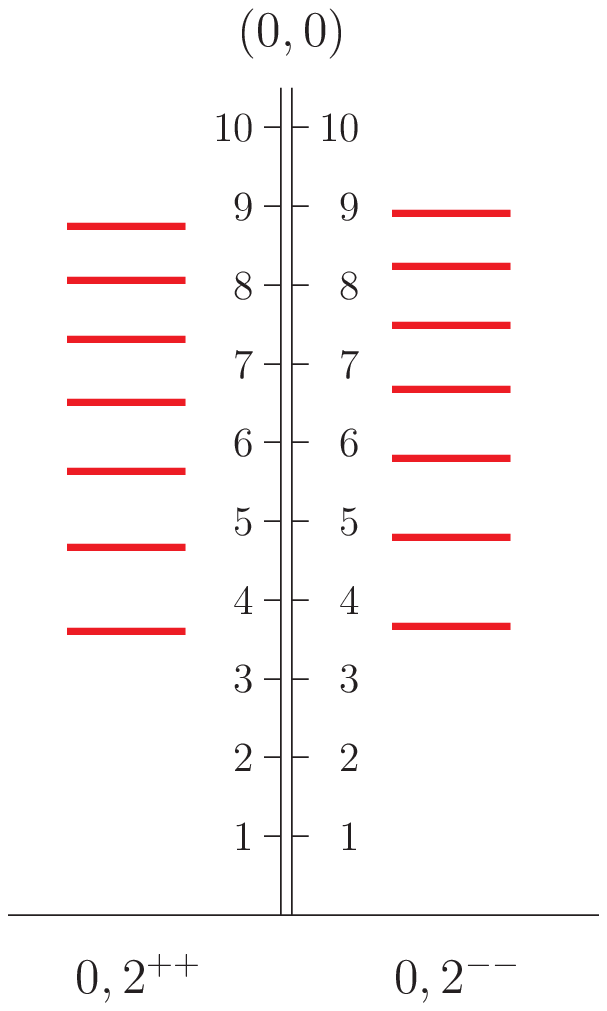}
\hspace*{0.5cm}
\includegraphics[height=5cm,clip=]{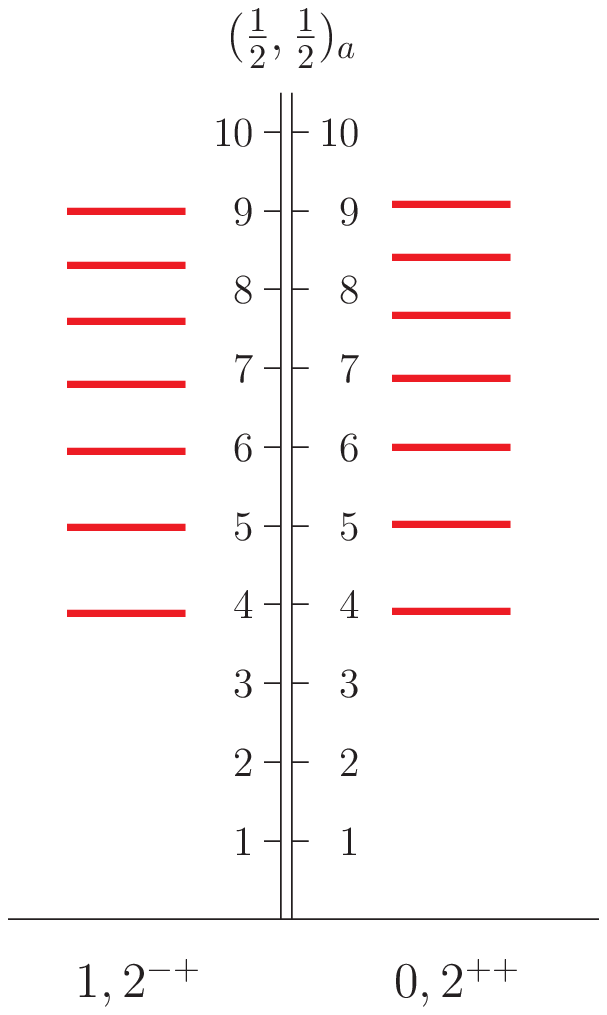}
\hspace*{0.5cm}
\includegraphics[height=5cm,clip=]{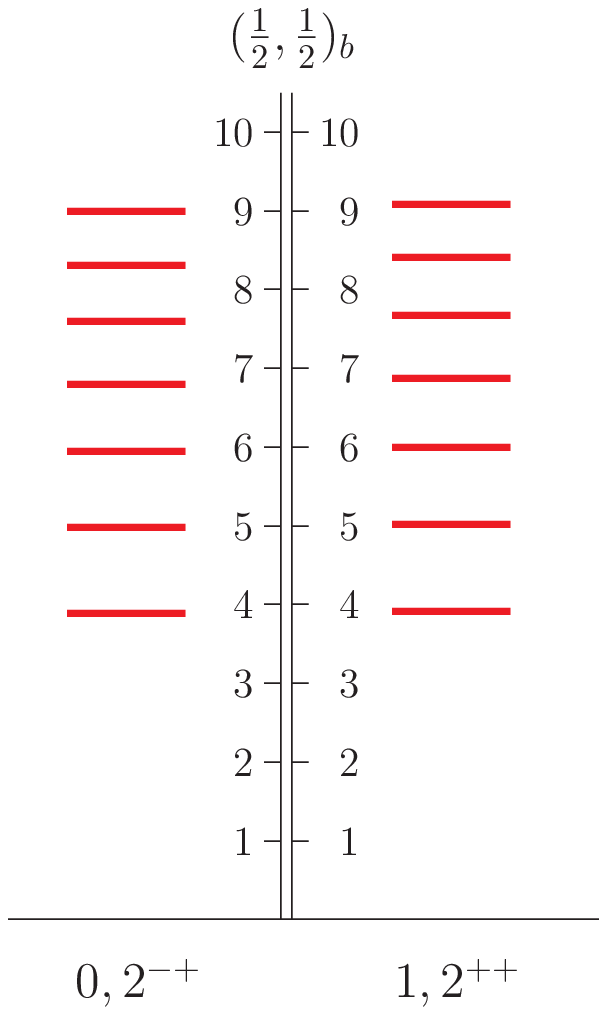}
\hspace*{0.5cm}
\includegraphics[height=5cm,clip=]{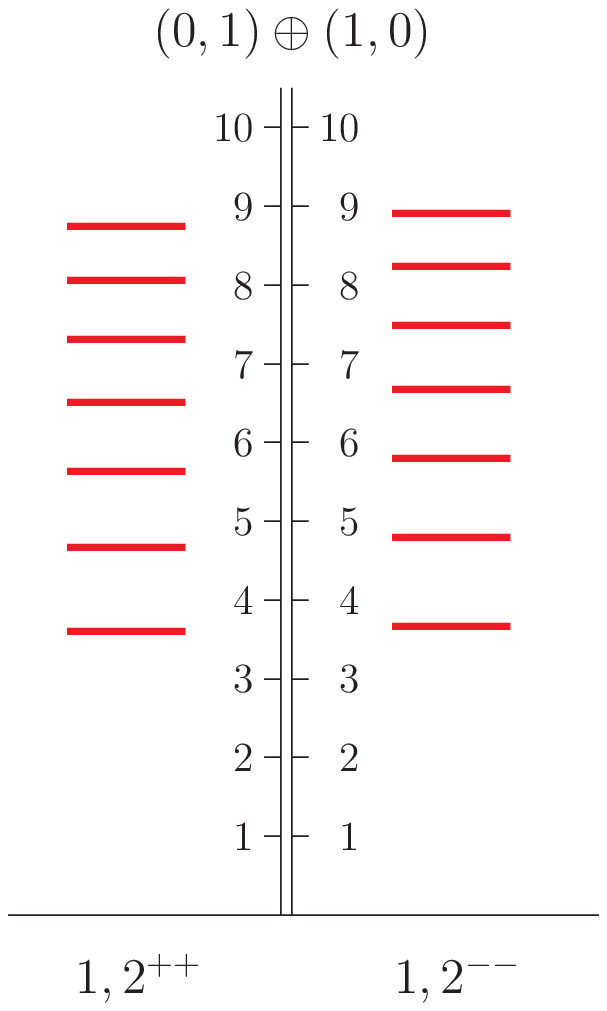}
\end{center}
\caption{ Spectra of $J=2$ mesons. Masses are in units of $\sqrt \sigma$.}
\label{s2}
\end{figure}

The spectrum of mesons in the chiral limit $m=0$
obtained in refs. \cite{WG1,WG2} is shown in 
Figs. \ref{s0} - \ref{s2}.
Note that within this model the axial anomaly is absent (there are no
vacuum fermion loops), hence both $SU(2)_L \times SU(2)_R$ and $U(1)_A$
symmetries are dynamically broken in the vacuum. In this case one expects
four Goldstone bosons with quantum numbers $I=1,0^{-+}$ and  $I=0,0^{-+}$,
as it is indeed seen in Fig. \ref{s0}. We see a very fast restoration
of chiral and $U(1)_A$ symmetries with increasing $J$ and rather slow 
restoration
with increasing of the radial quantum number $n$ with the fixed $J$.

\begin{figure}
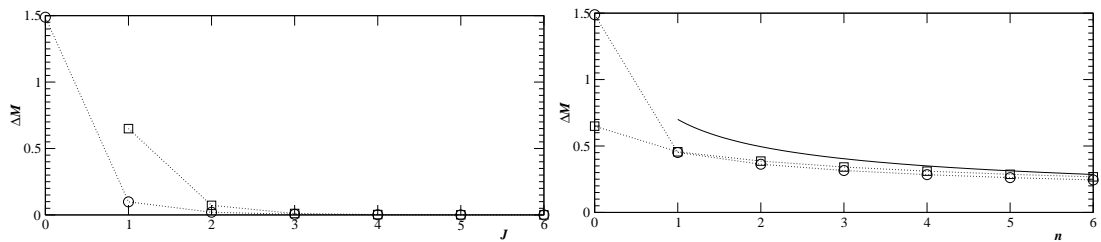

\includegraphics[width=0.45\hsize,clip=]{delta_m_spin.eps}
\includegraphics[width=0.45\hsize,clip=]{delta_m_radial.eps}
\caption{Mass splittings in units of $\sqrt{\sigma}$ for isovector mesons of the chiral
multiplets $(1/2,1/2)_a$ and $(1/2,1/2)_b$ (circles) and within
the multiplet $(0,1)\oplus(1,0)$ (squares) against $J$ for $n=0$ (left panel)
 and
against $n$ for $J=0$ and $J=1$, respectively (right panel).
The full line in the bottom plot is $0.7\sqrt{\sigma/n}$.}
\label{rate}
\end{figure}

In Fig. \ref{rate} the rates of the symmetry restoration against the radial
quantum number $n$ and spin $J$ are shown. It is seen that with  fixed
$J$ the splitting within the multiplets $\Delta \mu = \mu_+ - \mu_-$ 
decreases asymptotically as
$1/\sqrt n$, dictated by the asymptotic linearity of the radial Regge
trajectories with different intercepts.  
 Restoration of the chiral symmetry with increasing $J$  at a given $n$ proceeds 
much faster. 

\begin{figure}
\begin{center}
\includegraphics[width=0.5\hsize,clip=]{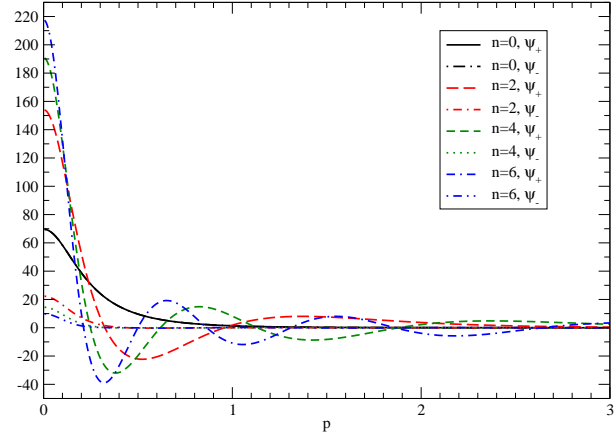}
\caption{Components $\psi_+(p)$
and $\psi_-(p)$
of the wave functions for ground and radially excited states
for mesons $0^{-+}$.
Notice that the ground state (the Goldstone boson)
 has $\mu=0$ and thus $\psi_-(p)=\psi_+(p)$, hence two curves for $n=0$
 coincide. Momenta are in units of $\sqrt{\sigma}$.}
\label{fig:radialcomponents}
\end{center}
\end{figure}

\begin{figure}
\begin{center}
\includegraphics[width=0.5\hsize,clip=]{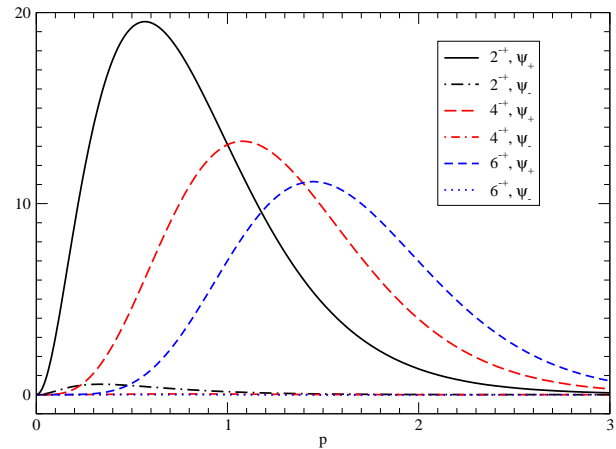}
\caption{Components $\psi_+(p)$ 
and $\psi_-(p)$
of the wave functions for mesons $J^{-+}$ 
with radial quantum number
$n=0$ and spin $J=2,4,6$. Momenta are in units of $\sqrt{\sigma}$.}
\label{fig:angularcomponents}
\end{center}
\end{figure}

In order to explain the origins of the fast chiral restoration rate
versus $J$ and rather slow rate versus $n$ we have to look into  meson
wavefunctions. The meson wave function with some fixed quantum numbers
can be decomposed into the coupled forward- and backward-propagating
quark-antiquark amplitudes, $\psi_+$ and $\psi_-$,
respectively. The leading Fock component is the forward-propagating
part  $\psi_+$, while the higher Fock components contain necessarily
in addition the backward-propagating components $\psi_-$. At large
$J$ or $n$ the semiclassical description requires that the higher
quark Fock components be suppressed relative the leading one and
asymptotically vanish, because the higher Fock components  represent
effects of quantum fluctuations. This property is well seen in
Figs. \ref{fig:radialcomponents} and \ref{fig:angularcomponents}.

Hence at large $J$ and/or $n$ only a "string-like" $\bar q q$
configuration survives and we can thus restrict ourselves to the $\psi_+$
component. In contrast, both the forward- and backward-propagating 
amplitudes are
equally important in the ground state Goldstone bosons which are  
highly-collective quark-antiquark modes.  

\begin{figure}
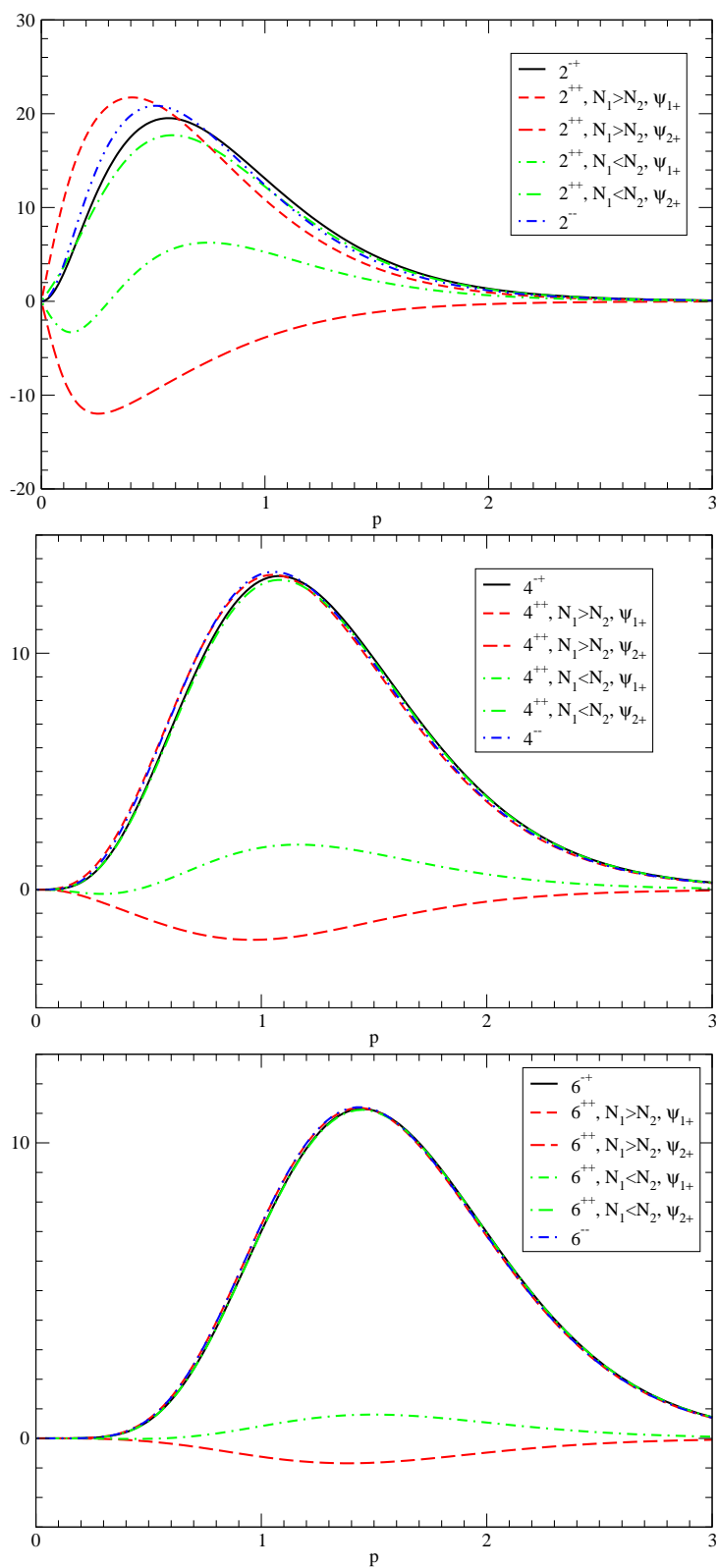

\begin{center}
\includegraphics[width=0.6\hsize,clip=]{J_2_wf_+.eps}
%
%
\includegraphics[width=0.6\hsize,clip=]{J_4_wf_+.eps}
%
%
\includegraphics[width=0.6\hsize,clip=]{J_6_wf_+.eps}
\caption{Components $\psi_+(p)$ of mesons with $J=2,4,6$ and $n=0$.
Four similar curves represent "large" components of $\psi_+(p)$,
while the other two represent "small" components of $\psi_+(p)$
for mesons $J^{++}, J=2,4,6$. $N_1$ and $N_2$ are the normalization
factors. See for details ref. \cite{WG2}.
Momenta are in units of $\sqrt{\sigma}$.
}
\label{fig:J_246_wf}
\end{center}
\end{figure}

The reason for the fast chiral restoration versus $J$ is that the radial wave
function $\psi_+$ at larger $J$ vanishes at small momenta due
to the centrifugal repulsion in the momentum space and hence the 
chiral symmetry breaking dynamical mass, $M(p)$, which is essential
only at small momenta, gets irrelevant. The valence quarks do not see
chiral symmetry breaking. These wave functions for all possible isovector
mesons with $n=0$ are shown in Fig. \ref{fig:J_246_wf}.
It is clear from these figures that the larger $J$, the stronger is the
wave function suppression at small momenta. This property provides a large
rate of the chiral restoration with increasing $J$. 

In contrast, at $J=0$
the  wave function $\psi_+$ of the pseudoscalar mesons does not vanish
at $p=0$, as it is well seen in Fig. \ref{fig:radialcomponents}. This is
because in this case the rotational motion of quarks is absent and at
the turning points the quarks are necessarily slow. When the quarks are
slow, their chiral symmetry breaking dynamical mass is important. 
Even though a typical momentum of quarks increases with $n$, the small
momenta always contribute to some extent.

A higher degeneracy is recovered  for $J \rightarrow \infty$ 
and/or $n\rightarrow \infty$. For a given $n$ but large $J$ or for a given
$J$ but large $n$ equations for all possible eight states (four states in 
case $J=0$) coincide. This is well seen in Fig. \ref{fig:J_246_wf},
where all "large" components of wave functions coincide. 
This means that the states fall into  reducible
representations

\begin{equation}
[(0,1/2) \oplus (1/2,0)] \times [(0,1/2) \oplus (1/2,0)],
\label{high}
\end{equation}

\noindent
which combine all possible chiral representations of the quark-antiquark
systems with the same $J$. In such limits the quantum loop effects become
irrelevant and all possible states with different quark chiralities
become equivalent.

\begin{figure}
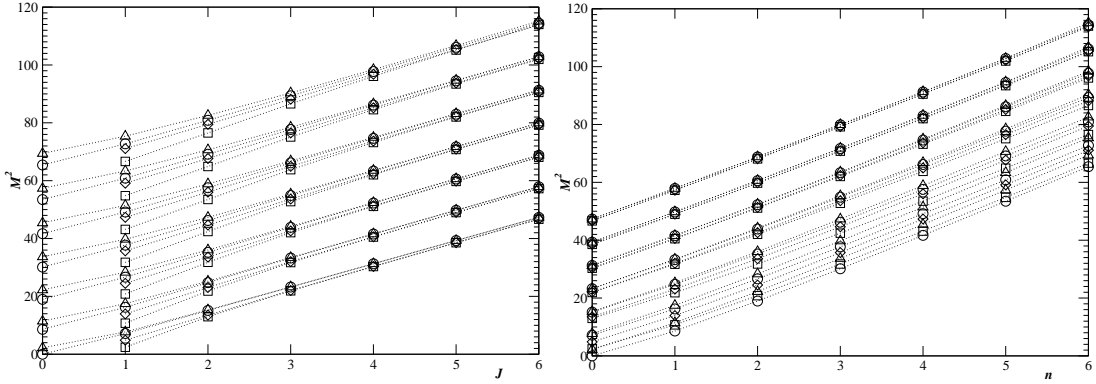

\includegraphics[width=0.45\hsize,clip=]{regge_spin.eps}
\includegraphics[width=0.45\hsize,clip=]{regge_radial.eps}
\caption{Angular (left panel) and radial (right panel) Regge trajectories 
for isovector 
mesons
with $M^2$ in units of $\sigma$.
Mesons of the chiral multiplet $(1/2,1/2)_a$ are indicated by circles,
of $(1/2,1/2)_b$ by triangles, and of $(0,1)\oplus(1,0)$
by squares ($J^{++}$ and $J^{--}$ for even and odd $J$, respectively)
and diamonds ($J^{--}$ and $J^{++}$ for even and odd $J$, respectively).
}
\label{regan}
\end{figure}

In Fig. \ref{regan}  the angular and radial Regge trajectories are shown.
Asymptotically both kinds of trajectories are linear.
This is expected a-priori, because with the pure linear potential,
neglecting the backward-propagating components
and without the self-energy loops the radial and angular Regge trajectories
are linear, see, e.g.  \cite{morg}.
Both kinds of trajectories exhibit deviations from the linear behavior
at smaller $J$ or $n$. 
This fact is obviously related to the
chiral symmetry breaking loop effects for lower mesons.
Each Regge trajectory is characterized by the proper chiral index and
hence the amount of the independent Regge trajectories coincides with the
amount of the chiral representations. Consequently, 
for many mesons there are two
independent Regge trajectories at a given $I,J^{PC}$ which belong to
different chiral representations.

\subsection{The quantum origin of chiral symmetry breaking and
the semiclassical origin of effective chiral restoration}

In the previous subsection we have obtained restoration of chiral
and $U(1)_A$ symmetry within the GNJL model. Here we want to illustrate
\cite{GNR} within this model the most general origin
of effective chiral restoration .
We have already discussed in Sec. 4 the quantum
nature of chiral symmetry breaking in QCD. Then, since in highly
excited hadrons the semiclassical regime must be manifest, all quantum
quark loop contributions get suppressed relative  the classical
contributions and the underlying chiral and $U(1)_A$ symmetries
should be approximately restored. 

The mass operator (\ref{Sigma03}) is the quark self-energy loop
integral. The loop integral is of the quantum origin and hence contains
as a factor the Planck constant $\hbar$. Typically the factors $\hbar$
are omitted in  formulas. In the present context it is important,
however, to restore this factor to see explicitly the quantum
nature of the chiral symmetry breaking\footnote{Here we do not restore, however,
the speed of light $c$.}. One gets

\begin{equation}
i\Sigma(\vec{p})= \hbar \int\frac{d^4k}{(2\pi)^4}V(\vec{p}-\vec{k})
\gamma_0\frac{1}{S_0^{-1}(k_0,\vk)-\Sigma(\vk)}\gamma_0. 
\label{SigmaH} 
\end{equation} 

The same is true, of course, for the functions $A_p$ and $B_p$:

\begin{eqnarray}
A_p & = & m+\frac{\hbar}{2}\int\frac{d^3k}{(2\pi)^3}V
(\vec{p}-\vec{k})\sin\vp_k,\quad  \\
B_p & = & p+\frac{\hbar}{2}\int \frac{d^3k}{(2\pi)^3}\;(\hat{\vec{p}}
\hat{\vec{k}})V(\vec{p}-\vec{k})\cos\vp_k. 
\label{ABH} 
\end{eqnarray} 

\noindent
They contain both classical, $m,p$, and quantum contributions. Then
it is obvious that in the classical limit, $\hbar =0$, there cannot
be any quantum contributions to the quark mass function and a nontrivial
solution of the gap equation (\ref{gap}), which signals dynamical
breaking of chiral symmetry in the vacuum, vanishes. In this case the chiral
angle becomes the free Foldy angle. In the chiral limit $m=0$ the quark
condensate as well as the dynamical mass of quarks are identically zero.

Consider now the gap equation,
\begin{equation}
pc\sin\vp_p-mc^2\cos\vp_p=\frac{\hbar}{2}\int\frac{d^3k}{(2\pi)^3}
{V}(\vpp-\vk)\left[\cos\vp_k\sin\vp_p-
(\hat{\vec{p}}\hat{\vec{k}})\sin\vp_k\cos\vp_p\right]. 
\label{mgh}
\end{equation}

\noindent
In the chiral limit  substituting $\hbar=0$ the only solution is
a trivial one $\varphi_p=0$. We want to
reconstruct an important scaling variable. 
In the chiral limit but with  finite $\hbar$
we have only four dimensional quantities in our
task, $\hbar, ~c, ~\sigma, ~p$.
The chiral angle $\varphi_p$ is dimensionless and hence can depend only
on  dimensionless parameter. Hence the chiral angle is a function
 of the variable 

\begin{equation}
\varphi_p =  \varphi_p\left ( \frac{p}{c \cdot \sqrt{\sigma \hbar c}/c^2} 
\right ),
\label{scaling}
\end{equation}

\noindent
because with the  dimensional quantities 
$\hbar, ~c,~\sigma$   one can construct only one  quantity,
$\sqrt{\sigma \hbar c}/c^2$, with the dimension of a mass.
This simple result shows an important property: increasing 
the quark momentum $p$ is equivalent to decreasing  the Planck
constant, $\hbar$. The chiral symmetry breaking effect vanishes
at large quark momenta, which is equivalent approaching to the classical
limit. The larger is the spin of the hadron, $J$, or its radial excitation
number $n$, the larger is a typical momentum of the valence quark. Then
one gradually approaches the classical regime at $J$ and/or $n$ 
$\rightarrow \infty$. Classically both $SU(2)_L \times SU(2)_R$ and
$U(1)_A$ are manifest. 

Beyond the chiral limit, $m \neq 0$, one can define two different
regimes according to the value of the parameter $m/\sqrt {\sigma}$ \cite{GNR}:
Chiral symmetry and its spontaneous breaking are relevant for
$m \ll \sqrt {\sigma}$, while "heavy-quark physics" is adequate
in the opposite limit $m \gg \sqrt {\sigma}$.

\subsection{The Lorentz structure of the effective confining
single quark potential}

It is an old question what is the Lorentz structure of the single-quark
confining potential. The bag model \cite{BAG} 
adopts the view that this potential is of the Lorentz-scalar nature and
hence it manifestly breaks chiral symmetry. In the naive potential
constituent quark model the Lorentz-scalar confinement was needed to
cancel at least partly a spin-orbit force from the one-gluon-exchange
interaction between constituent quarks \cite{IK}. A Lorentz-scalar nature
of the effective confining interaction appears quite naturally {\it at
very low momenta} as it follows from the analysis of the quasi-static
Wilson loop potential \cite{WL}. However, the original gluonic
interaction in QCD is of  Lorentz-vector nature, so if the
confining potential could be of  Lorentz-scalar nature, it would
be only at the level of the {\it effective} potential.

Certainly the effective confining potential cannot be of  Lorentz-scalar
nature in the regime where chiral symmetry is approximately restored,
because a Lorentz-scalar potential manifestly breaks chiral symmetry.
This question has been addressed in ref. \cite{KNR}, where the
Lorentz structure of the potential has been studied within the GNJL
model for the heavy-light system.

If one of the two valence quarks is infinitely heavy and
if the only interaction
between quarks is an instantaneous potential, then there is
no backward-propagating quark-antiquark component. 
Hence such a system
can be exactly reduced to the effective Dirac equation

\be 
({\vec \alpha}\vpp+\beta m){\Psi}(\vx)+\frac{\sigma}{2}\int d^3z
\left(\vphantom{A^2}|\vx|+|\vz|-|\vx-\vz|\right)U(\vx-\vz){\Psi}(\vz)=
E{\Psi}(\vx),
\label{DS2} 
\ee 
with the unitary matrix 
\be
U(\vx)=\int \frac{d^3p}{(2\pi)^3}U(\vpp)e^{-i{\vpp\vx}},\quad
U(\vpp)=\beta\sin\vp_p+({\vec\alpha}\hat{\vpp})\cos\vp_p.
\label{L2} 
\ee

 The Lorentz nature of the effective
inter-quark interaction in the Dirac-like Eq.~(\ref{DS2}) is
governed by the structure of the matrix $U$, that is, by the value
of the chiral angle $\vp_p$. Indeed, for
$\vp_p\approx\frac{\pi}{2}$, the effective interaction is scalar,
chiral symmetry is strongly broken
so that no parity doublers can appear. Even though the original
inter-quark potential is  a Lorentz-vector, loops induce
also  Lorentz-scalar component and this Lorentz-scalar component
becomes dominant at very low  momenta. This reflects a self-consistency
of the approach: the chiral symmetry breaking happens due to 
the Lorentz-scalar part of the quark self-energy, or due to the
Lorentz-scalar part of the {\it effective} potential, which is actually
the same. 

On the contrary, for a
vanishing chiral angle, the effective interaction becomes spatial-vectorial,
Eq.~(\ref{DS2}) respects chiral symmetry and, as a result, this
symmetry manifests itself in the spectrum, in the form of
approximate chiral multiplets. This
obviously happens to highly excited mesons, since the mean
relative inter-quark momentum in such states is large, and,
consequently, the corresponding value of the chiral angle is small.

However, the form of this effective potential is not universal
for a given excited state. The reason is that for the pseudoscalar
states with zero angular momentum a rotation of quarks is impossible and
 there are necessarily points during the
quark motion where the quark is slow - these are the turning points. At these
turning points the dynamical mass of quarks is large and
 chiral symmetry is broken. As was discussed in the next
to the preceeding subsection, this is a reason for a slow chiral
restoration rate with the radial quantum number $n$. Exactly at such
points the semiclassical expansion fails - it is a well known effect
in  nonrelativistic quantum mechanics \cite{LANDAU3}. Hence the
effective potential is a Lorentz-scalar near the turning points, and
becomes a Lorentz-vector when the quark is fast within the given
hadron. This remark makes the concept of an effective confining
potential somewhat  limited.

\subsection{Goldstone boson decoupling from the high-lying states}

The most general chiral symmetry reason for the Goldstone boson
decoupling from hadrons in the effective symmetry restoration regime
has been discussed in secs. 9 and 10. Here we overview a microscopical
reason for such a decoupling \cite{G4,GN}.

The coupling of the Goldstone bosons to the valence quarks
is regulated by the conservation of the axial vector current. This 
conservation results in a Goldberger-Treiman relation, taken at the
"constituent quark" level, giving

\begin{equation}
g_\pi \sim M.
\label{qgt}
\end{equation}

\noindent
Here $M$ is the quark Lorentz-scalar dynamical mass that can be associated
with the momentum-dependent effective dynamical mass of quarks (\ref{dyna}).
A general feature of this dynamical mass is that it results from 
quantum fluctuations of the quark field and vanishes at large momenta
where  classical contributions dominate. Then, since the average
momentum of the valence quarks in excited hadrons increases, the valence
quarks decouple from the quark condensate and their dynamical chiral
symmetry breaking mass decreases and asymptotically vanishes. This implies,
via the Goldberger-Treiman relation (\ref{qgt}), that the valence quarks,
as well as the whole hadron, decouple from the Goldstone bosons.
This is a qualitative mechanism behind the decoupling.

\begin{figure}
\begin{center}
\includegraphics[width=10cm]{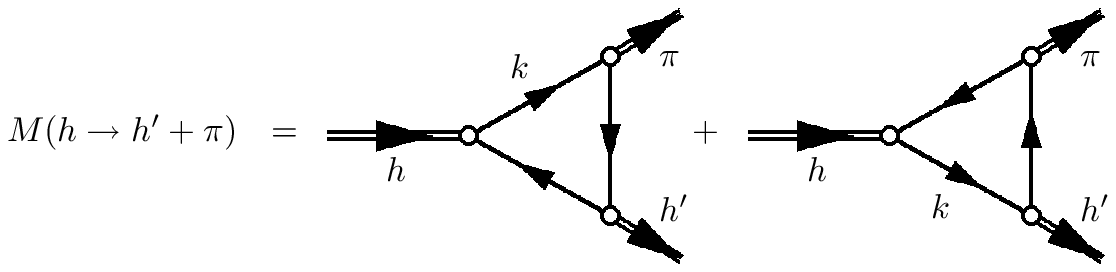}
\caption{The amplitude of the decay $h\to h'+\pi$.}
\label{decay}
\end{center}
\end{figure}

Microscopically the coupling of the given hadron $h$ to the Goldstone boson
and the hadron $h'$ (we consider for simplicity only mesons
below) is regulated by the quark loop diagram in Fig. \ref{decay}.
Each of the three verteces depends on the momentum circulating
in the loop, so the maximal overlap is achieved for all three
meson wave functions localized at comparable values of this momentum.
Clearly the pion vertex is dominated by the
low momentum $k$, because this vertex is proportional to the
pion wave function, i.e. to $\sin \varphi_k$. On the other hand
the $h$-vertex is suppressed at small momenta, as was illustrated
in Sec. 11.4, if the hadron $h$ has a large angular momentum $J$ or
radial quantum number $n$. Therefore, the
decay amplitude vanishes
with the increase of the hadron $h$ and/or $h'$ excitation, and so
does the effective coupling constant for the decay. We emphasize that
it is the pion wave function that suppresses the Goldstone boson
coupling to highly excited hadrons.

\section{Chiral multiplets and the string}

In the preQCD time a string description of hadrons was
one of the directions of a search of the fundamental theory
of strong interaction. When QCD, a local
gauge invariant field theory, was established as such a fundamental
 theory, it has become clear that the
string description of hadrons could represent only a limiting
case of QCD. Indeed, a picture of a hadron as a dynamical flux 
tube between moving color charges can be correct only if all
higher quark-antiquark Fock components are small. These higher
quark Fock components are  small at larger $L$ and/or $n$,
as it follows from the semiclassical expansion,
and hence the string picture should be indeed applicable higher
in the spectrum.

The most important achievement of the string description of
hadrons  is the celebrated linear Regge trajectories

\begin{equation}
M^2 (n,L) = c_n n + c_L L + corrections,
\label{rtr}
\end{equation}

\noindent
where $n$ and $L$ are the radial quantum number and angular momentum
of the string, respectively.
The slope of the angular Regge trajectories, $c_L$, is fixed by the
string tension $\sigma$, i.e. by the energy accumulated by the string 
per unit length,
which is a fundamental parameter of the Nambu-Goto action,
$c_L = 2 \pi \sigma$. The presence of angular daughter Regge 
trajectories that are parallel to the parent (leading) trajectory is a
main property of the Veneziano dual amplitude \cite{VEN}. This requires the
slope of the radial Regge trajectories to be the same as for the angular
ones, $c_n = c_L$.  If  the slopes of the radial and angular Regge 
trajectories coincide, then there appears a degeneracy between 
states with different angular momentum, because 
a mass is determined only by the quantum number $N = n + L$. Actually 
this is a general property of the open bosonic string \cite{POL}.
In this
case the states with different angular momenta $L$ that belong to a
band with a given $N$ must be degenerate. Such a property is clearly seen
in Fig. \ref{lear}. 

What is still missing in this description is a degeneracy of states
with opposite parity, i.e. a presence of the chiral multiplets in the
spectrum. This is because the
spin degree of freedom of the quarks at the ends of the string
 is missing in the standard open bosonic string description. 
 
\begin{figure}
\begin{center}
\includegraphics[width=5cm,angle=-90]{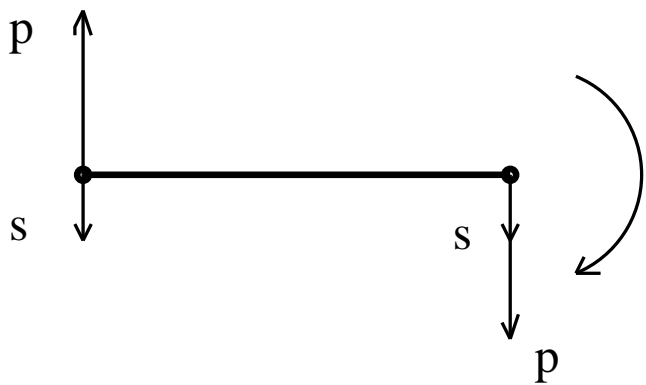}
\caption{The rotating string with quarks with definite chirality
at the ends.}
\label{str}
\end{center}
\end{figure}

The ends of the rotating string move with the velocity of light. It is this
property that ensures the linear angular Regge trajectories \cite{PGRT}.
Hence it is natural to view a highly excited hadron as a string
with massless quarks at the ends that have definite chiralities \cite{G4},
see Fig. \ref{str}, because only a fermion with a fixed
chirality can move with the  speed of light.

Then we automatically incorporate the $U(2)_L \times U(2)_R$ 
chiral symmetry of QCD on the top of the dynamical symmetry of the string.
Namely, (i) all hadrons with different chiral configurations of
the quarks at the ends of the string that belong to the same intrinsic quantum
state of the string must be degenerate; (ii) the total parity of  the hadron
is determined by the product of the parity of the string in the given quantum
state and the parity of the specific parity-chiral configuration of the
quarks at the ends of the string. This means that  the given intrinsic quantum
state of the string must have an
additional degeneracy of the $SU(2)_L \times SU(2)_R \times U(1)_A$ group.
 Then the total symmetry
group of the string can be schematically written as

\begin{equation}
Full ~ string ~ symmetry = Dynamical ~ string ~ symmetry \times 
 SU(2)_L \times SU(2)_R \times U(1)_A.
\end{equation}

It is worth to emphasize that the chiral symmetry restoration requires
that asymptotically every independent Regge trajectory is characterized by the
proper chiral index and hence the amount of independent Regge trajectories
coincides with the amount of chiral representations. This crucial property
is missing in the traditional Regge description. For example, there
must be two independent rho-trajectories; one of them with the chiral
index $(0,1) + (1,0)$ and the other one with the index $(1/2,1/2)_b$.

There are important implications of this picture. The first is that
 the spin-orbit  interactions of quarks vanish at the classical level. 
 It is easy to understand. If
the quark has a definite chirality, then its spin is necessarily parallel
(or antiparallel) with its momentum, see Fig. \ref{str}. Hence the
spin-orbit force, $\sim \vec L \cdot \vec S$, is necessarily zero. 
This is also true for the spin-orbit force due to the Thomas precession.
At the quantum level the notion of the spin-orbit force is absent
once the chiral symmetry is restored. This is because the chirality
operator does not commute with the orbital angular momentum operator
$\vec L$. Hence in the state with a definite chirality there are no
conserved quantum numbers $L$ and $S$. The total spin $J$ is a conserved
quantity, however \cite{G4}.

For the pure rotating $\bar q q$ string the tensor force also vanishes. 
Indeed, the tensor force consists of the scalar products 
$\vec S_i \cdot \vec R_j$, where $\vec R_j$ is the radius-vector
of the given quark in the center-of-mass frame.
Note, that it is the spin-orbit and tensor interactions that split
the states with the same $L$ and $S$, but different $J$.
 Such a splitting is indeed absent higher in the spectrum, 
see  Fig. \ref{lear}.

A relativistic quantum string has three different modes of 
excitation: rotational, radial vibrational and transverse vibrational.
The first two represent mesons with the "usual" quantum numbers, i.e.
quantum numbers that can also be obtained within the 
nonrelativistic potential two-body model. The third one is the transverse
excitation, i.e. an intrinsic excitation of the gluonic flux tube. This type
of excitation is typically referred to as "hybrid". The hybrids can
have "usual" and "unusual" quantum numbers. According to the standard
string theory \footnote{A consistent relativistic and quantum
description of the open string is possible only
in 26 space-time dimensions.} the energy of a string is given only by the principal
quantum number, $n = n_{rot} + n_{rad} + n_{tr}$,
 which can be obtained by different combinations of
the rotational, radial vibrational and transverse vibrational quanta.
This implies that there should be in addition a lot of hybrid mesons,
$n_{tr} > 0$,
 in the
high-lying clusters of mesons, seen in Fig. \ref{lear} - a striking prediction
awaiting for the experimental verification!

Actually there are two well established mesons  with 
hybrid quantum numbers, $\pi_1(1400)$  and   $\pi_1(1600)$. Their
interpretation is umbiguous, however, because the states with such
quantum numbers can also be constructed as 4q states (i.e. without
any intrinsic excitation of the gluonic flux tube). Indeed, in the low-lying
states a quantum interference of all possible  amplitudes should
be significant, so it could  well be that these states are either pure 4q
states or mixtures of the real hybrid with the 4q state. There is no
way to separate these two kinds of amplitudes. However, high in the spectrum
the contributions of the 4q component should be suppressed and consequently
practically a pure string mode should be visible. And it is indeed well visible.
If it is indeed a string, then one should see there also hybrids.

A consistent relativistic and quantum description of a hadron as a string with 
quarks at the ends that have definite chiralities is an open question.

\section{Chiral symmetry selection rules for heavy quarkonium decay into
two mesons}

If chiral symmetry is approximately restored in a given
 hadron state, there should be additional implications of chiral
 symmetry, beyond the spectroscopic patterns and Goldstone boson
 decoupling. One of the implications, reviewed in this section, are
 the chiral symmetry selection rules for the Okubo-Zweig-Iizuka (OZI)
 favored mechanism of the $J/\Psi$ decay into two mesons \cite{GOZI}.

\begin{figure}
\begin{center}
\includegraphics[width=0.6\hsize,clip=]{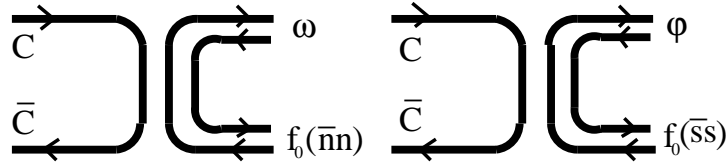}
\caption{The OZI-favored decays into two mesons}
\label{figOZI1}
\end{center}
\end{figure}

\begin{figure}
\begin{center}
\includegraphics[width=0.45\hsize,clip=]{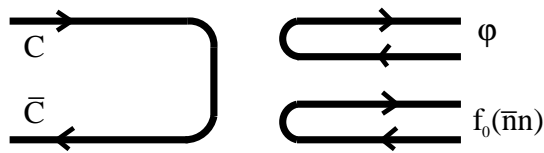}
\caption{The OZI-forbidden decay into two mesons}
\label{figOZI2}
\end{center}
\end{figure}

Consider the OZI-favored and forbidden decays of  charmonium into
two mesons. Let us choose one of the final state
mesons to be $f_0$. Then the OZI-favored and forbidden decay mechanisms are
presented in Fig. \ref{figOZI1} and Fig. \ref{figOZI2}.

Consider for clarity a limiting case, where chiral symmetry is
completely restored in the  $\bar n n$ $f_0$ final state meson. 
This means that its valence
quark wave function is fixed and belongs to the $(1/2,1/2)_a$ representation 
and
along with its chiral partner $\pi$ has the following chiral content

\begin{equation}
 f_0:   \frac{\bar R L + \bar L R}{\sqrt 2},
 \label{f0OZI}
\end{equation}

\begin{equation}
\pi:   \frac{\bar R \vec \tau L - \bar L \vec \tau R}{\sqrt 2},
 \label{piOZI}
\end{equation}

\noindent
where $\vec \tau$ are isospin Pauli matrices,
$L$ is a column consisting of the left-handed $u$ and $d$
quarks, while $R$  is a similar column for the right-handed 
quarks.

\begin{figure}
\begin{center}
\includegraphics[width=0.6\hsize,clip=]{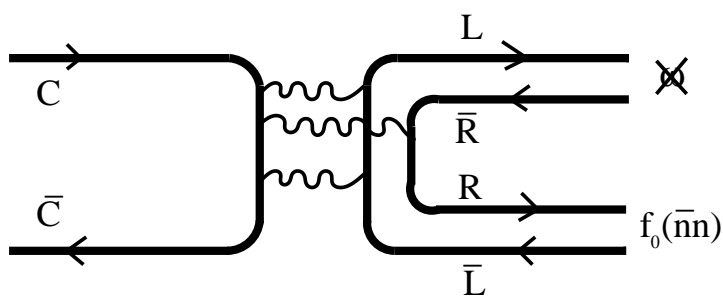}
\caption{The OZI-favored decay which is forbidden by chiral symmetry.}
\label{figOZI3}
\end{center}
\end{figure}

The OZI-allowed decay of $J/\Psi$ or some other heavy quarkonium into two
mesons proceeds via two or three perturbative gluons, depending on the
C-parity of the decaying state, see Fig. \ref{figOZI3}.

\noindent
The $gqq$ perturbative vertex conserves chirality. Then, since 
the final state meson $f_0$ has a definite chiral structure of its
wave function, the $\bar q q$ source for the recoil meson must have
the same chiral structure and the quantum numbers of this source are
fixed. They must be the same as for $f_0$. Certainly, the non-perturbative
quark-gluon dressing in the final state recoil meson can essentially
violate this chiral structure, because chiral symmetry is strongly
broken in the low-lying mesons. However, this non-perturbative interaction
cannot change the quantum numbers and the quantum numbers of the final
state recoil meson must coincide with the quantum numbers of its source.
Then the recoil meson for the OZI-favored mechanism can be only one of the $f_0$
mesons with the dominant $\bar n n$ content and it cannot be 
$\omega,\phi$ or other
mesons. This prediction is true if the final state interaction 
between both final state mesons is inessential.

There are recent experimental data that do confirm this prediction.
The two-meson decay of the charmonium with one of the final state
mesons be $\omega$ or $\phi$, and the other meson be $f_0(1370)$ or
the new $ \bar n n$ state $f_0(1790)$ (seen previously at 1770 MeV 
in $\bar p p$  \cite{BUGG2}) has been studied at BES \cite{BES}. It is
claimed that both states, $f_0(1370)$ and $f_0(1790)$,  are not
seen in the OZI-allowed channel  of Fig. \ref{figOZI1}. In contrast, they are
seen in the OZI-forbidden channel  of Fig. \ref{figOZI2}. This means
that some dynamics must suppress the OZI-allowed mechanism of Fig. 
\ref{figOZI1}. This dynamics, as explained above, 
is the conservation of chirality in the perturbative quark-gluon vertex,
provided that the given $f_0$ meson has a definite chiral structure
(with a small possible breaking),
as illustrated in Fig. \ref{figOZI3}. If this explanation is true, then
the dominant two-meson channels when one of the final
states is $f_0(1790)$ or  $f_0(1370)$  must be 
$J/\Psi \rightarrow f_0 + f_0(1790)$; $J/\Psi \rightarrow f_0 + f_0(1370)$,
because there is no OZI suppression in this case and no restriction from the
chiral symmetry.

Certainly, similar chiral selection rules should be expected in other
decays with one of the final state mesons being in the chirally restored
regime.

\section{Suppression of the glueball - $\bar n n$ mixing}

There is still a controversy, which of the states, $f_0(1500)$ or
$f_0(1710)$ is mostly a glueball, for different scenarios see Ref.
\cite{glue}. A generic reason for this controversy is that a
low-lying glueball must be necessarily mixed with  $\bar n n$
and $\bar s s$ components. It is not possible to avoid such a
mixing for  low-lying glueball. Then a question arises whether
it is possible or not to have a glueball with only a small
admixture of the quark component? The answer is positive, for this
we need to observe a highly-excited glueball in scalar, tensor or
other channels. Since the chiral symmetry breaking
must be suppressed in the highly-excited hadrons, and the mixing
of the glueball with the $\bar q q$ components is proportional to
such a breaking, the mixing of the glueball with the "usual" mesons
must be suppressed \cite{Gglue}. Similar chiral symmetry arguments
are given also in ref. \cite{cha}.

\begin{figure}
\begin{center}
\includegraphics[width=0.4\hsize,clip=]{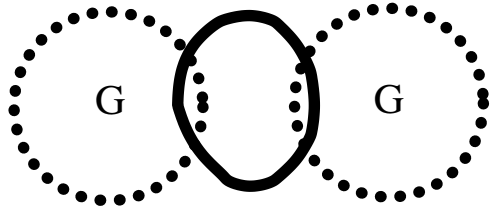}
\caption{The glueball self-energy due to the mixing with the
quark-antiquark components.}
\label{glqq}
\end{center}
\end{figure}

In the highly-excited hadrons the quark loop contributions must
 be suppressed \cite{G}.   Such a suppression is a fundamental reason for
 the chiral and $U(1)_A$ restorations. However, it is the quark loops
 which provide a self-energy contribution to the glueball mass from 
 the coupling of the glueball with the $\bar q q$ components,
 see Fig. \ref{glqq}.  This
 simple argument provides a physical mechanism for "decoupling".

At the formal group-theoretical language the argument goes as follows. The
pure glueball does not contain any $\bar q q$ component
and hence must be the chiral scalar $(0,0)$. Once chiral symmetry
is restored, hadrons belonging to  different representations of the 
chiral group are not mixed. Usual $\bar n n$ $f_0$ states belong to
the $(1/2,1/2)_a$ representations, and hence cannot be mixed with the 
glueball. A similar argument applies to glueballs with other quantum
numbers.

Certainly, the chiral symmetry is not completely "restored" and hence
some amount of mixing is still possible. There are no means at the moment,
unfortunately, to calculate the precise amount of the mixing. What we can
only claim is that asymptotically such a mixing must vanish.

\section{Conclusions and outlook}

We have discussed many different aspects of the effective chiral and $U(1)_A$
restorations in excited hadrons. Among them the experimental evidence,
the chiral classification of excited hadrons and a generic mechanism of chiral
and $U(1)_A$ restoration. This physics has been demonstrated in detail within
the exactly solvable confining and chirally symmetric model. An effective
Lagrangian approach can be used as a phenomenological tool to study
approximate chiral multiplets. Certainly there are other implications
of chiral restoration, beyond the spectroscopical patterns. The most
important one is a decoupling of the approximate chiral multiplets
from the Goldstone bosons. The other ones are the
OZI violations, suppression of the glueball-usual meson mixing, etc.

The most intriguing aspect is that the physics responsible
for the low- and high-lying states is very different. In the former
case the quark loop effects and spontaneous breaking of chiral symmetry
are crucial, the quasiparticle degrees of freedom and the pion cloud
are essential elements of the physics. In essence, the physics
of the low-lying states is a complicated many-channel problem. From
this point of view
 a transition to the string regime
in highly excited states looks very interesting. Indeed, while the 
higher quark Fock components
are very important in the low-lying hadrons, it is the leading one which
determines the physics of the high-lying states. This physics is "simpler"
and we obtain an access to the regime of the dynamical strings. It is
this regime which could shed a light on the problem of confinement.

Does it mean that the question is solved and the problem is closed? Certainly
not. It is only the beginning of the story.
A decisive conclusion can be made only upon discovery of still missing
states. For that a vigorous experimental program of the highly excited
hadrons should be established. Certainly implications of the approximate
chiral symmetry for decays and other observables must be worked out
and confronted with the future experiments.

Among other challenges one has to establish  theoretical tools to study
highly excited hadrons. Lattice will not help us. It is essentially
Minkowskian problem. Analytical methods are required including modeling.
A mathematical description of a hadron as a dynamical string with quarks at the
ends that have definite chirality is a completely open problem. There
must be some "simple" solutions to the problem in view of a large symmetry
which we observe. Certainly these solutions will help us to understand
QCD, the origins of confinement and its interrelation with the chiral symmetry
dynamical breaking. This perspective
allows one to look into the future of the field with optimism.

\section{Acknowledgments}

It is a great pleasure to thank Tom Cohen for a fruitful collaboration,
for stimulating and challenging discussions and for sharing  insights.
I am very grateful to Alexei Nefediev and Robert Wagenbrunn for our
common efforts to get  microscopic insights.

Different topics  have been discussed with many people,
among them notably W. Broniowski,
G. Brown, D. Bugg,  
D. Diakonov, A. Gal, T. DeGrand, R. Jaffe, T. Hatsuda, A. Hosaka,
Yu. Kalashnikova, C. Lang, M. Oka, S. Peris, J. Ribeiro, G. Ripka, D. Riska,
M. Rosina, M. Shifman,  E. Shuryak, Yu. Simonov, A. Vainstein, J. Wambach,
 A. Zhitnitskii.
I am thankful to all of them. 

This work was supported by the Austrian Science Fund (project P19168-N16).


\begin{thebibliography}{9}



\bibitem{GL} M. Gell-Mann, M. Levy, Nuovo Cimento, {\bf 16} (1960) 705.
\bibitem{NJL} Y. Nambu and G. Jona-Lasinio, Phys. Rev. {\bf 122} 
 (1961) 345 ; {\bf 124} (1961) 246.
\bibitem{ioffe} B. L. Ioffe, Nucl. Phys. {\bf B188} (1981) 317;
E: {\bf B191} (1981) 591.
\bibitem{collins} P. D. B. Collins, An introduction to Regge
Theory and High Energy Physics (Cambridge University Press,
Cambridge, 1977)
\bibitem{iachello} F. Iachello, Phys. Rev. Lett. {\bf 63} (1989) 1891.
\bibitem{marianna} M. Kirchbach, Mod. Phys. Lett. {\bf A12} (1997) 2373;
Int. J. Mod. Phys. {\bf A15} (2000) 1435.
\bibitem{G1} L. Ya. Glozman, Phys. Lett. {\bf B475} (2000) 329.
\bibitem{CG1} T. D. Cohen and L. Ya. Glozman, Phys. Rev. {\bf D65} (2002)
016006.
\bibitem{CG2} T. D. Cohen and L. Ya. Glozman,  Int. J. Mod. Phys. 
{\bf A17}  (2002) 1327.
\bibitem{DETAR} C. DeTar and T. Kunihiro, Phys. Rev. {\bf D 39} (1989) 2805.
\bibitem{TIT} D. Jido, M. Oka, A. Hosaka, Progr. Theor. Phys.
{\bf 106} (2001) 873.
\bibitem{JIDO} D. Jido, T. Hatsuda, T. Kunihiro,
 Phys. Rev. Lett. {\bf 84} (2000) 3252. 
\bibitem{novak} M. A. Nowak, M. Rho, I. Zahed, 
Phys. Rev. {\bf D 48} (1993) 4370.
\bibitem{bardeen} W. A. Bardeen, C. T. Hill, 
Phys. Rev. {\bf D 49} (1994) 409.
\bibitem{jaffe} R. L. Jaffe, D. Pirjol, A. Scardicchio,
Phys. Rev. {\bf D 74} (2006) 057901; Phys. Rev. Lett.
{\bf 96} (2006) 121601.
\bibitem{JPS} R. L. Jaffe, D. Pirjol, A. Scardicchio,
Phys. Rep. {\bf 435} (2006) 157.
\bibitem{DM} A. V. Manohar, in: At the frontier of Particle Physics.
Handbook of QCD., Ed. M. Shifman,  v. 1 p. 507 (World Scientific, 2001). 
\bibitem{GR} L. Ya. Glozman, D. O. Riska, Phys. Rep. {\bf 268} (1996) 263;
L. Ya. Glozman, Nucl. Phys.  {\bf A 663} (2000) 103.
\bibitem{BUGG1} A. V. Anisovich et al, Phys. Lett. {\bf B491} (2000) 47;
{\bf B517} (2001) 261; {\bf B542} (2002) 8; {\bf B542} (2002) 19;
 {\bf B513} (2001) 281.
\bibitem{BUGG2} D. V. Bugg, Phys. Rep. {\bf 397} (2004) 257.
\bibitem{G2} L. Ya. Glozman, Phys. Lett. {\bf B539}  (2002) 257.
\bibitem{G3} L. Ya. Glozman, Phys. Lett. {\bf B587} (2004) 69.
\bibitem{afonin} S. S. Afonin, Eur. Phys. J. {\bf A 29} (2006) 327.
\bibitem{G4} L. Ya. Glozman, Phys. Lett. {\bf B541}  (2002) 115.
\bibitem{G}  L. Ya. Glozman, Int. J. Mod. Phys {\bf A 21} (2006) 475. 
\bibitem{GNR} L. Ya. Glozman, A. V. Nefediev, J. E. F. T. Ribeiro,
 Phys. Rev. {\bf D72} (2005) 094002.
\bibitem{orsay} A. Amer, A. Le Yaouanc, L. Oliver, O. Pene, and J.-C.
Raynal, Phys. Rev. Lett. {\bf 50} (1983) 87; A. Le Yaouanc, L. Oliver,
O. Pene, and J.-C. Raynal, Phys. Lett. {\bf 134B} (1984) 249; Phys.
Rev. D {\bf 29} (1984) 1233; Phys. Rev. D {\bf 31}  (1985) 137.
\bibitem{KNR} Yu. S. Kalashnikova, A. V. Nefediev, J.E.F.T. Ribeiro,
 Phys. Rev. D {\bf 72} (2005)  034020.
\bibitem{WG1}  R. F. Wagenbrunn, L. Ya. Glozman,  Phys. Lett.  {\bf
B 643}  (2006) 98.
\bibitem{WG2}  R. F. Wagenbrunn, L. Ya. Glozman, Phys. Rev. D 
{\bf 75} (2007)  036007.
\bibitem{CG3} T. D. Cohen, L. Ya. Glozman, Mod. Phys. Lett. , {\bf A 21}
(2006) 1939.
\bibitem{GN}  L. Ya. Glozman, A. V. Nefediev, Phys. Rev.  {\bf D 73}  
(2006)  074018.
\bibitem{Gglue} L. Ya. Glozman, Eur. Phys. J. {\bf A 19} (2004) 153.
\bibitem{GOZI} L. Ya. Glozman,  Phys. Rev.  {\bf D 73} (2006) 017503.
\bibitem{DEGRAND} T. DeGrand, Phys. Rev. {\bf D 69} (2004) 074024.
\bibitem{coh} T. D. Cohen, Nucl. Phys. {\bf A 775} (2006) 89.
\bibitem{BURCH} T. Burch, C. Gattringer, L. Ya. Glozman, R. Kleindl,
C. B. Lang, A. Sch\"afer, Phys. Rev. {\bf D 70} (2004) 054502;
T. Burch, C. Gattringer, L. Ya. Glozman, C. Hagen,
C. B. Lang, A. Sch\"afer, Phys. Rev. {\bf D 73} (2006) 094505;
T. Burch, C. Gattringer, L. Ya. Glozman, C. Hagen, D. Hierl,
C. B. Lang, A. Sch\"afer, Phys. Rev. {\bf D 74} (2006) 014504;
\bibitem{LHC} S. Basak, R. G. Edwards, G. T. Fleming, J. Juge,
A. Lichtl, C. Morningstar, D. G. Richards, I. Sato, S. J. Wallace,
hep-lat/0609052
\bibitem{ANOMALY} S. L. Adler, Phys. Rev. {\bf 177} (1969) 2426;
J. S. Bell and R. Jackiw, Nuovo Cimento, {\bf A 60} (1969) 47.
\bibitem{FU} K. Fujikawa, Phys. Rev. Lett., {\bf 42} (1979) 1195.
\bibitem{WE1} S. Weinberg, Phys. Rev. {\bf D12} (1975) 3583.
\bibitem{H} G. 't Hooft, Phys. Rev. {\bf D14} (1976) 3432; 
E: {\bf D18} (1978) 2199.
\bibitem{WV} E. Witten, Nucl. Phys. {\bf B149} (1979) 285; 
{\it ibid}, {\bf B156} (1979) 269; G. Veneziano,{\it ibid},
 {\bf B159} (1979) 213.
\bibitem{VW} C. Vafa and E. Witten, Nucl. Phys. {\bf B234} (1984) 173. 
\bibitem{GOR} M. Gell-Mann, R. J. Oakes and B. Renner, 
Phys. Rev. {\bf 175} (1968)  2195.
\bibitem{L} H. Leutwyler, "Chiral dynamics", in: At
the Frontier of Particle Physics/Handbook of QCD, vol. 1, ed. M. Shifman,
World Sc., Singapore, 2001, p. 271-316.
\bibitem{VWNJL} U. Vogl and W. Weise, Progr. Part. Nucl. Phys.,
{\bf 27} 195 (1991); S. P. Klevansky, Rev. Mod. Phys. {\bf 64}
649 (1992); T. Hatsuda and T. Kunihiro, Phys. Rep. {\bf 247}
221 (1994).
\bibitem{SD} T. Sch\"afer and E. V. Shuryak, Rev. Mod. Phys.,
{\bf 70} 323 (1998); D. Diakonov,  Progr. Part. Nucl. Phys.,
{\bf 51} 173 (2003).
\bibitem{A} R. Alkofer and L. von Smekal,  Phys. Rep. {\bf 353}, 281 (2001);
A. Holl, C. D. Roberts, S. V. Wright, nucl-th/0601071.

\bibitem{BES} BES Collab. (M. Ablikim et al), Phys. Lett. {\bf B607}
(2005) 243. 
\bibitem{SVZ} M. A. Shifman, A. I. Vainstein, and V. I. Zakharov,
Nucl. Phys. {\bf B147} (1979) 385.
\bibitem{CJ} T. D. Cohen and X. Ji, Phys. Rev. {\bf D55}  (1997) 6870.
\bibitem{swanson} N. Ligterink, E. Swanson, Phys. Rev. {\bf C69} (2004) 025204.
\bibitem{LANDAU} L.D. Landau and E. M. Lifshitz, Course
of Theoretical Physics, vol. IV, $\S 69$, Pergamon, 1982
\bibitem{PDG} Particle Data Group,  Phys. Lett. {\bf B592} (2004)
1.
\bibitem{WE} S. Weinberg, Phys. Rev. {\bf 166} (1968) 1568.
\bibitem{C} S. R. Coleman, J. Wess and B. Zumino, Phys. Rev. {\bf 177} (1969)
2239.
\bibitem{C1}  C. C. Callan, S. R. Coleman, J. Wess and B. Zumino,
Phys. Rev. {\bf 177} (1969)  2247.
\bibitem{LEE} B. W. Lee, Chiral dynamics, Gordon and Breach, 1972
\bibitem{WSR} S. Weinberg, Phys. Rev. Lett., {\bf 18} (1967)  507.
\bibitem{CGHJO} T. D. Cohen, L. Ya. Glozman, A. Hosaka, D. Jido, M. Oka,
in preparation.
\bibitem{HOOFT1} G. 't Hooft, Nucl. Phys. {\bf B75} (1974) 461.
\bibitem{CCG} C. G. Callan, N. Coote, and D. J. Gross, 
Phys. Rev. {\bf D13} (1976) 054502.
\bibitem{BG} I. Bars and M. B. Green, Phys. Rev. {\bf D17} (1978) 537.
\bibitem{LI} M. Li, Phys. Rev. {\bf D34} (1986) 3888; M. Li, L. Wilets
and M. C. Birse, J. Phys. {\bf G13} (1987) 915.
\bibitem{LENZ} F. Lenz, M. Thies, S. Levit and K. Yazaki,
Ann. Phys. {\bf 208} (1991) 1.
\bibitem{ZHI} A. R. Zhitnitskii, Phys. Rev. {\bf D53} (1996) 5821.
\bibitem{NK} Yu. S. Kalashnikova and A. V. Nefediev,
Phys.Usp. {\bf 45} (2002) 347 (Usp.Fiz.Nauk {\bf 172} (2002)378)
\bibitem{Finger}
  J.~R.~Finger and J.~E.~Mandula,
  Nucl.\ Phys.\ B {\bf 199}  (1982) 168.
\bibitem{BR}  P. Bicudo and J. E. Ribeiro,
Phys. Rev. D {\bf 42}  (1990) 1611; {\bf 42} (1990)  1625.
\bibitem{Gribov} V. N. Gribov, Nucl. Phys. {\bf B 139} (1978) 1.
\bibitem{Zwanziger:2003de}
  D.~Zwanziger,
  Phys.\ Rev.\ D {\bf 70} (2004)  094034.
\bibitem{Greensite:2004ke}
J. Greensite, {\v S}. Olejn{\'{\i}}k, and D. Zwanziger, Phys. Rev. D {\bf 69}
 (2004) 074506.
\bibitem{SS} A. P. Szczepaniak and E. Swanson, Phys. Rev. D {\bf 65} (2001)
025012.
\bibitem{R} C. Feuchter and H. Reinhardt, Phys. Rev. D {\bf 70} (2004) 105021;
 H. Reinhardt and C. Feuchter, Phys. Rev. D {\bf 71} (2005) 105002.
\bibitem{Nakamura:2005ux}
A. Nakamura and T. Saito,
  Prog. Theor. Phys. {\bf 115} (2006)  189.
\bibitem{AL}
R. Alkofer, M. Kloker, A. Krassnigg, and R. F. Wagenbrunn,
Phys. Rev. Lett. {\bf 96}  (2006) 022001.
\bibitem{Adler:1984ri}
S.~L. Adler and A.~C. Davis, Nucl. Phys. B {\bf 244}  (1984) 469.
\bibitem{Alkofer:1988tc}
R. Alkofer and P.~A. Amundsen, Nucl. Phys. B {\bf 306}  (1988) 305.
\bibitem{COT} F. J. Llanes-Estrada and S. R. Cotanch, Phys. Rev. Lett.,
{\bf 84} (2000) 1102.
\bibitem{BN} P. J. A. Bicudo and A. V. Nefediev, 
Phys. Rev. D {\bf 68} (2003) 065021.
\bibitem{morg} V. L. Morgunov, A. V. Nefediev, Yu. A. Simonov, Phys. Lett.
{\bf B 459} (1999) 653. 
\bibitem{glue} C. Amsler, F. Close, Phys. Lett. {\bf B342} (1995) 433;
 N. A. T\"ornqvist, Z. Phys. {\bf C68} (1995) 647;
M. Boglione and M. R. Pennington, Phys. Rev. Lett. {\bf 79} (1997) 1998;
A. V. Anisovich, V. V. Anisovich, A. V. Sarantsev,  
Z. Phys. {\bf A357}  (1997) 123; P. Minkowski and W. Ochs, Eur. Phys. J,
{\bf C9}  (1999) 283; D. Weingarten, Nucl. Phys. Proc. Suppl.,
{\bf 73}  (1999) 249; C. Amsler, Phys. Lett. {\bf B451}  (2002) 22;
V.V. Anisovich, hep-ph/0208123; F. E. Close, Q. Zhao,  Phys. Rev. D {\bf 71}
(2005) 094022.
\bibitem{cha} M. Chanowitz, Phys. Rev. Lett. {\bf 95} (2005) 172001;
Int. J. Mod. Phys. {\bf A21} (2006) 5535.
\bibitem{beane} S. R. Beane,  Phys. Rev. {\bf D 64} (2001) 116010.
\bibitem{afon} S. S. Afonin et al, J. High. Energy Phys., {\bf  04} (2004) 039.
\bibitem{shifm} M. Shifman, hep-ph/0507246, published in: A. Fantoni, S. Liuti,
O. Rondon (Eds.), Quark-Hadron Duality and the Transition to pQCD,
World Scientific, Singapore, 2006 
\bibitem{cata} O. Cata, M. Golterman, S. Peris,   
Phys. Rev. {\bf D 74} (2006) 016001.
\bibitem{BAG} A. Chodos, R. L. Jaffe, K. Jonson, C. B. Thorn, V. Weisskopf,
Phys. Rev. {\bf D 9} (1974) 3471.
\bibitem{IK} N. Isgur, G. Karl, Phys. Rev. {\bf D 18} (1978) 4187;
{\bf D 19} (1979) 2653.
\bibitem{WL} D. Gromes, Z. Phys. {\bf C 26} (1984) 401.
\bibitem{LANDAU3} L. D. Landau and E. M. Lifshitz, v. 3 "Quantum mechanics",
(Pergamon Press, Oxford, 1965).
\bibitem{VEN} G. Veneziano, Nuovo Cimento, {\bf 57A} (1968) 190.
\bibitem{POL} J. Polchinski, String Theory, Cambridge University Press, 1998.
\bibitem{PGRT} P. Goddard, J. Goldstone, C. Rebbi, C. B. Thorn,
Nucl. Phys. {\bf B 56} (1973) 109.

\end{thebibliography}
\end{document}